\documentclass{jfm}

\usepackage{graphicx}
\usepackage{newtxtext}
\usepackage{newtxmath}
\usepackage[round]{natbib}
\usepackage{hyperref}
\hypersetup{
    colorlinks = true,
    urlcolor   = blue,
    citecolor  = black,
}

\newcommand{\RomanNumeralCaps}[1]
\linenumbers

\usepackage{xcolor}
\usepackage{siunitx}
\usepackage{bm}
\usepackage[noabbrev]{cleveref}
\usepackage{subcaption}
\usepackage[normalem]{ulem}

\graphicspath{
  {figures/}
}

\renewcommand{\vec}[1]{\bm{{#1}}}
\newcommand{\tensor}[1]{\bm{{#1}}}
\newcommand{\Dev}[1]{\mathrm{Dev}\left(#1\right)}
\newcommand{\Trace}[1]{\mathrm{Tr}\left(#1\right)}
\newcommand{\Dim}{\mathtt{d}}
\newcommand{\ID}{{\mathbb{I}_\Dim}}

\shorttitle{Revisiting the role of friction coefficients in granular collapses}
\shortauthor{G. Rousseau, T. Metivet, H. Rousseau, G. Daviet, F. Bertails-Descoubes }

\title{
  Revisiting the role of friction coefficients in granular collapses: confrontation of 3-D non-smooth simulations with experiments}

\author{
  Gauthier Rousseau\aff{1,2,4}\corresp{gauthier.rousseau@gmail.com},
  Thibaut M\'etivet\aff{1}\corresp{thibaut.metivet@inria.fr},
  Hugo Rousseau\aff{2,3,5}\corresp{hugo.rousseau@geo.uzh.ch},  
  Gilles Daviet\aff{1}\corresp{gdaviet@gmail.com}, 
  and Florence Bertails-Descoubes\aff{1}\corresp{florence.descoubes@inria.fr}
  }

\affiliation{\aff{1}Univ. Grenoble Alpes, Inria, CNRS, Grenoble INP, LJK, 38000 Grenoble, France
\aff{2} Environmental Hydraulics Laboratory, \'Ecole Polytechnique F\'ed\'erale de Lausanne, Lausanne, Switzerland
\aff{3} Univ. Grenoble Alpes, INRAE, UR ETNA, 38000 Grenoble, France
\aff{4}Institute of Hydraulic Engineering and Water Resources Management, TU Wien, Karlsplatz 13, 1040, Vienna, Austria
\aff{5}Department of Geography, University of Zurich, CH-8057 Zurich, Switzerland
}

\begin{document}

\maketitle

\begin{abstract}
In this paper, transient granular flows are examined both numerically and experimentally. Simulations are performed using the continuous three-dimensional (3-D) granular model introduced in~\citet{daviet2016semi}, which represents the granular medium as an inelastic and dilatable continuum subject to the Drucker--Prager yield criterion in the dense regime. One notable feature of this numerical model is to resolve such a non-smooth rheology without any regularisation.

We show that this non-smooth model, which relies on a constant friction coefficient, is able to reproduce with high fidelity various experimental granular collapses over inclined erodible beds, provided the friction coefficient is set to the avalanche angle - and not to the stop angle, as generally done. 

In order to better characterise the range of validity of the fully plastic rheology in the context of transient frictional flows, we further revisit scaling laws relating the shape of the final collapse deposit to the initial column aspect ratio, and accurately recover established power-law dependences up to aspect ratios in the order of 10.

The influence of sidewall friction is then examined through experimental and simulated collapses with varying channel widths. The analysis offers a comprehensive framework for estimating the effective flow thickness in relation to the channel width, thereby challenging previously held assumptions regarding its estimation in the literature.

Finally, we discuss the possibility to extend the constant coefficient model with a hysteretic model in order to refine the predictions of the early-stage dynamics of the collapse.
This illustrates the potential effects of such phenomenology on transient flows, paving the way to more elaborate analysis.
\end{abstract}


\section{Introduction and related work}
\label{sec:intro}
Granular materials exhibit state-transitions: under specific pressure and shear
rates, the grain assembly may either behave as a fluid, a solid or a gas~\citep{andreotti2013granular}. As an
example, avalanches first flow over complex topographies as viscoplastic fluids
and ultimately stop their course as a static solid deposit. Similarly, during granular
impact cratering, the falling object ejects grains outwards, forming a gaseous
entity for a brief instant. Performing efficient and reliable predictions of
such flows remains challenging and raises several issues, from the fundamental underlying theory to applied industrial or geophysical processes. 

In this context, two families of numerical approaches have been employed: the \textit{Discrete Element Method (DEM)}, which
consists in modelling the dynamics of each grain and its interactions, and the
\textit{continuum-based} method, which describes the granular material at
a macroscopic scale, relying on rheological closures to capture the phase transitions and relate the state of stress to the material deformation.
By modelling inter-grain forces such as friction, adhesion or elasticity, DEM is the
most \textit{ab initio} approach to simulate the behaviour of granular
materials, and has been used extensively to simulate various model experiments
in unsteady~\citep{staron2005study,lacaze2008planar} or steady~\citep{silbert2001granular, da2005rheophysics,azema2014internal} regimes, yielding valuable insights about the relation between the microstructure and the macroscopic properties.
However, DEM inherently suffers from high computational costs, which in practice restricts its use to small-scale systems.
In contrast, continuum methods provide opportunities to simulate larger scale scenarios using depth-averaged~\citep{naaim2003dry, balmforth2005granular, moretti2012numerical} or
full two-dimensional (2-D) or three-dimensional (3-D) ~\citep{lagree2011granular, dunatunga2015continuum, gaume2018dynamic} fluid models, factoring most of the microstructure complexity in macroscopic constitutive laws.

Many studies in the last two decades have therefore focused on the formulation of such rheological closures, using model experiments such as the plane shear or inclined plane set-ups. The main feature of a granular material is the presence of a pressure-dependent yield stress, characterised by an internal friction coefficient $\mu$, which sets the threshold on the maximal sustainable ratio of the shear stress over the pressure before flow, $\tau / p \leq \mu$. Experimental methods found in the literature to estimate the internal friction coefficient of a material are diverse, ranging from the use of the avalanche angle or the angle of repose of a wedge-type pile \citep{hutter1991motion, balmforth2005granular}, to triaxial tests \citep{ancey2001dry,adjemian2004experimental}.

Further analyses of dense granular flows (see \citep{midi2004dense} for a review) has led to the emergence of the $\mu(I)$ rheology introduced by \citet{jop2006constitutive} to describe granular matter in the liquid -- flowing -- regime. This viscoplastic rheology features a simple Drucker--Prager yield criterion, albeit with a non-constant friction coefficient depending on the non-dimensional inertial number $I = \dot{\varepsilon} d \sqrt{\rho/p}$, which naturally arises from dimensional analysis when considering strictly hard grains of size~$d$ and density $\rho$ subjected to some pressure $p$, and strain rate $\dot{\varepsilon}$. 
As such, the $\mu(I)$ rheology is able to capture two crucial features of granular flows: (i)~the existence of a minimal shear stress for the material to keep flowing, described by the \emph{stop friction coefficient} $\mu_{stop}= \tan \theta_{stop}$ (with $\theta_{stop}$ the corresponding \emph{stop angle}, also called \emph{repose angle}) and (ii)~the presence of shear-rate-dependent viscous dissipation at large inertial number $I > I_0 \sim 0.3$, precisely responsible for the existence of steady flows in a range of applied shear-rates.

While extensive testing of the $\mu(I)$ rheology has been performed in steady-flow contexts, the monotonic increase of the friction coefficient with $I$ is unable to describe the transition from rest to flow characterised by the existence of a static friction coefficient larger than the dynamic friction coefficient \citep{hutter1991motion,pouliquen2002friction,da2002viscosity,perrin2019interparticle}. This friction gap results in an hysteresis effect in avalanches: when inclining a container initially filled with static grains, the material starts to flow at a higher \emph{start angle} or \emph{avalanche angle} $\theta_a$ (corresponding to a friction $\mu_a= \tan \theta_a$) than the stop angle.

As suggested in \citet{pouliquen2002friction}, the $\mu(I)$ law could be extended as a decreasing function of $I$ for small $I \in [0, I_*]$ in order to account for a higher start angle while recovering the classical $\mu(I)$ rheology at larger $I$. Exploring the $I<10^{-2}$ behaviour, \citet{degiuli2017friction} pointed out that $\mu(I)$ is indeed non-monotonic, exhibiting a decrease from $\mu(I=0)$ to $\mu( I_* \sim 10^{-3}) = \mu_{stop}$ which amplitude $\Delta_{\mu,hyst}$ is induced by endogenous acoustic noise depending on grain rigidity and applied pressures.

Among the different model experiments available to study granular flows at low inertial number $I$, the simple collapse of a granular column onto a flat or inclined surface is one of the most studied yet distinctive cases \citep{lajeunesse2004spreading,lube2005,balmforth2005granular, lacaze2009axisymmetric, lagree2011granular,  farin2014fundamental,ionescu2015viscoplastic}. Closely reminiscent of natural avalanches or landslides, it challenges both the accuracy of constitutive laws in \emph{friction-dominated}, \emph{unsteady} and \emph{hysteretic} regimes, and the sensitivity of granular flows to boundary conditions (through the presence of frictional side walls or a lifting gate for instance).

Granular collapse features were first studied experimentally \citep{boutreux1997evolution,daerr1999sensitivity,lajeunesse2004spreading,lube2005,balmforth2005granular,mangeney2005use}. In particular, \citet{lajeunesse2005granular,lube2005,balmforth2005granular} focused on the final deposit shape of collapses by studying the impact of the initial aspect ratio $a=H_0/L_0$ (i.e. the ratio between the initial height $H_0$ and length $L_0$ of the granular column) on the final height $H_f$ and run-out $L_f$, unravelling simple power-law relations between the normalised final heights or run-outs, and the aspect ratio $a$, however with different regimes for the exponents, depending on the initial aspect ratio (with a regime transition around $a \approx 3$) or the channel width.

Following these first experimental studies, new numerical models demonstrated a good agreement between DEM simulations and the local $\mu(I)$-rheology for axisymmetric collapses on horizontal planes \citep{lacaze2009axisymmetric}, triggering a series of numerical studies based on continuous models to explore the relevance of the $\mu(I)$ viscoplastic rheology in the context of granular collapses \citep{lagree2011granular,ionescu2015viscoplastic,dunatunga2015continuum,martin2017continuum,chupin2021pressure}, and more generally transient flows. In particular, \citet{lagree2011granular} highlighted the ability of the $\mu(I)$-rheology to recover power-law dependences of the final shape on the initial aspect ratio of the column, albeit with a regime transition around $a \sim 7$, slightly higher than \citet{lajeunesse2005granular} or \citet{lube2005}. Similar results were obtained by \citet{dunatunga2015continuum} with a $\mu(I)$-rheology implemented in a \emph{Material Point Method} (MPM) algorithm.

Although these numerical studies have shown an ability to reproduce collapses with the $\mu(I)$-rheology, it is still however not established whether the well-known aforementioned power-law scaling relations originate from the viscoplastic $\mu(I)$-rheology or could be obtained from a purely plastic rheology.
The performance of the $\mu(I)$-rheology for transient collapses was even questioned by \citet{lagree2011granular}, whose observations indicate that the final deposits of granular collapses could also be recovered with a constant friction coefficient, but at the expense of adjusting the value of the coefficient for large initial aspect ratios.
More generally, while purely plastic rheologies can undeniably not account for some well-known characteristic effects of granular flows (such as for instance the Bagnold velocity profiles observed in inclined plane geometries \citep{bagnold1954experiments,silbert2001granular}), disentangling the respective roles of plastic friction and viscosity appears essential to devise future discriminatory transient flow experiments.

Beyond rheological matters, the sensitivity of granular collapses to frictional sidewalls has also raised concerns, largely disrupting the conclusions about the physical origins of observations \citep{lajeunesse2005granular,balmforth2005granular,ionescu2015viscoplastic,martin2017continuum}.
Containing walls have been shown to play an important role in the analysis of steady inclined plane experiments, owing to the additional lateral frictional condition they involve, which can strongly affect the free-surface flow velocity \citep{jop2005crucial}. In particular, sidewall friction is responsible for the existence of super-stable static inclined piles \citep{taberlet2003superstable} in narrow channels, and can substantially affect the apparent rheological parameters for channels of width $W$ below $\sim 200$ grain diameters \citep{jop2005crucial}. In this context, \citet{ionescu2015viscoplastic} have used the linear scaling law relating the incline flow thickness to the channel width \citep{savage1979gravity,taberlet2003superstable}, originally designed in steady configurations, to support an increase of the effective 2-D friction coefficient and better match the 3-D collapse experiments, albeit without proper definition of a flow thickness in such transient setups. To our knowledge, this two-dimensional rescaling has however never been verified on granular collapses, and an advanced numerical study including direct account of sidewalls effects in 3-D simulations could provide valuable insight into the exact role of the walls in this unsteady configuration.

\subsubsection*{Contributions}

In this paper, we perform 3-D \emph{numerical simulations} at the continuum level, as well as \emph{experiments} of granular column collapses over erodible beds with slopes varying from $\SI{0}{\degree}$ to $\SI{20}{\degree}$. 
We model the granular material as a dilatable, fully plastic continuum with a Drucker--Prager yield surface and non-associated flow rule. We solve the resulting non-smooth rheology without any regularisation by leveraging the {\sc Sand6} software (\href{https://gitlab.inria.fr/elan-public-code/sand6}{https://gitlab.inria.fr/elan-public-code/sand6}), a numerical scheme introduced by \citet{daviet2016nonsmooth,daviet2016semi} and inspired from non-smooth optimisation algorithms originally developed in the context of rigid body contact dynamics~\citep{moreau1994some}. 
In such framework, we can model granular collapses either as viscoplastic or as purely plastic, allowing us to investigate the role of viscosity as well as the relevance of a unique constant friction coefficient.

To this end, we perform numerical simulations of the inclined collapses using the $\mu(I)$-rheology or a constant friction coefficient, set either from the experimental \emph{avalanche} or \emph{stop} angles, and compare them with the experiments, faithfully accounting for the boundary conditions and the lifting door.

Our results confirm the weak impact of a $\mu(I)$-rheology in this particular context, as suggested by \citet{ionescu2015viscoplastic}, but also show that experimental granular collapses with various slopes, flume widths and aspect ratios, are better predicted by using the constant friction coefficient corresponding to the (higher) \emph{avalanche} angle, as opposed to the usual (lower) stop angle provided by the \citet{pouliquen1999scaling} or \citet{pouliquen2002friction} experimental protocols. 

In order to explore the domain in which this result applies, we perform simulations with a constant friction coefficient for various aspect ratios and show that we recover the various regimes in the power-law dependences of the final collapse heights and run-outs proposed by \citet{balmforth2005granular,lajeunesse2005granular,lube2005,lagree2011granular,dunatunga2015continuum}, suggesting that a strain-rate dependant model is not necessary to recover the final run-out for a large range of aspect ratios.

The need for a higher avalanche friction coefficient $\mu_a$ instead of the stop friction coefficient $\mu_{stop}$ used in previous studies is then investigated in light of the results of \citet{ionescu2015viscoplastic}.  Analysing the effect of wall friction on collapses in narrow flumes with variable width, we show that a linear rescaling of the two-dimensional friction coefficient as observed by \citet{savage1979gravity,taberlet2003superstable,jop2005crucial} and used by \citet{ionescu2015viscoplastic} to simulate collapses in two-dimensions remains valid. However, this finding challenges previous assumptions in the literature that overestimated the effective flow thickness using the maximum flow thickness, thus attributing excessive importance to the influence of walls in the considered flume \citep{ionescu2015viscoplastic,martin2017continuum}.

Finally, seeking a way to improve the correspondence between experiment and numerical scenarios at low inertial numbers $I$, we show that a simplified hysteresis rule built in light of previous knowledge \citep{pouliquen2002friction,degiuli2017friction} improves predictions in the incipient motion of the granular collapse, paving the way to more elaborate analysis.

Overall, our conclusions significantly deviate from those of previous works, which model granular collapses by using a $\mu(I)$-rheology based on the stop friction coefficient \citep{lagree2011granular,dunatunga2015continuum,ionescu2015viscoplastic}. Our results may suggest that unsteady collapses are more likely governed by local transitions from rest to flow, and highlight the irrelevance of viscous effects in the context of granular collapses, while steering further studies toward a better quantification of hysteretic laws in such configurations. Along the paper, we support our observations by validating carefully our simulator against experiments and/or prior research in a number of well-controlled scenarios.

The paper is organised as follows: we present in \cref{sec:model} our modelling strategy combining an exact Drucker--Prager yield criterion with usual fluid conservation equations, along with our constraint-based approach to solve for the corresponding non-regularised discrete equations (c.f. \cref{subsec:numericalMethod}). We show in \cref{subsec:DruckerPragerFrictionCoefficientAndYieldTransition} that our numerical model is self-consistent using a numerical avalanche test.
We then describe our different experimental set-ups for studying granular collapses and measuring macroscopic parameters in \cref{sec:experimental}, and validate our numerical model with a constant friction coefficient on model experiments in \cref{sec:validation}, before showing its predictive potential on collapses of natural sediments. 
We further analyse experimentally and numerically the impact of the lateral walls -- in combination with the initial column aspect ratio -- during granular collapses  in \cref{sec:walls}.
Finally, we discuss the possibility to extend the constant coefficient model with a hysteretic model in order to refine the predictions of the early-stage dynamics of the collapse in \cref{sec:hysteresis}.

The maintained C++ {\sc Sand6} software implementing the semi-implicit scheme to solve the continuum conservation equations with the Drucker--Prager rheology, used to study the role of the unique friction coefficient is available at \href{https://gitlab.inria.fr/elan-public-code/sand6}{https://gitlab.inria.fr/elan-public-code/sand6}. Simulations of the experimental collapses were performed using a fork of the {\sc Sand6} code available at \href{https://gitlab.com/groussea/sand6py}{https://gitlab.com/groussea/sand6py} which provides a python binding of the C++ library along with analysis scripts.

\section{A non-smooth numerical model for Drucker--Prager flows}
\label{sec:model}

  In this section, we present our continuum dilatable Drucker--Prager plastic fluid model for granular flows, along with the non-smooth numerical discretisation used to robustly and efficiently simulate true-scale collapses. The numerical model is based on the original reformulation of the Drucker--Prager rheology as a conic constraint introduced in \citet{daviet2016nonsmooth,daviet2016semi}, and implemented in the open-source {\sc Sand6} software.

  In order to assess the physical relevance of this model in the context of transient granular flows, we numerically reproduce a classical avalanche experiment in two and three dimensions, and analyse the role of the Drucker--Prager friction coefficient in regards to the usual experimental avalanche and stop friction coefficients.

\subsection{Continuum modelling}

Focusing on the large deformation regimes, we model the granular material as a continuous yield-stress fluid and introduce the corresponding strain-rate tensor 
\begin{equation}
  \dot{\tensor{\varepsilon}} = \frac{1}{2} \left( \nabla\vec{u} + \nabla\vec{u}^{T} \right)
  \label{eq:strainRateTensor}
\end{equation}
with $\vec{u} \equiv \vec{u}(\vec{x},t)$ the velocity of the material in Eulerian coordinates. 
We also introduce the volume fraction field $\phi(\vec{x},t)$ and assume that all the grains composing the material have a constant density $\rho_g$, so that the density field is 
\begin{equation*}
  \rho(\vec{x},t) = \rho_g \phi(\vec{x},t).
\end{equation*}

The mass conservation equation can then be written in terms of the volume fraction field as:
\begin{equation}
  \frac{\partial\phi}{\partial t} + \nabla \cdot \left( \phi \vec{u} \right) = 0.
  \label{eq:massConservation}
\end{equation}

The momentum conservation equation is 
\begin{equation}
  \rho \left( \frac{\partial \vec{u}}{\partial t} + \left( \vec{u} \cdot \nabla \right) \vec{u} \right) - \nabla \cdot \tensor{\sigma} = \vec{f}
  \label{eq:momentumConservation}
\end{equation}
with $\tensor{\sigma} \equiv \tensor{\sigma}(\vec{x},t)$ is the Cauchy stress tensor and $\vec{f}$ denotes the external volumic forces. 

Note that the volume fraction field $\phi$ takes by definition values between $0$ and some critical value $\phi_c$, which accounts for the maximal ``packing'' fraction of the grains. As such, the granular material is not supposed uniformly dense, and the model can describe the different phases of granular matter.
The varying volume fraction thus tightly couples the mass and momentum conservation equations \labelcref{eq:massConservation,eq:momentumConservation} through the phase-dependent constitutive relation $\tensor{\sigma}\left[\dot{\tensor{\varepsilon}},\phi\right]$ required to close the system.

\subsection{The plastic Drucker--Prager rheology}

The central feature of granular materials is the existence of a pressure-dependent yield stress, somehow reminiscent of the Amontons-Coulomb law for solid friction.
As mentioned above, we consider the material as perfectly plastic, following the Drucker--Prager rheology proposed in \citet{drucker1952soil}. 
Note that our choice for such perfectly plastic modelling naturally allows to overcome the numerical stiffness inherent to the very small time scales resulting from granular elasticity. This is in contrast with the elasto-plastic models of \citet{dunatunga2015continuum,mast2015simulating} or \cite{klar2016drucker}, which are theoretically able to account for more complex physics but require in practice artificially lower elastic stiffnesses in order to ensure numerical workability, thereby weakening the physical meaning of elastic modelling for hard granular material.

Introducing the usual isotropic-deviatoric decomposition of tensors
\begin{equation*}
  \forall{\tensor{\tau}}, \quad
  \tensor{\tau}
  = \frac{\mathrm{Tr}(\bm{\tau})}{\Dim} \,\ID +
  \mathrm{Dev}(\bm{\tau}) \textrm{,}
  \label{eq:isotropicDeviatoricDecomposition}
\end{equation*}
where $\Dim$ is the dimension of the space, we assume that the \emph{dense} state ($\phi = \phi_c$) is governed by the constitutive relation
\begin{subequations}
  \begin{align}
    \lVert \Dev{\tensor{\sigma}} \rVert & \leq \mu p 
    & \quad \textrm{if} \quad \mathrm{Dev}(\tensor{\dot{\varepsilon}}) &=  0
    \quad \textrm{(static regime)} \label{eq:DruckerPragerStatic}
    \\
    \Dev{\tensor{\sigma}} & = \mu p \,
    \frac{\Dev{\tensor{\dot{\varepsilon}}}}{\lVert\Dev{\tensor{\dot{\varepsilon}}}\rVert} 
    & \quad\textrm{if} \quad \mathrm{Dev}(\tensor{\dot{\varepsilon}}) &\neq 0
    \quad \textrm{(flowing regime)}\label{eq:DruckerPragerDynamic}
  \end{align}
\end{subequations}
where we have defined the pressure $p = - \frac{\mathrm{Tr}(\tensor{\sigma})}{\Dim}$
and the Frobenius tensor norm $\lVert \bm{\tau} \rVert \equiv \sqrt{\frac{\tensor{\tau} :\tensor{\tau}}{2}}$ associated to the natural inner product $\langle \tensor{\tau}, \tensor{\upsilon} \rangle = \frac{\tensor{\tau} : \tensor{\upsilon}}{2} = \frac{\Trace{\tensor{\tau}^T \tensor{\upsilon}}}{2}$.
The Drucker--Prager yield surface thus defines a second-order cone in the space of principal stresses (c.f. \cref{eq:DruckerPragerStatic}), which leads to a simpler numerical treatment as compared to the hexagon-like Mohr-Coulomb model. To the best of our knowledge, the phenomenological differences between the Drucker--Prager or Mohr-Coulomb yield models are generally application-dependent, and both surfaces provide very similar results for granular collapses \citep{rauter2020granular}.

We also assume that for $\phi < \phi_c$, the material is in a \emph{disconnected stress-free} state $\tensor{\sigma} = 0$, as proposed in \citet{narain2010free} and \citet{dunatunga2015continuum}. In our case, this can be naturally imposed through the Drucker--Prager rheology by simply constraining the pressure to vanish for $\phi < \phi_c$, which can be written concisely as a complementarity condition
\begin{equation}
  0 \leq \phi_{c} - \phi  ~ \perp ~ p \geq 0
  \label{eq:complementarityRelation}
\end{equation}
or equivalently $\phi \leq \phi_c$, $p \geq 0$ and $(\phi_c - \phi)\,p = 0$.
When the volume fraction is below the packing fraction $\phi_c$, the complementarity relation constrains the pressure to vanish, which imposes $\Dev{\tensor{\sigma}} = 0$ and as a result $\tensor{\sigma} = \Dev{\tensor{\sigma}} - p \ID = 0$.

The resulting constitutive relation is a constrained multi-valued and not everywhere differentiable functional. Solving for the whole system of equations \labelcref{eq:massConservation,eq:momentumConservation} with \labelcref{eq:DruckerPragerStatic,eq:DruckerPragerDynamic,eq:complementarityRelation} is therefore numerically challenging, and is often handled through a regularisation of the rheology to reformulate the problem as a complex fluid as in \citet{lagree2011granular}, \citet{chauchat2014three} or \citet{franci20193d}, or solved using a full elasto-plastic model with explicit or implicit return-mapping projections as \citet{dunatunga2015continuum,mast2015simulating,klar2016drucker}.

The former approach suffers from viscous artefacts such as effective creeping flows, which can affect some physical observations.
The latter approach is able to recover accurately the yield condition at the expense of small time-steps imposed by the small elastic time scale of hard granular material, thereby strongly increasing the computational costs, and intrinsically opposes recompaction after plastic expansion, leading to eventual volume gains. 

One can also mention the approach of \citet{ionescu2015viscoplastic} which is based on an augmented Lagrangian formulation to solve for the corresponding variational inequality, and could therefore in theory deal with the non-differentiability of the rheology but requires in practice a viscous regularisation to ensure viable convergence of the iterative fixed point algorithm.
  
In contrast, our numerical approach is entirely based on non-smooth optimisation tools, and exploits in particular the similarities between the Drucker--Prager yield criterion and the Amontons-Coulomb friction law to leverage recent developments in the field of contact dynamics.

\subsection{Numerical method}
\label{subsec:numericalMethod}

As mentioned above, our simulation framework is based on the 3-D non-smooth numerical model introduced in \citep{daviet2016semi}, which rewrites the plastic Drucker--Prager equations as a non-smooth root-finding problem, and leverages efficient Gauss-Seidel algorithms originally developed in the context of \emph{contact dynamics}~\citep{moreau1994some,jean1999nonsmooth} to solve for the resulting non-linear equation.
Details about this approach can be found in \citet{daviet2016semi,daviet2016nonsmooth}, and are briefly described here for the sake of completeness.

Introducing the material derivative
\begin{equation*}
  \frac{D\bullet}{Dt} \equiv \frac{\partial \bullet}{\partial t} + \vec{u} \cdot \nabla \bullet \textrm{,}
\end{equation*}
we can rewrite the mass conservation equation \labelcref{eq:massConservation} as
\begin{equation*}
    \frac{D\phi}{Dt} + \phi \, \nabla \cdot \vec{u} = 0 \textrm{,}
\end{equation*}
which can be discretised in time with a first-order Euler scheme,
\begin{equation}
  \label{eq:discreteMassConservation}
  \phi^{(n+1)} = \phi^{(n)} - \delta t \, \phi^{(n)} \nabla\cdot\vec{u} \textrm{,}
\end{equation}
where $\phi^{(n+1)} \equiv \phi(\vec{x}(t+\delta t), t+\delta t)$ and $\phi^{(n)} \equiv \phi(\vec{x}(t), t) = \phi(t)$. This allows us to linearise the complementarity relation \labelcref{eq:complementarityRelation} by enforcing the $\phi \leq \phi_c$ constraint on the Lagrangian transported volume fraction, namely
\begin{equation*}
  0 \leq \phi(t) \nabla\cdot\vec{u} + \frac{\phi_{c} - \phi(t)}{\delta t}  ~ \perp ~ p \geq 0 \textrm{,}
\end{equation*}
which is now linear in $\vec{u}$.

Defining the auxiliary tensor fields 
\begin{equation*}
  \begin{aligned}
    \tensor{\lambda} &\equiv -\tensor{\sigma}
    \\
    \tensor{\gamma} &\equiv \phi(t) \, \dot{\tensor{\varepsilon}} + \frac{\phi_{c} - \phi(t)}{\Dim \, \delta t} \,\ID \textrm{,}
  \end{aligned}
\end{equation*}
the whole rheology can be written as
\begin{equation}
  \mathcal{DP}(\mu) \equiv \left\lbrace
  \begin{aligned}
    \lVert \Dev{\tensor{\lambda}} \rVert & \leq \mu \, \frac{\Trace{\tensor{\lambda}}}{\Dim}
    &&\textrm{if} \quad \mathrm{Dev}(\tensor{\gamma}) =  0
    \quad \textrm{(static regime)}
    \\
    \Dev{\tensor{\lambda}} & = -\mu \, \frac{\Trace{\tensor{\lambda}}}{\Dim} \,
    \frac{\Dev{\tensor{\gamma}}}{\lVert\Dev{\tensor{\gamma}}\rVert} 
    &&\textrm{if} \quad \mathrm{Dev}(\tensor{\gamma}) \neq 0
    \quad \textrm{(flowing regime)}
    \\
    0 &\leq \Trace{\tensor{\gamma}} ~ \perp ~ \Trace{\tensor{\lambda}} \geq 0.
  \end{aligned}
  \right.
  \label{eq:fullDruckerPragerRheology}
\end{equation}
As shown in \citep{daviet2016semi,daviet2016nonsmooth}, this rheological law can be recast as a normal cone inclusion in $\mathbb{R}^{s_\Dim}$ with $s_\Dim \equiv \frac{\Dim(\Dim+1)}{2}$ the dimension of $S(\Dim)$ the space of symmetric $\Dim \times \Dim$ matrices. It is thus equivalent to a root-finding problem on a generalised Fischer-Burmeister non-smooth function introduced in \citep{fukushima2002smoothing} and modified in \citep{daviet2011hybrid} to handle non-symmetric and non-associated second-order cone complementarity problems.
We denote here $f_{\textrm{MFB}}$ this modified Fischer-Burmeister function. Additional details regarding the definition of $f_{\textrm{MFB}}$ is given in \cref{appendix:fischerBurmeisterFunction}.
We then have
\begin{equation}
  (\tensor{\gamma},\tensor{\lambda}) \in \mathcal{DP}(\mu) 
  \iff 
  f_{\mathrm{MFB}}(\tensor{\gamma},\tensor{\lambda}) = 0.
\end{equation}

The conservation equations are then discretised in space using an hybrid Finite-Element/Material-Point Method (FEM-MPM) described in \cref{appendix:spatialDiscretization}, finally giving the algebraic problem
\begin{equation}
  \begin{gathered}
  \textrm{Find } \tensor{\gamma}, \tensor{\lambda} \in S(\Dim) \times S(\Dim) \textrm{ s.t.}
  \\
  \left\lbrace\begin{aligned}
    &\tensor{\gamma} = \tensor{W} \tensor{\lambda} + \tensor{b}
    \\
    &(\tensor{\gamma},\tensor{\lambda}) \in \mathcal{DP}(\mu) \textrm{,}
  \end{aligned}\right.
  \end{gathered}
  \label{eq:dual}
\end{equation}
which is solved by minimising the functional 
\begin{equation}
  \mathcal{F}[\tensor{\lambda}] \equiv \frac{1}{2} \lVert f_{\mathrm{MFB}}(\tensor{W} \tensor{\lambda} + \tensor{b},\tensor{\lambda}) \rVert^2
\end{equation}
using a Gauss-Seidel iterative procedure with a generalised Newton algorithm to solve the local problems. The overall algorithm is summarised in \cref{appendix:gaussSeidelAlgorithm}.

Note that the constraint $(\tensor{\gamma},\tensor{\lambda}) \in \mathcal{DP}(\mu)$ in \cref{eq:dual} actually denotes a vector concatenation of local constraints: for each finite-element interpolation node $\vec{x}_i$, we require that the local auxiliary stress and strain rate tensors satisfy the Drucker--Prager yield constraint, i.e. $\forall i, (\tensor{\gamma}(\vec{x}_i),\tensor{\lambda}(\vec{x}_i)) \in \mathcal{DP}(\mu)$. 
The yield constraint is thus only imposed strongly at interpolation points, but not at the continuous nor integral level. In practice, due to the strong non-linearity of the constraint and the MPM interpolation, this can induce small but noticeable constraint violations, which manifest themselves in particular as potential volume losses over the course of long simulations (note that the \textit{mass} is however accurately conserved as imposed by \cref{eq:massConservation}). 
In the following simulations, we have carefully adjusted the numerical parameters to ensure that the total relative volume losses always stay below $3\%$.

\subsection{Drucker--Prager friction coefficient and yield transition}
\label{subsec:DruckerPragerFrictionCoefficientAndYieldTransition}
As shown above, our model solves for a perfectly plastic Drucker--Prager rheology, without regularising the yield criterion. As such, it features a \emph{single} friction coefficient $\mu$ which defines the yield surface, and thereby the transition between the static and flowing regimes. 

To assess the numerical behaviour of our model close to the yield point, we consider a simple avalanche experiment: a two- or three-dimensional dense bed of granular material with height $h$ in a closed box under gravity is quasi-statically inclined, and we measure for each bed inclination $\theta$ the corresponding stabilised granular free-surface $\varphi$.

This simple ``inclined chute'' configuration usually involves two characteristic angles: the \emph{start} -- or \emph{avalanche} angle $\theta_{a}$, which is the maximal possible angle attainable with a static granular bed, and the \emph{stop} angle $\theta_s$, which is the minimal angle needed to sustain flow in the same granular bed (c.f. \citet{savage1979gravity,daerr1999two,pouliquen2002friction,artoni2011hysteresis}). In our model however, we only parameterise the yield transition using a \emph{single} friction coefficient. As such, we do not expect to recover the full hysteretic phenomenology of such experiments, but rather seek to characterise our Drucker--Prager rheology from a macroscopic point of view, and to provide a way to set the friction coefficient $\mu$ in the context of granular collapses.

We consider in practice a 2-D (or 3-D) box of dimensions $\SI{1}{m} \times \SI{0.3}{m}$ ($\times\, \SI{0.1}{m}$), discretised with a cartesian mesh with resolution $250 \times 150$ (resp. $100 \times 30 \times 10$), filled with a $\SI{0.1}{m}$-high bed of granular material with slip boundary conditions on the sides and stick boundary conditions at the bottom of the box.
To minimise the effect of boundary conditions, we measure the angle of the free surface between $x = 0.3 \,\si{m}$ and $0.7 \,\si{m}$ using a $\tan^{-1}$ on the end-point coordinates of a line fit. The simulation time-step is chosen as $\delta t = 1 \,\si{ms}$ and we increase $\theta$ by steps of $\SI{0.5}{\degree}$ ensuring equilibrium is always reached between each change of inclination.
Note that the simulated box is actually kept fixed and we vary the inclination $\theta$ by rotating the external gravity force.

\begin{figure}
  \centering
  \includegraphics[width=\textwidth]{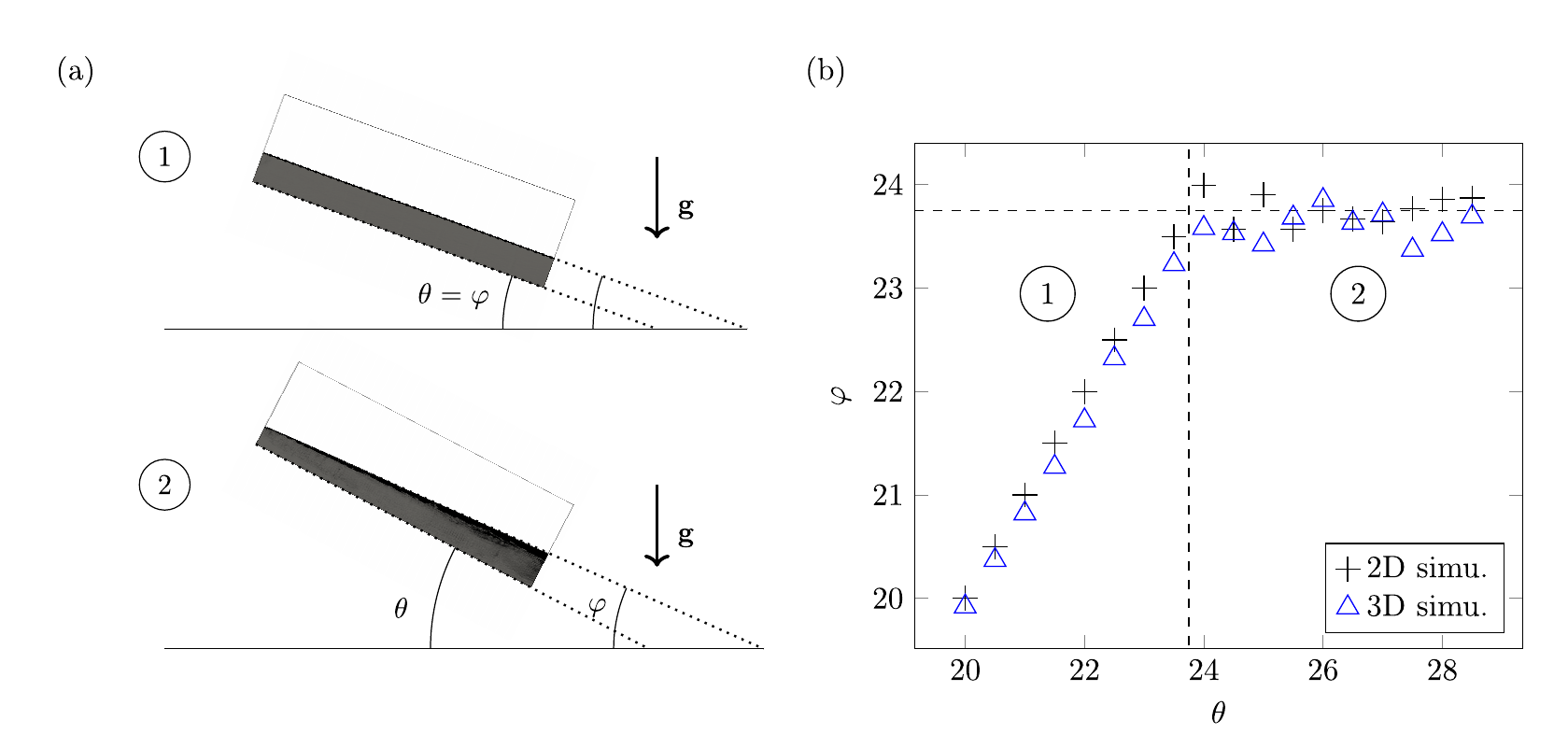}

  \caption{Inclination $\varphi$ of the stabilised granular bed free surface as a function of the box inclination $\theta$ for the avalanche numerical setup obtained in our 2-D and 3-D simulations. (a) Simulation snapshots of the 2-D granular bed for $\theta=23^\circ$ and $27^\circ$. (b) $\varphi$ vs $\theta$, the vertical and horizontal dashed lines indicate the friction angle used in the simulation $\tan^{-1}(0.44) \approx 23.75^{\circ}$.}
  \label{fig:avalanchePhiVSTheta}
\end{figure}
\Cref{fig:avalanchePhiVSTheta} shows the results for the free surface angle $\varphi$, once the surface has stabilised, as a function of $\theta$ for a simulation with $\mu=0.44$. 
As we increase the bed inclination $\theta$, we observe that the free surface first remains aligned with the bed ($\varphi = \theta$); the material is jammed and remains in static equilibrium. After some critical inclination $\theta_a$ (the so-called \textit{avalanche} angle), the material starts to flow before stopping. If we again increase quasi-statically the bed inclination, the same happens again, and the angle of the stabilised free surface basically takes a constant value, corresponding to the so-called \textit{stop} angle $\varphi_s = \theta_s$ for $\theta > \theta_a$.

As expected from our simple non-hysteretic Drucker--Prager model, we observe that the avalanche and stop angles are indistinguishable. Furthermore, as shown in \cref{fig:avalanchePhiVSTheta} (b), both angles correspond precisely to the friction angle prescribed by the friction coefficient $\mu$ of the Drucker--Prager law, as $\theta_a = \theta_s = \theta_{\mu} = \tan^{-1}(\mu)$. 

On the one hand, these two observations demonstrate that our numerical model provides a consistent discretisation of the Drucker--Prager law, and gives us confidence in using the corresponding {\sc Sand6} code for further experiments. On the other hand, this elementary numerical experiment makes it clear that modelling granular material with a simple plastic Drucker--Prager rheology requires to be careful in the choice of the -- constant -- friction coefficient.

In the following, we show that our numerical model, which features a single friction parameter, is sufficient to predict the macroscopic flowing of a real granular collapse, provided the friction coefficient is chosen as the \emph{avalanche} angle $\theta_a$ of the real granular material. This observation, which departs from the usual choice for setting the friction parameter (the stop angle), is supported by complementary validations and comparisons of our model throughout the paper, and eventually discussed. 

\section{Experimental collapses and measure of macroscopic parameters}
\label{sec:experimental}

\begin{figure}
  \begin{center}
    \includegraphics[width=\linewidth]{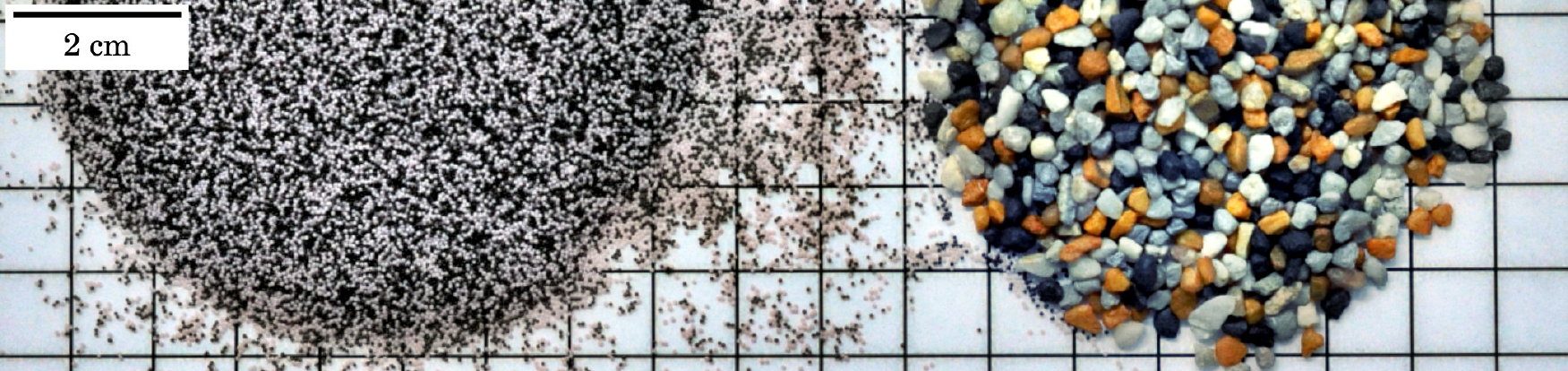}
    \caption{Photograph of the two granular materials used for our validation experiments: glass beads on the left, and natural irregular granules on the right. Multicoloured material improves feature detection, hence velocimetry performance. }
    \label{fig:materials}
  \end{center}
\end{figure}

In order to validate our non-smooth simulator in transient flow configurations, we have conducted various experiments of granular column collapses in a $\SI{6}{cm}$-wide channel. Our experiments are performed on various inclinations of the channel bed and using two different materials, which we describe in the following. To relate our physical experiments to our numerical simulations, we also need to measure appropriately, from the experiments, all the -- macroscopic -- parameters at play in the continuum model: in particular the yield friction coefficient $\mu$ of the Drucker--Prager rheology, but also the friction coefficient between the granular medium and the lateral walls. These measurement processes are discussed in the next section.

\subsection{Materials}
We consider two different materials: \SI{0.5}{mm} glass beads (Sigmund Linder - SiliBeads - 45015) and \SI{2.7}{mm} natural granules. Their aspects are shown in \cref{fig:materials}, and their physical parameters -- which we have experimentally measured -- are summarised in \cref{tab:materialsParameters}.

Our \SI{0.5}{mm} glass beads share the same geometric and physical features as the glass beads used in \citet{pouliquen1999scaling,pouliquen2002friction,jop2005crucial}. The latter references represent seminal experimental works for calibrating the $\mu(I)$ constitutive law of \citet{jop2006constitutive}, relying on a steady granular flow in a wide channel with various inclinations. In \cref{subsec:measureMacro}, we verify experimentally that our beads behave in accordance with the results reported in the aforementioned literature, when subjected to a similar setting. We note that our beads are also similar to those used for the granular collapses studied in \citet{mangeney2005use,farin2014fundamental}, whose experimental results are exploited to test modelling assumptions formulated by \citet{ionescu2015viscoplastic} and later by \citet{martin2017continuum}. Aligning our choice on this material thus makes it possible to check such assumptions and better transfer insights. For this reason we have used this well-calibrated model material preferably for extensive and systematic validation, as reported in \cref{subsec:validationInclinations}.

In order to test the robustness of our model on natural materials --- generally exhibiting larger friction coefficients than spherical glass beads, we have also used coarse natural granules (\SI{2.7}{mm} irregular grains). 
The extension of our validation study to the collapse of such natural granules is shown in \cref{fig:validationInclinationsGranules}.

\begin{table}
  \begin{center}

    \begin{tabular}{l@{\hskip 0.2in}|cc@{\hskip 0.3in}c@{\hskip 0.3in}c@{\hskip 0.3in}c}
        type        & $d$ (mm) & $\rho_c$ (kg m$^{-3}$) & $\mu_a$ & $\mu_{stop}$ & $\mu_w$ \\
        \hline
        glass beads (B) & 0.5  & \num{1.47e3}   & 0.44 $\pm$ 0.03  &  0.37 $\pm$ 0.03    & 0.23  $\pm$ 0.05   \\
        granules (G)   & 2.7  & \num{1.52e3}  &  0.75 $\pm$ 0.1  &  0.65 $\pm$ 0.1    & 0.3  $\pm$ 0.1    \\
      \end{tabular}

    \caption{Measured material and rheological parameters for the two granular materials used in our experiments: $d$ is the median grain diameter, $\rho_c$ the density of the material in the dense state, $\mu_a$ the avalanche friction coefficient, $\mu_{stop}$ the stop friction coefficient, and~$\mu_w$ the estimated friction coefficient between the granular material and the lateral walls.
    Note that the density $\rho_c$ is measured on the granular media in its dense state and not on the composing grain, and corresponds as such to a direct measurement of $\rho_c = \rho_g \phi_c$.}
    \label{tab:materialsParameters}
  \end{center}
\end{table}

\subsection{Collapse set-up}
\label{subsec:collapseSetup}

\begin{figure}
  \begin{center}
  
  \includegraphics[width=\textwidth]{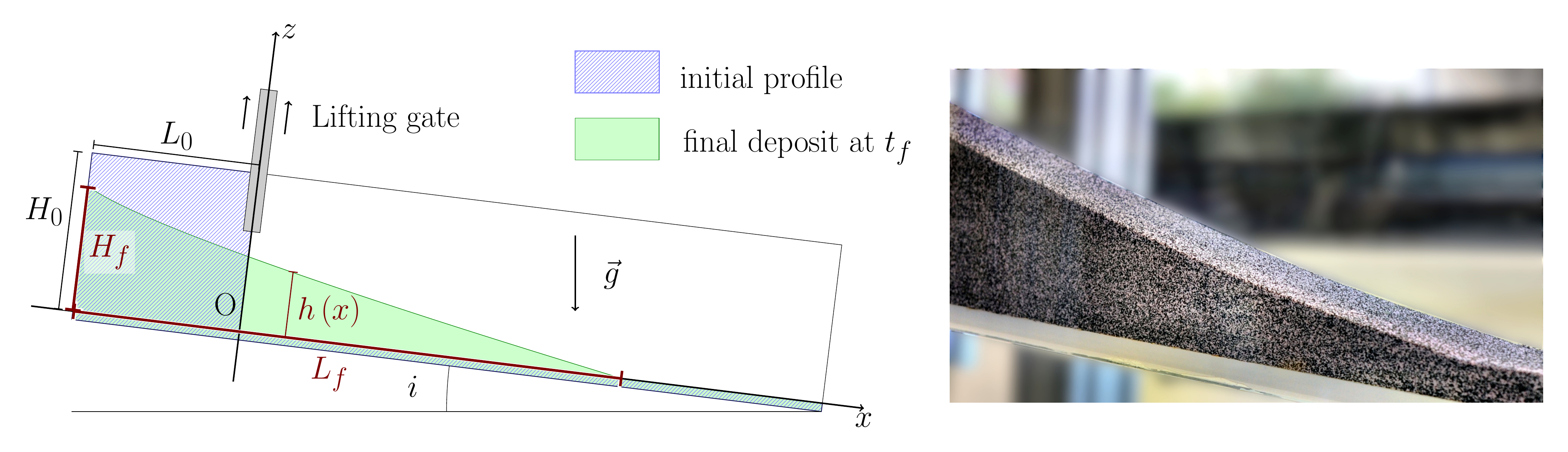}

  \caption{[left] Sketch of our granular column-collapse experimental configuration showing the initial column dimension ($L_0$,$H_0$) and final deposit state (run-out  $L_f$ and final upslope height $H_f$). Collapses are triggered by a pneumatic lifting gate. [right] Photograph of the final deposit.}
  \label{fig:sketch-collapse}
  \end{center}
  
  \end{figure}

  \begin{figure}
    \begin{center}
    
      \includegraphics[scale=0.9]{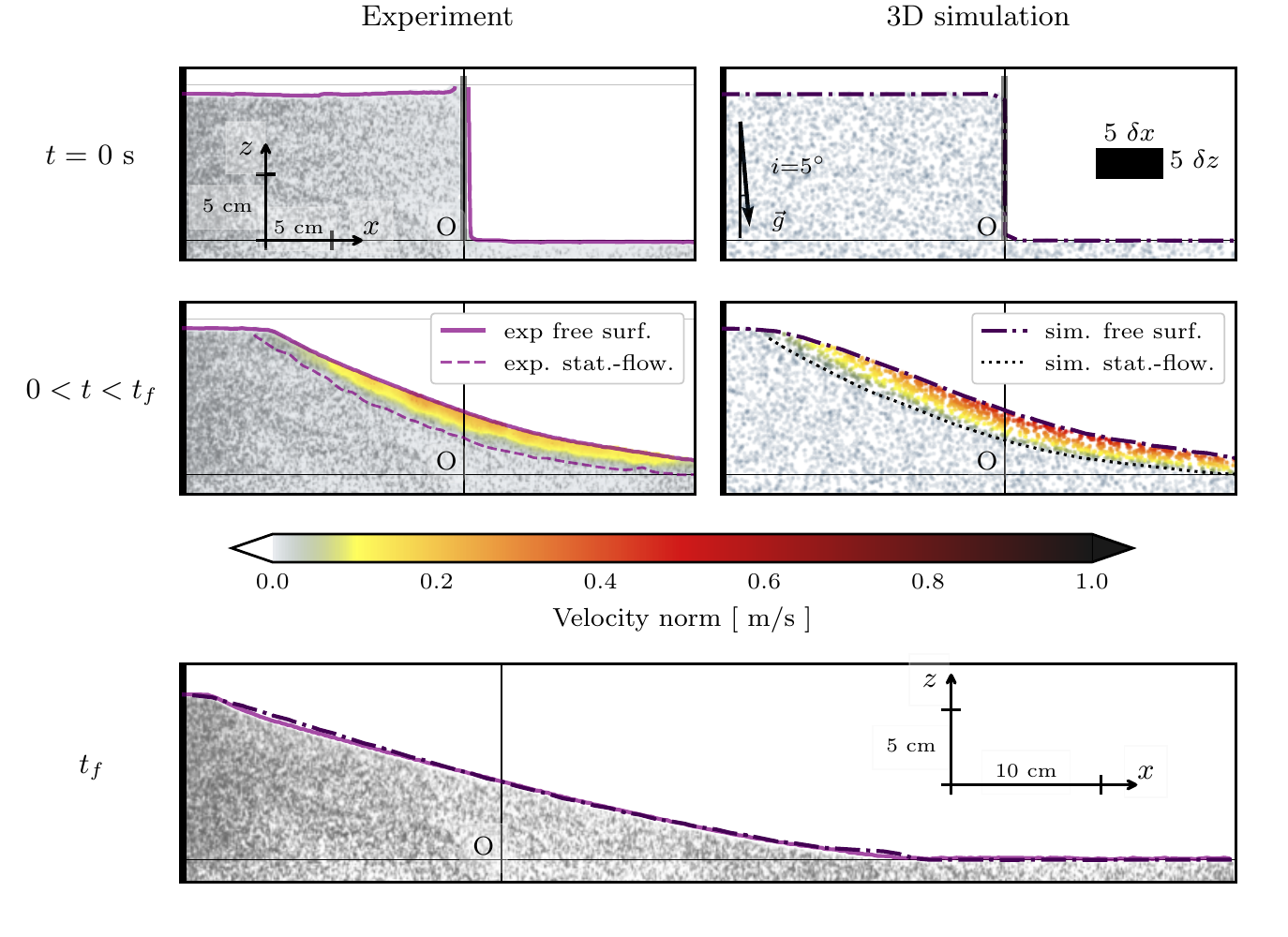}

    \caption{Capturing the free surface and velocity field from experimental and simulated granular collapses at different instants. Camera snapshots are shown in left column and bottom figures where experimental velocities are extracted using image velocimetry and shown on the snapshots foreground.  The static-flowing transition corresponds to the \SI{0.01}{m/s} contour of the experimental and numerical velocity fields. The simulated material points are shown on the right column and coloured according to their velocity magnitude. The black rectangle on the top left figure has dimensions $5 \, \delta_x \times 5 \, \delta_z$ with $\delta_x$ and $\delta_z$ the respective horizontal and vertical MPM resolutions. Data correspond to the B05 collapse (\SI{0.5}{mm} beads -- $i=\SI{5}{\degree}$). See movie~1 in the online supplementary material for a visualisation of the experimental and simulated evolution in time.}

    \label{fig:expProcedureVsNumeric}

    \end{center}
    
    \end{figure}

The experimental apparatus is inspired by \citet{balmforth2005granular} and sketched in \cref{fig:sketch-collapse}. An initial granular pile with aspect ratio $a=H_0/L_0\sim 0.5$, where $H_0$ is the depth and $L_0$ is the length along the channel, is confined in a $W = \SI{6}{cm}$ wide flume with a lifting gate preventing the pile to collapse. The experiment consists in opening the pneumatic lifting gate to trigger the granular collapse. Similarly to the second set of experiments of \citet{farin2014fundamental}, the base of the box is covered with an horizontal erodible granular bed of $\sim \SI{2}{cm}$ height made of the same material. 

Free surface profiles as well as velocities are collected using a high-speed camera from the transparent side walls with an interframing time of \SI{5.3}{ms}. We estimate the local optical flow of good feature to track \citep{shi1994good,miozzi2008performances}, i.e. features having optimal contrasts to be tracked (enhanced by the presence of multicoloured materials as shown in \cref{fig:materials}). We interpolate feature velocities on a mesh to produce the velocity fields using the \textit{opyf} python package. The reader may refer to Annex~1 of \citet{rousseau2020scanning} for details on the velocimetry procedure.

\Cref{fig:expProcedureVsNumeric} illustrates the methodological data used to compare experiments and simulations in this article: the upper left column shows snapshots of the recorded experimental collapse for the initial and some intermediate state, coloured with the velocity magnitude as measured by velocimetry, while the upper right column shows the corresponding snapshots of the simulation, where the background medium and velocity are displayed by the Material Points quantities. We superpose on all these snapshots the post-processed profile height and static-flowing transition lines, respectively determined as the $0.5$ contour on the volume fraction $\phi$ and $\SI{0.01}{m/s}$ contour on the velocity. The bottom figure illustrates the direct comparison of the height profiles in the final state.

\subsection{Measurement of macroscopic parameters}
\label{subsec:measureMacro}

As mentioned in \cref{subsec:DruckerPragerFrictionCoefficientAndYieldTransition}, the flowing threshold in our Drucker--Prager constitutive law is determined by a single friction coefficient $\mu = \tan \theta$. Yet, there is still no clear consensus regarding which of the avalanche angle $\theta_a$ and the stop angle $\theta_s$ should be used as the yield angle. In order to discriminate between these two options, we first need to measure both parameters for the granular media used in our experiments. Since the role of sidewalls is also investigated, we furthermore evaluate the friction coefficient between the granular medium and the lateral walls $\mu_w$.

\subsubsection{Measurement of the avalanche friction coefficient}
\label{subsec:avalancheAngle}

\begin{figure}
  \begin{center}
  
  \includegraphics[width=\textwidth]{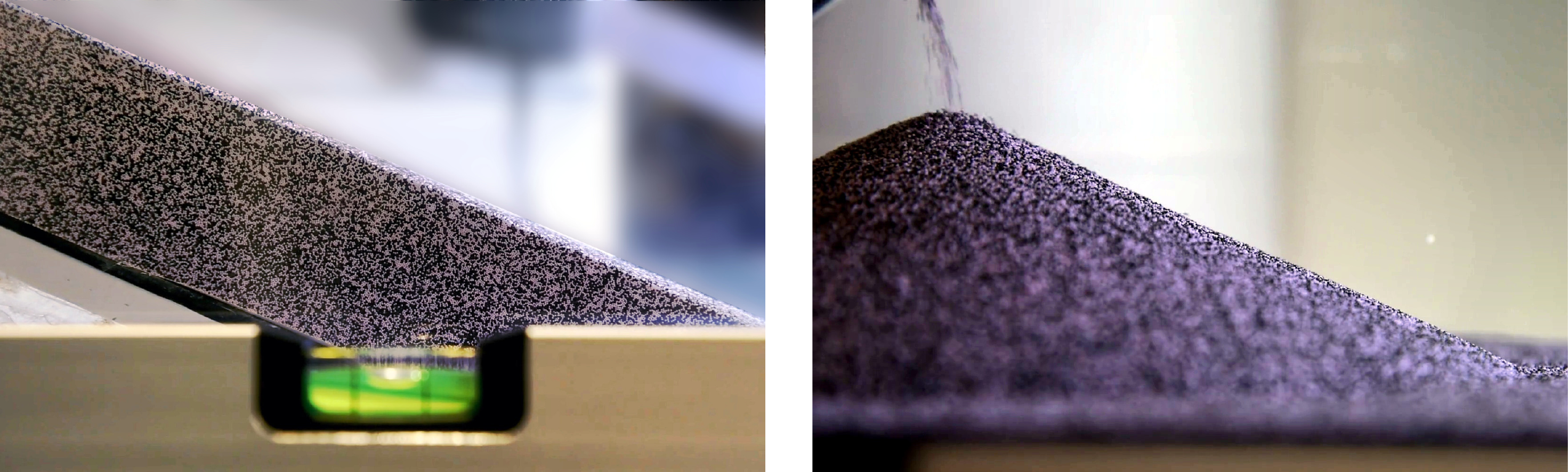}

  \caption{The avalanche angle $\theta_a$, used to set the yield friction coefficient $\mu$ of the Drucker--Prager model as $\mu = \mu_a = \tan \theta_a$ using either the large channel inclination test (left), or the conical heap test (right). For our 0.5 mm glass beads depicted here, both protocols converged to a same average measurement value ($ \tan \theta_a = \tan \SI{23.7}{\degree} = 0.44 \pm 0.03$) }
  \label{fig:measurementAvalancheAngle}
  \end{center}
  
\end{figure}
Our first experiment for measuring the avalanche angle $\theta_a$, is based on the configuration used by \citet{balmforth2005granular} for measuring the ``internal'' angle of friction -- a terminology which is then equivalent to our avalanche angle from an experimental point of view. Similarly to the numerical set-up presented in \cref{subsec:DruckerPragerFrictionCoefficientAndYieldTransition}, we consider a \SI{20}{cm} wide flume filled with the granular material and inclined the flat granular bed until motion downslope begins (see \cref{fig:measurementAvalancheAngle}, left). Our second experiment consists in pouring slowly granular material upon a conical heap (see \cref{fig:measurementAvalancheAngle}, right).  In both experiments, we measure $\theta_a$ as the critical angle corresponding to the maximal slope of the material, just before an avalanche occurs and causes the slope to decrease (see movie~2 in the online supplementary material for an illustration of the process).
As emphasised by \citet{balmforth2005granular} or \citet{russell2019retrogressive}, there is a large spread in the $\theta_{a}$ measurements owing to the sensitivity of the protocol to small perturbations. 

For the \SI{0.5}{mm} glass beads material, no matter the experimental protocol used (inclined channel or conical heap), we found the avalanche angle $\theta_{a} = \SI{23.7}{\degree} \pm 1$, which corresponds to the friction coefficient $\mu_a = \tan \theta_{a} \approx 0.44 \pm 0.03$. 
This value is consistent with those made in the literature for similar materials, such as the \SI{0.7}{mm} or \SI{0.8}{mm} glass beads of respectively \citet{farin2014fundamental} and \citet{balmforth2005granular} which give $\tan \theta_a = \tan \SI{25}{\degree} = 0.47 \pm 0.01$ and $\tan \theta_a = \tan \SI{24.5}{\degree} = 0.46 \pm 0.04$. 
For the \SI{2.7}{mm} granules, we have only conducted the experiment with the first protocol, and found $\theta_{a} = \SI{36.8}{\degree} \pm 5$ ($\mu_a = 0.75 \pm 0.1$).

\subsubsection{Measurement of the stop friction coefficient}
\label{subsubsec:stopFrictionCoeff}

As mentioned above, another feature of granular flows is the existence of a minimal angle necessary to sustain motion in an already flowing material. In the case of a granular layer with constant height flowing on rough inclines, this \textit{stop} angle depends on the height $h$ as a function $\Theta(h)$ \citep{pouliquen1999scaling,pouliquen2002friction}, which naturally converges toward a constant value as the height increases, namely $\theta_s = \Theta_s(h \to \infty)$.

This phenomenology has led to nowadays well-established protocols to measure $\theta_s$ in a steady granular flow, in which either the bed with steady flow of known height is progressively lowered until flow ceases~\citep{pouliquen1999scaling}, or the initially static bed is inclined up until it starts to flow, and then stops~\citep{pouliquen2002friction}.

\begin{figure}
  \centering
  \includegraphics[width=\textwidth]{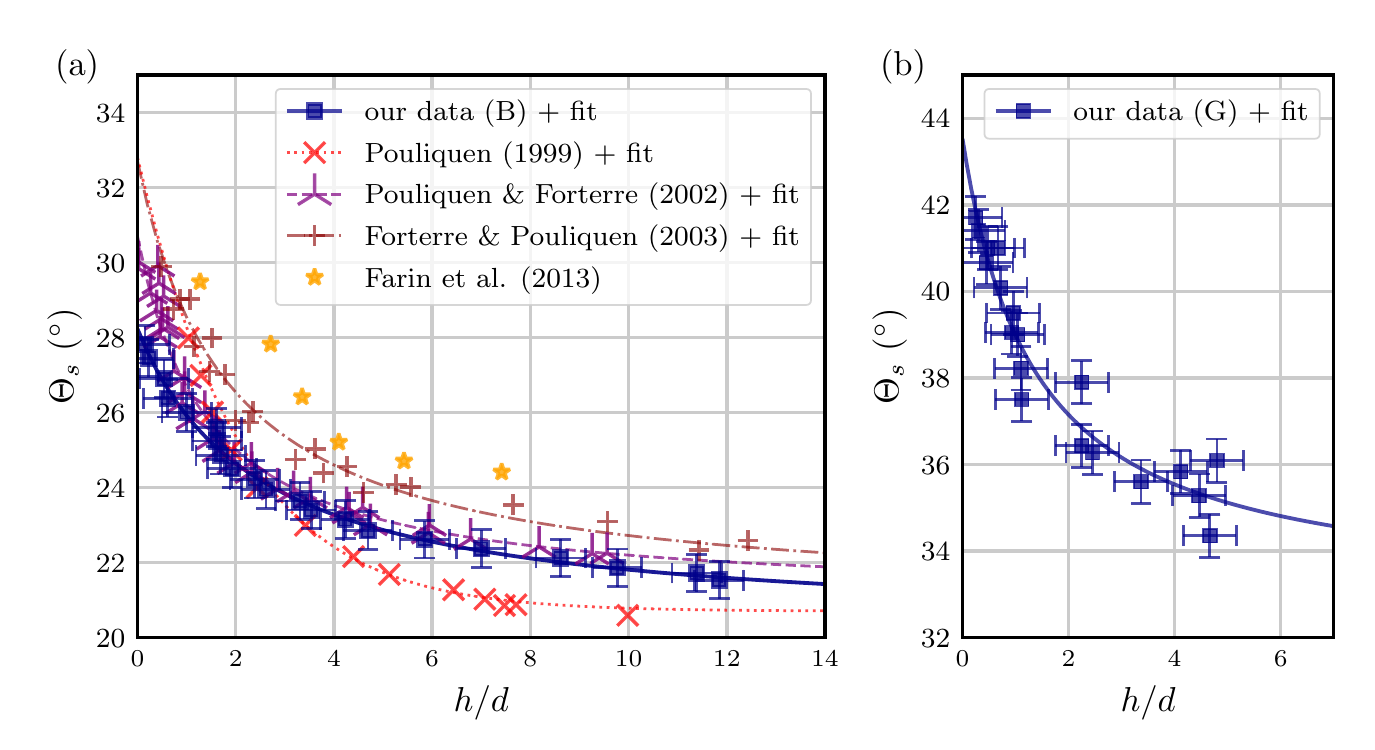}

  \caption{(a) Variation of the stop angles $\Theta_s$ as a function of the non-dimensional thickness $h/d$ for our glass beads (B) of diameter $d=\SI{0.5}{mm}$, measured using the protocol described in \citet{pouliquen2002friction}. The plain curve corresponds to the best fit of our data with the function $\tan(\Theta_s(h)) = \mu_1 + (\mu_2 - \mu_1)/(1+h/L)$, obtained for $\mu_1 = 0.37$, $\mu_2 = 0.54$ and $L = \SI{1.16}{mm}$. We also report the data and fits of \citet{pouliquen2002friction}, \citet{pouliquen1999scaling}, \citet{forterre2003long} and \citet{farin2014fundamental} for the sake of comparison. Curves do not all collapse perfectly, which may be explained by slight differences between experimental conditions, protocols or materials. Overall, our data look consistent with the previous studies performed with $\SI{0.5}{mm}$ glass beads. (b) Same measurements made for our granules (G) with a best fit obtained for $\mu_1 = 0.65$, $\mu_2 = 0.95$ and $L = \SI{3.1}{mm}$.}
  \label{fig:thetastop}
\end{figure}

We have replicated the steady inclined bed protocol experimentally with our \SI{0.5}{mm} beads and \SI{2.7}{mm} granules, and have measured the resulting stop angle as in~\citet{pouliquen2002friction} using a \SI{3}{m}-long, \SI{10}{cm}-wide channel ($W/d = 200$) for beads and a \SI{2}{m}-long, \SI{12}{cm}-wide channel for granules ($W/d = 44$).

While our channels are narrower than the one used in \citet{pouliquen2002friction}, the maximal characteristic flowing heights $h$ are systematically small w.r.t the channel $W$, with $h/W \sim 0.06$ for both beads and granules. Following \citet{jop2005crucial}, side wall friction can thus result in a small overestimation of the stop friction coefficient of order $\mu_w h/W \sim 0.02$ in both cases.

Our results for beads are shown in \cref{fig:thetastop} (a), together with the experimental data and corresponding fits of \citet{pouliquen1999scaling,pouliquen2002friction, forterre2003long,farin2014fundamental}. We observe that despite some spreading in all the collected data, which might be attributed to the different experimental protocols used -- \citet{pouliquen2002friction} vs. \citet{pouliquen1999scaling} -- and/or to the slightly different beads used -- \citet{pouliquen2002friction} vs. \citet{farin2014fundamental} --, we obtain a good agreement with \citet{pouliquen2002friction}, in particular regarding the $h/d_p \to +\infty$ asymptote, which corresponds to the $I \to 0$ stop angle. 

Indeed, we retrieve a stop friction coefficient $\mu_{stop} = \tan(\Theta_s(\infty)) = \mu_1 = 0.37$ which is very close to the values for the same type of beads, $\mu_{stop} = 0.38$~\citep{pouliquen1999scaling,pouliquen2002friction,jop2006constitutive}. Accounting for the $\SI{10}{cm}$ width of our channel, we thus deduce that our actual stop friction coefficient is slightly less than $0.37$. In practice, for our comparisons we shall stick to $\mu_{stop} = 0.38$ (see \cref{subsec:comparisonMuIRheology}).

Following the same measurement protocol with our granules, we find a stop friction coefficient of $\mu_{stop} = 0.65 \pm 0.1$ for this natural sediment (see \cref{tab:materialsParameters} and \cref{fig:thetastop} (b)).

These results confirm that our beads behave fairly similarly to the ones from \citet{pouliquen2002friction}, which enables direct comparison of our results with other works. 

It is noteworthy that whatever the material used, the estimated stop friction coefficient ($\mu_{stop} = 0.38$ and $\mu_{stop} = 0.65$ for the beads and granules, respectively) is significantly lower compared to the measured avalanche coefficient ($\mu_a = 0.44$ and $\mu_a = 0.75$ respectively). In \cref{sec:validation}, we find that using $\mu = \mu_a$ in our simulations yields more realistic collapses than taking $\mu = \mu_{stop}$.

\subsubsection{Measurement of the friction coefficient between the granular medium and the lateral walls}
\label{subsubsec:wallsFrictionCoeff}

Following \citet{balmforth2005granular, hutter1991motion}, we estimate the friction coefficient $\mu_w$ between the glass walls and the granular materials by measuring the critical angle above which a block of particles held together within a rigid light plastic cylinder begins to slide on an inclined glass surface. The cylinder is \SI{4}{cm} in diameter and \SI{6}{cm} high. 
We obtain $\mu_w = \tan(\SI{13}{\degree}) = 0.23 $ for the $\SI{0.5}{mm}$ glass beads and $\mu_w = \tan(\SI{17}{\degree}) =  0.3$ for the $\SI{2.7}{mm}$ granules. These values are reported in \cref{tab:materialsParameters} along with the corresponding estimated standard deviations.

\section{Validation: numerics vs. experiments}
\label{sec:validation}

To evaluate the performance of our simulation approach and assess the validity of the Drucker--Prager rheological law -- featuring only one friction coefficient -- in the case of transient collapse flows, we perform direct 3-D simulations of the collapse setups presented in \cref{subsec:collapseSetup}, with full account of frictional interactions with the encasing walls and the lifting door.

We first compare the simulation results with the experiments for the glass beads (B), in order to validate our numerical model with a controlled material, and then highlight its robustness by considering the case of natural irregular granules (G) (c.f. \cref{fig:materials}).

\subsection{Simulation parameters}
\label{subsec:physicalParameters}

The rheological parameters used in our simulations are derived from independent experimental measurements performed with the same materials. As discussed in~\cref{subsec:avalancheAngle}, the Drucker--Prager friction coefficients are determined using the respective avalanche angles, measured consistently with both the channel and heap setups.
To reduce uncertainties on the actual granular dense packing fraction, the density is directly measured for the granular media in the dense state, and thus corresponds to the critical density $\rho_c = \rho_g \phi_c$. This amounts to rescaling the volume fraction field $\phi$ by its dense critical value $\phi_c$, so as to ensure $\rho_c \frac{\phi}{\phi_c} = \rho_g \phi = \rho$. As such, we take in the simulations $\phi_c = 1$. The values for the material parameters and the friction coefficients are summarised in \cref{tab:materialsParameters}.

Gravity is set to its standard value $g = \SI{9.81}{m.s^{-2}}$, and aligned with a rotated downward unit vector to account for the various respective inclinations of the collapse beds.

The $152 \times 15 \times 6$ \SI{}{cm^3} simulation domain is discretised with a $152 \times 30 \times 6$ mesh of regular rectangular cuboids, which gives a $\SI{1}{cm}$ length and width resolution, and a $\SI{0.5}{cm}$ resolution for the height, in order to better capture the free surface. Top right plot in \cref{fig:expProcedureVsNumeric} illustrates the resolution of the background MPM-grid: the black rectangle has dimensions $5 \, \delta_x \times 5 \, \delta_z$ where $\delta_x$ and $\delta_z$ are the horizontal and vertical resolutions respectively.

\begin{figure}
  \centering
  \includegraphics[scale=0.8]{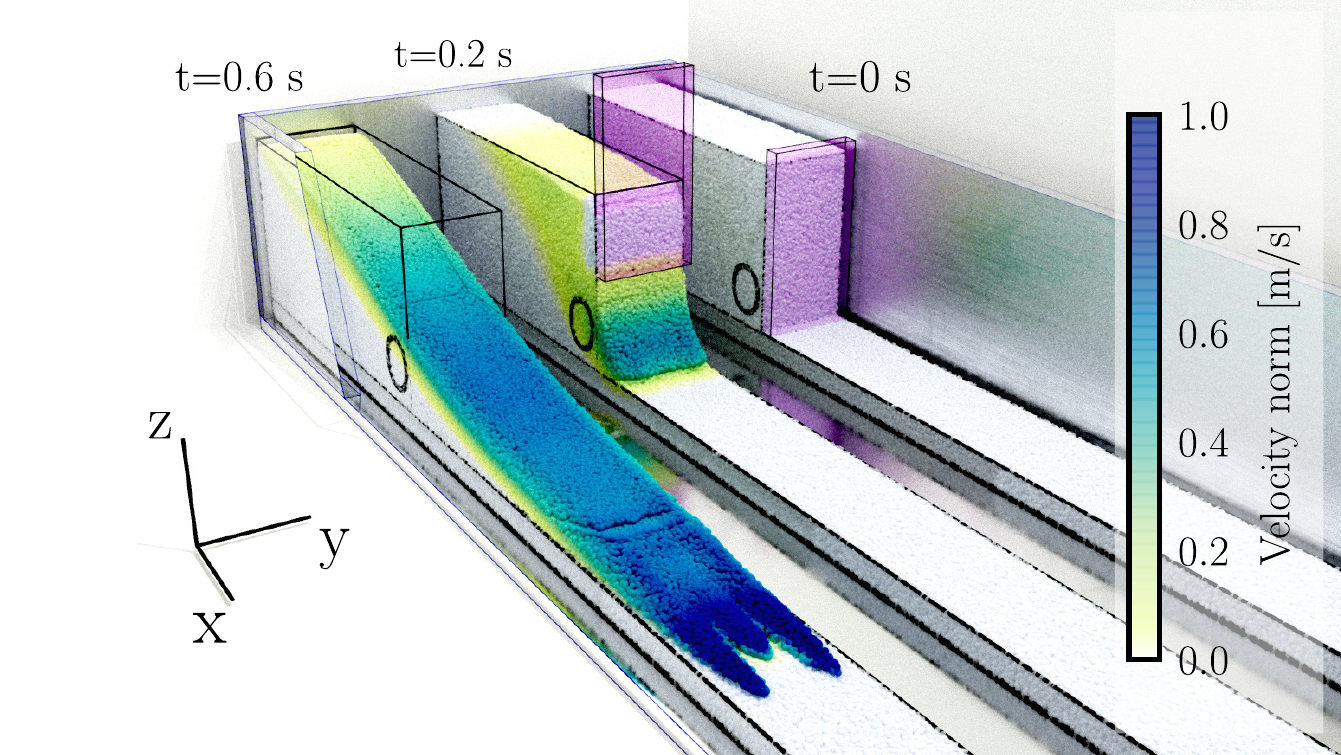} 
  \caption{3-D MPM simulations of the B15 granular column collapses over a $\SI{15}{\degree}$-tilted bed at three different times. The motion due to the lifting gate is particularly visible at $t=\SI{0.2}{s}$. A three-dimensional animated visualisation is provided in the online supplementary material (c.f. movie~3).}
  \label{fig:3Dsim}
\end{figure}

We impose frictional boundary conditions with the lateral walls and the left upstream wall -- the bottom wall being covered with a $\SI{2}{cm}$ bed of erodible granular material to ensure the same bulk friction conditions.
As mentioned in \cref{subsec:collapseSetup}, the experimental collapses are initiated by the lifting of a pneumatic gate, which can interact frictionally with the granular material and affect the flow in the initial phase. We therefore directly account for the lifting of the gate by introducing a rigid plane with imposed motion mimicking the gate. This ``numerical'' door also interacts frictionally with the material, with the same friction coefficient $\mu_w$ as the lateral walls. \Cref{fig:3Dsim} illustrates the numerical setup with the lifting of the gate. Note that while the upward lifting motion of the door influences the early dynamics, frictional interaction between the material and the door plays a negligible role on the flow and the final deposit, as discussed in \cref{subsec:impactLiftingGate}.

The simulations are run with about $216$ MPM particles per mesh cell, with random initial positions, and a time-step of $\SI{8.3e-4}{s}$.

\begin{table}
  \begin{center}

    \begin{tabular}{l@{\hskip 0.2in}|cccccccccc}
      Run  & B00 & B05 & B10 & B15 & B20 \\
      \hline
      $i$ ($^{\circ}$) & 0 $\pm$ 0.05 & 5 $\pm$ 0.05 & 10 $\pm$ 0.05 & 15 $\pm$ 0.05 & 20 $\pm$ 0.05\\
      $t_{f,\textrm{exp}}$ (s)  & 0.99 $\pm$ 0.02 & 1.23 $\pm$ 0.02 & 1.63 $\pm$ 0.03 &1.72 $\pm$ 0.04 &2.79 $\pm$ 0.05	\\
      $t_{f,\textrm{sim}}$ (s)   & 0.87 $\pm$ 0.02 & 1.00 $\pm$ 0.02 & 1.30 $\pm$ 0.03 & 1.47 $\pm$ 0.04 & 2.10 $\pm$ 0.04\\
      $H_0$ (m)	&0.11 $\pm$ 0.005 	&0.11 $\pm$ 0.005  &0.11 $\pm$ 0.005  &0.11 $\pm$ 0.005  &0.11 $\pm$ 0.005  \\
      $a$	&0.5&0.5&0.5&0.5&0.5\\
      $H_{f,\textrm{exp}}$ (m) & 0.11 $\pm$ 0.005  & 0.11 $\pm$ 0.005  & 0.09 $\pm$ 0.005  & 0.07 $\pm$ 0.005  & 0.04 $\pm$ 0.005  \\
      $H_{f,\textrm{sim}}$ (m) &  0.11 $\pm$ 0.005	& 0.11 $\pm$ 0.005 &0.09 $\pm$ 0.005  &0.07 $\pm$ 0.005  &0.04 $\pm$ 0.005 \\
      $L_{f,\textrm{exp}}-L_0$ (m) &  0.18 $\pm$ 0.01 & 0.22 $\pm$ 0.02 &0.31 $\pm$ 0.03  &0.49 $\pm$ 0.03 & -  \\
      $L_{f,\textrm{sim}}-L_0$ (m) &  0.19 $\pm$ 0.01 &0.21 $\pm$ 0.02 &0.30 $\pm$ 0.03  & 0.46 $\pm$ 0.03  & 0.88 $\pm$ 0.05 \\
    \end{tabular}
    \caption{Metrics of the experimental and numerical runs for the beads.  The experimental and simulation rest times ($t_{f,\textrm{exp}}$ and $t_{f,\textrm{sim}}$) correspond to the times when we detect no velocity above \SI{0.01}{m/s} in the domain. $H_0$ is the initial pile height and $a$ the initial collapse aspect ratio. $H_{f,\textrm{exp}}$ and $H_{f,\textrm{sim}}$ are the experimental and simulated final pile height while $L_{f,\textrm{exp}}-L_0$ and $L_{f,\textrm{mod}}-L_0$ are the experimental and simulated run-out distances measured from the gate position ($L_{f}$ is defined as the final run-out distance from the left collapse wall). The run-out distance is defined as the length of the continuous deposit extent which height is above a $\SI{2}{mm}$ threshold, following \cite{balmforth2005granular}. Note that the missing final experimental run-out distance for the B20 collapse is due to restrictions on the field of view of the camera.
    }
    \label{tab:runs}

  \end{center}
\end{table}

\subsection{Validation on various materials for different inclinations}
\label{subsec:validationInclinations}

\begin{figure}
  \centering
  \includegraphics[width=\textwidth]{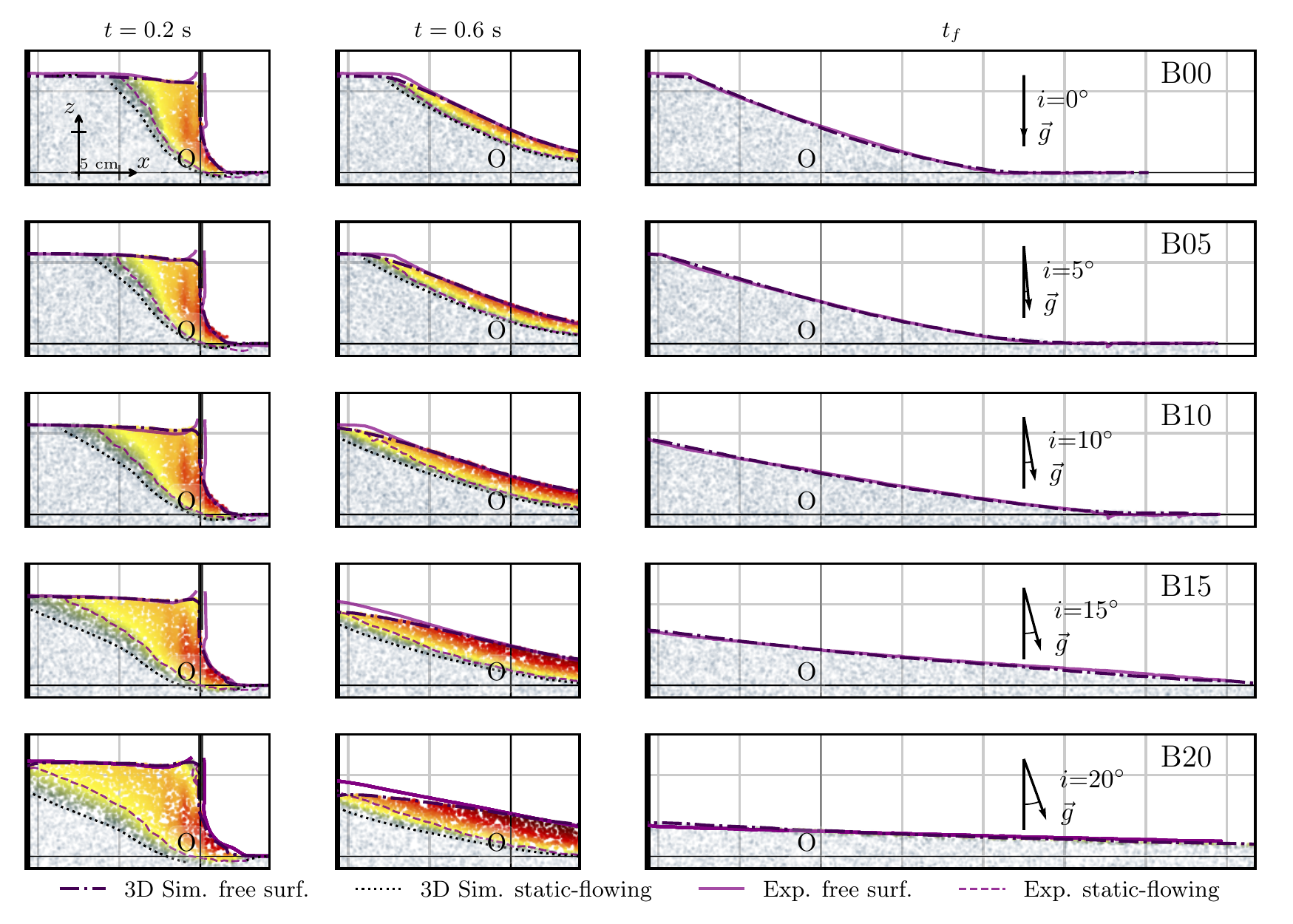}
    \caption{Comparison between collapse experiments with the \SI{0.5}{mm} glass beads and 3-D simulations  for various bed inclinations ranging from $0^{\circ}$ (top) to $20^{\circ}$ (bottom) (B00, B05, B10, B15 and B20 runs). The profiles are extracted next to the side wall position. We compare both the thickness profiles (solid pink line for the experiment, dash-dotted black line for the simulation) and the static-flowing transition (pink dashed line for the experiment, black dotted line for the simulation). The velocity heatmap in the background is computed from the simulation. In the simulations, the constant friction $\mu$ was set to $\mu_a = 0.44$, and the wall friction to $\mu_w = 0.23$.}  
    \label{fig:validationInclinationsBeads}
\end{figure}

\Cref{fig:validationInclinationsBeads} shows a comparison between profiles for the collapse experiments and simulations performed on our \SI{0.5}{mm} glass beads for 5 bed inclinations, ranging from \ang{0} to \ang{20}.
Profiles are plotted at three different times, to highlight the behaviour of the collapse in the starting, flowing and stopped phases. Although numerical profiles presented a negligible difference along the flume width ($y$-axis), they were extracted next to the side wall position to faithfully compare with measurements.  The right-most column shows the final state of the deposit, when the mass has stopped moving, i.e. when we detect no velocity above \SI{0.01}{m/s} in the whole domain, with corresponding respective experimental and simulation final -- stop -- times $t_{f,exp}$ and $t_{f,sim}$ which naturally depends on the bed inclination, as summarised in \cref{tab:runs}. 
Note that the differences observed for the final time $t_f$ between the experiment and the simulation are mainly caused by the local nature of the ``rest'' criterion we use (no velocity above \SI{0.01}{m/s} in the whole domain), which artificially delays the final time of the experiment due to marginal grain movement at the free surface.
We also plot the experimental and simulation static-flowing transition lines determined as the \SI{0.01}{m/s} contour of the velocity field in order to highlight the bulk dynamics of the collapse.

We first observe an excellent agreement between the experimental and computed final thickness profiles for all the inclinations, highlighting the ability of our numerical method to capture threshold effects and large strains.
The $t=\SI{0.2}{s}$ snapshots also highlight the role of simulating the lifting of the door, which influences the initial dynamics, as also observed by \citet{ionescu2015viscoplastic}. 

The dynamics of the collapses at early steps also appears to be well described by the simulation for small inclinations. Above \ang{15} inclination, we observe some difference in the thickness profiles, with in particular a systematic under-estimation of the profile height near the left-hand wall, which was also observed by \citet{martin2017continuum}. 
The static--flowing transition lines follow the same overall behaviour, with a fair agreement for small inclinations, degrading for larger ones.

\begin{figure}
  \centering

    \includegraphics[width=\textwidth]{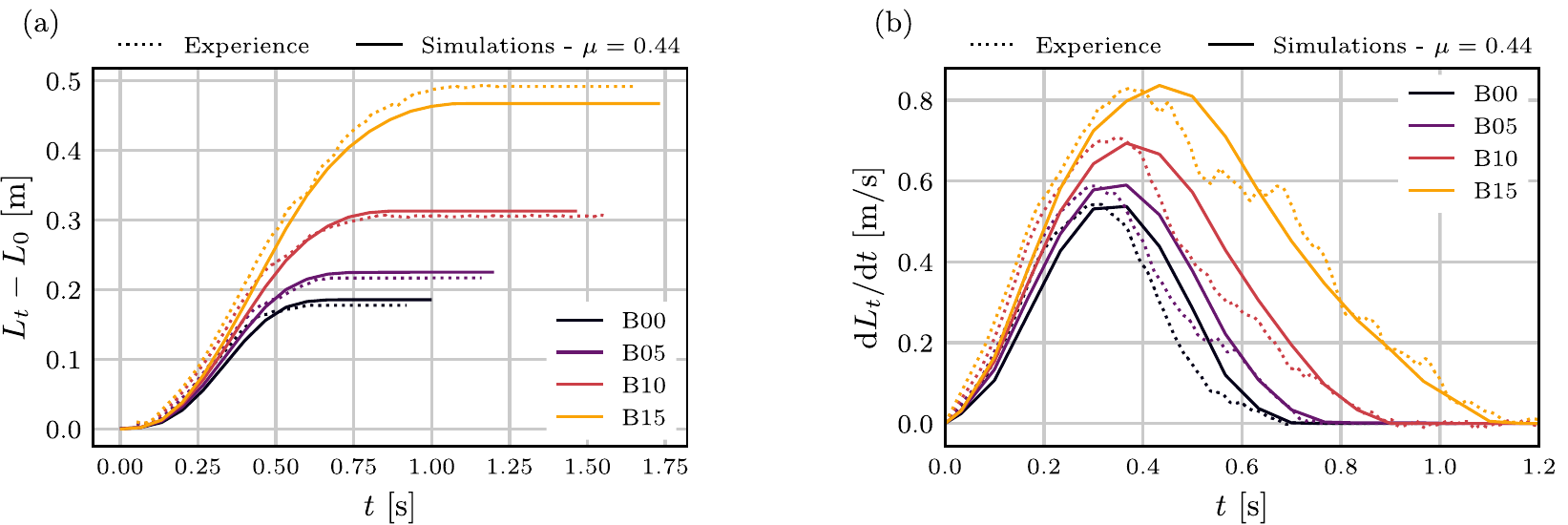}
  \caption{Front position (a) and velocity (b) for the experimental and simulated granular collapses, for bed inclinations from $\SI{0}{\degree}$ (top) to $\SI{15}{\degree}$. Note that the run at $\SI{20}{\degree}$ inclination was left out since our experimental recording setup could not frame the run-out during the whole collapse.
  The 3-D simulations were performed with the constant friction $\mu = 0.44$, and the wall friction $\mu_w = 0.23$. 
  }
  \label{fig:frontsPositionVelocity}
\end{figure}

We also show the position and velocity of the granular front as a function of time in \cref{fig:frontsPositionVelocity}. We can first observe that our simulations accurately capture the stopping time of the front, with a rather good prediction of the front velocity for each collapse inclination. Note that the front always stops before the final stopping time $t_f$ of the whole flow. 

Overall, while the early dynamics of the collapses seem to depend on the inclination, these results validate our simulation approach, and in particular support the use of a simple Drucker--Prager law with the constant avalanche friction coefficient to capture final deposit profiles.

\begin{figure}
	\centering
	\includegraphics[width=\textwidth]{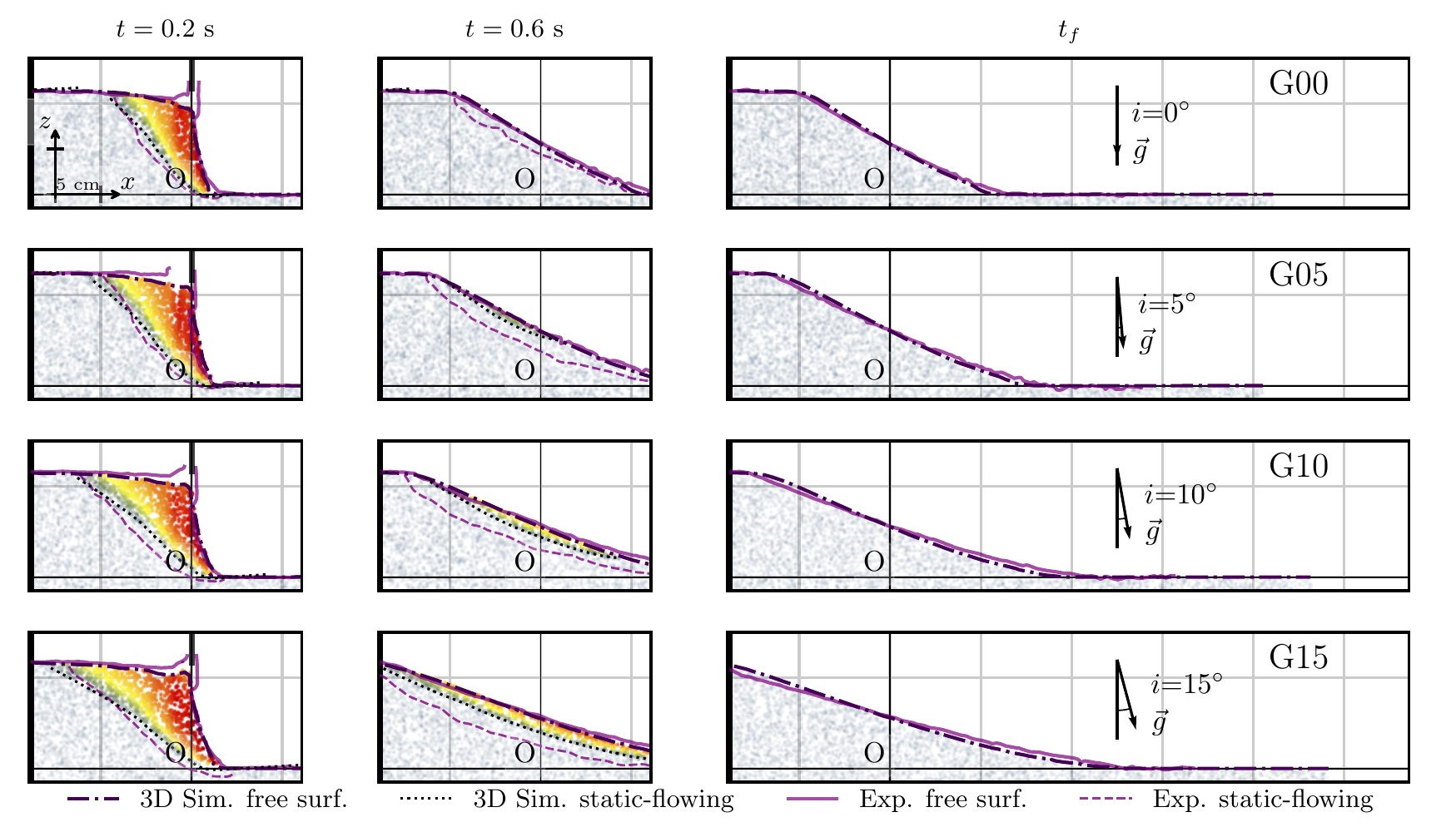}
	\caption{Comparison between collapse experiments with the natural granules and 3-D simulations for various bed inclinations ranging from $\SI{0}{\degree}$ (top) to $\SI{15}{\degree}$ (bottom) (G00, G05, G10 and G15 runs). The profiles are extracted next to the side wall position.} We compare both the thickness profiles (solid pink line for the experiment, dash-dotted black line for the simulation) and the static-flowing transition (pink dashed line for the experiment, black dotted line for the simulation). The velocity heatmap in the background is computed from the simulation. In the simulations, the constant friction $\mu$ was set to $\mu_a = 0.75$, and the wall friction to $\mu_w=0.3$. See movie~4 in the online supplementary material for an animated evolution of the G15 simulated collapse against experiment. 
	\label{fig:validationInclinationsGranules}
\end{figure}

The robustness of these results has been evaluated using a more natural and irregular material (denoted \textit{granules} above, c.f. \cref{fig:materials}, \cref{tab:materialsParameters}).
The corresponding profiles are presented in \cref{fig:validationInclinationsGranules}, and the final pile heights and run-out values in \cref{tab:runsG}. We again observe a good agreement between the experiments and simulations.

\begin{table}
  \begin{center}

    \begin{tabular}{l@{\hskip 0.2in}|cccccccccc}

      Run &  G00& G05& G10 & G15\\
      \hline
      $i$ ($^{\circ}$)  & 0 $\pm$ 0.05 & 5 $\pm$ 0.05 & 10 $\pm$ 0.05 & 15 $\pm$ 0.05 \\
      $t_{f,\textrm{exp}}$ (s)	&0.79 $\pm$ 0.01  &0.89 $\pm$ 0.01 &0.99 $\pm$ 0.02 &1.09 $\pm$ 0.03\\
      $t_{f,\textrm{sim}}$ (s)  & 0.47 $\pm$ 0.01 & 0.53 $\pm$ 0.01 & 0.60 $\pm$ 0.02 & 0.73 $\pm$ 0.03\\
      $H_0$ (m)	& 0.12 $\pm$ 0.005 &0.12 $\pm$ 0.005 &0.12 $\pm$ 0.005 & 0.12 $\pm$ 0.005 \\
      $a$	&0.6 &0.6	&0.4	&0.4&\\
      $H_{f,\textrm{exp}}$ (m) & 0.12 $\pm$ 0.005 & 0.12 $\pm$ 0.005 & 0.12 $\pm$ 0.005 & 0.11 $\pm$ 0.005 \\
      $H_{f,\textrm{sim}}$ (m) & 0.12 $\pm$ 0.005 &0.12 $\pm$ 0.005 &0.12 $\pm$ 0.005 &0.12 $\pm$ 0.005 \\
      $L_{f,\textrm{exp}}-L_0$ (m) &0.11 $\pm$ 0.01 &0.15  $\pm$ 0.01  &0.10 $\pm$ 0.02  & 0.24 $\pm$ 0.03  \\
      $L_{f,\textrm{sim}}-L_0$ (m) &0.10 $\pm$ 0.01 &0.13 $\pm$ 0.01  &0.15 $\pm$ 0.02 &0.19 $\pm$ 0.03 \\
    \end{tabular}
    \caption{Metrics of the experimental and numerical runs for the granules. The letter G indicates granules. Please refer to the definition of the quantities in \cref{tab:runs}}.

    \label{tab:runsG}
  \end{center}
\end{table}

The small discrepancies observed close to the lifting gate at the beginning of the flow suggest that the frictional interaction of the material with the door is higher. We also notice, as was also the case for the glass beads, that the material flows less in the early dynamics, suggesting that more subtle rheological effects are at play in this stage. However, again this does not impact the final run-outs and profiles.

We should stress that the measurement protocols for the avalanche and stop friction coefficients are subjected to important uncertainties (as reported in \cref{tab:materialsParameters}) in the case of such gritty material, due to the inherently large inclination angles required to reach the yield and stop transitions, and to the increased impact of boundary effects in the resulting very inclined channels. 
As such, we do not try to discriminate accurately against the use of either the avalanche or stop friction coefficient, but rather interestingly illustrate the reliability of the plain Drucker--Prager rheology in the context of transient collapse flows of natural materials.

\subsection{Comparison with the $\mu(I)$ rheology}
\label{subsec:comparisonMuIRheology}

The $\mu(I)$ rheology, initially proposed by \cite{jop2006constitutive} in the context of steady inertial flows, introduces a local dependence of the Drucker--Prager coefficient on the inertial number 
\begin{equation}
  I = \frac{d \lVert \dot{\varepsilon} \rVert}{\sqrt{p/\rho}}
  \label{eq:inertialNumber}
\end{equation}
with $d$ the particle diameter, $\dot{\varepsilon}$ the strain-rate tensor introduced in \cref{eq:strainRateTensor}, $p$ the pressure and $\rho$ the material density. This inertial number characterises the ratio between the inertial timescale $d \sqrt{\rho / p}$ and the deformation timescale $1/\lVert \dot{\varepsilon} \rVert$.

The $\mu(I)$ rheology then writes
\begin{equation}
  \mu(I) = \mu_{stop} + \frac{\mu_2 - \mu_{stop}}{1 + \frac{I_0}{I}}
  \label{eq:muIrh}
\end{equation}
where $\mu_{stop}$, $\mu_2$ and $I_0$ are material-dependent coefficients which can be calibrated using the steady inclined plane experiment as described in \cite{jop2005crucial}.

This model has recently been used with success for transient collapse prediction in several studies \citep{lagree2011granular,ionescu2015viscoplastic,martin2017continuum}.

The $\mu(I)$ rheology \cref{eq:muIrh} can be straightforwardly implemented in our simulation framework by adding an explicit friction coefficient update in the yield criterion and solving for the Drucker--Prager law with this new coefficient as before.
As discussed in \cref{subsec:measureMacro}, we have performed the steady inclined flow protocol with our $\SI{0.5}{mm}$ beads, which are similar to the one from \citet{jop2005crucial}, and extracted the $\mu_{stop}$ and $\mu_2$ coefficients from the fit of $\Theta_s(h/d)$ (c.f. \cref{fig:thetastop}). As expected, we retrieve values close to the ones from \citet{jop2005crucial}; for the sake of comparison, and owing to the materials similarities, we have not determined the characteristic inertial number $I_0$, and have chosen to use the parameters from \citet{jop2005crucial} to perform our $\mu(I)$ simulations: $\mu_{stop} = 0.38$, $\mu_2 = 0.64$ and $I_0 = 0.279$.

\begin{figure}
  \centering
  \includegraphics[width=\textwidth]{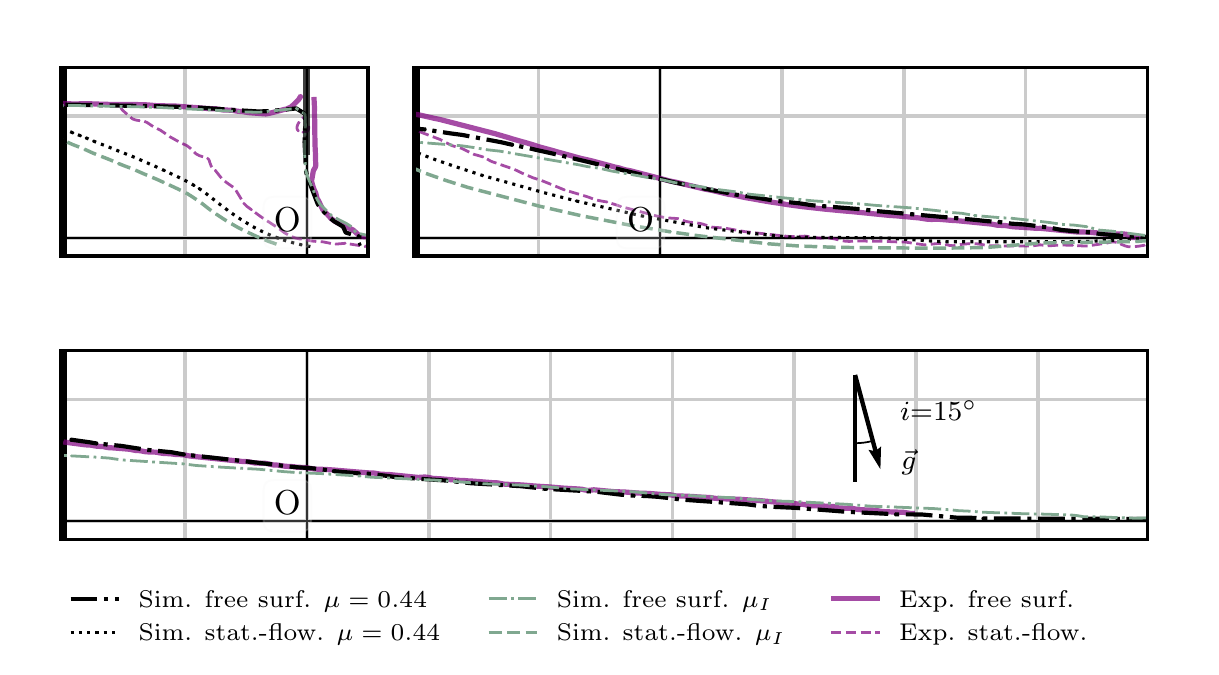}
  \caption{Comparison of 3-D simulations performed with $\mu = \mu_a = 0.44$ (black lines) and $\mu = \mu(I)$ (green lines) for the $\SI{15}{\degree}$ bead collapse (B15) at $t=\SI{0.2}{s}$, $\SI{0.6}{s}$ and $t_f$. Experimental curves (solid pink line) are provided for reference, and we show both the free-surface lines (resp. plain and dash-dotted) and the static-flowing transition contours (resp. dashed and dotted). The results for the other collapse inclinations can be found in \cref{appendix:frictionCoefficientComparison}.}  
  \label{fig:comparisonMuIRheology}
\end{figure}

\begin{figure}
  \centering
  \includegraphics[width=\textwidth]{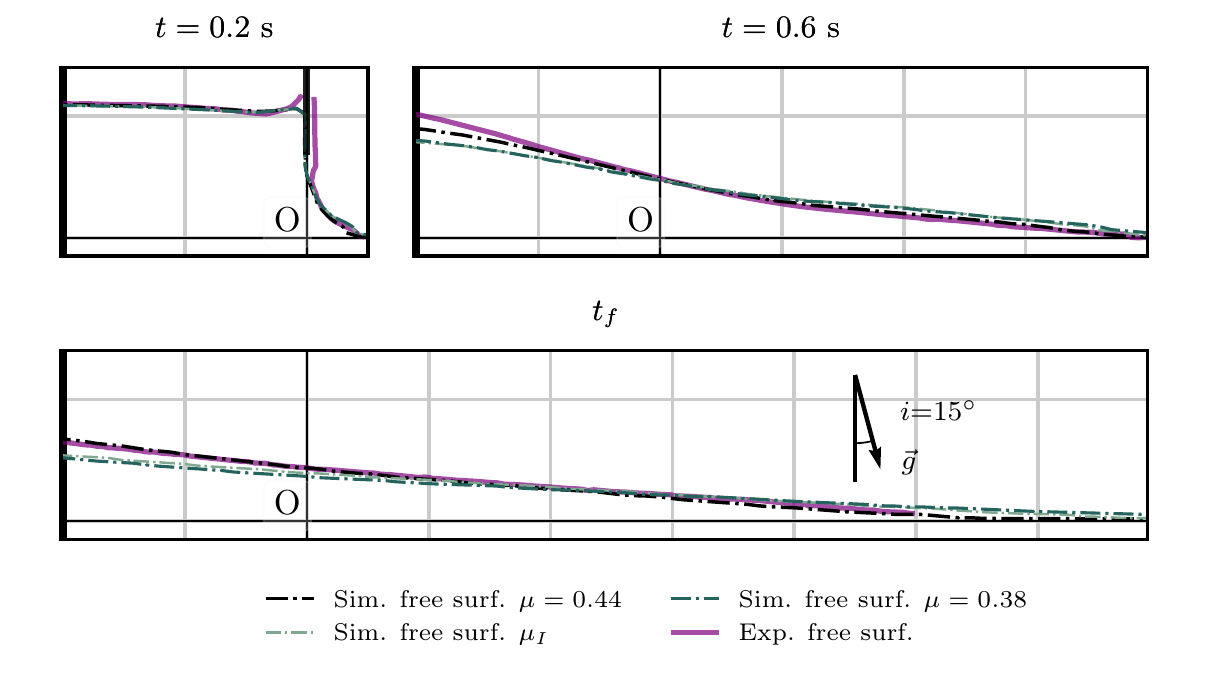}
  \caption{Comparison of 3-D simulations performed with $\mu = \mu_a = 0.44$ (black lines), $\mu=\mu(I)$ and $\mu = \mu_{stop} = 0.38$ (blue lines) for the $\SI{15}{\degree}$ bead collapse (B15) at $t=\SI{0.2}{s}$, $\SI{0.6}{s}$ and $t_f$. Experimental curves (solid pink line) are provided for reference, and we show both the free-surface lines (resp. plain and dash-dotted) and the static-flowing transition contours (resp. dashed and dotted). The results for the other collapse inclinations can be found in \cref{appendix:frictionCoefficientComparison}.}  
  \label{fig:comparisonMuAMuIMuS}
\end{figure}

\Cref{fig:comparisonMuIRheology} shows a comparison of the experimental and computed collapse profiles with the simple Drucker--Prager ($\mu_a=0.44$) and the $\mu(I)$ ($\mu_{stop}=0.38$) rheologies. 
For completeness, we also show the numerical results obtained with a constant $\mu_{stop}$ friction coefficient -- the lower bound of $\mu(I)$ -- in \cref{fig:comparisonMuAMuIMuS}.
For the sake of readability, we focus here on the results for the \SI{15}{\degree}-inclined configuration; the corresponding comparison figures are provided for all inclinations in \cref{fig:comparisonMuAMuIMuSAllInclinations}, and support the same observations.

We observe that both the $\mu(I)$ and the constant $\mu_{stop}$ rheologies fail to capture accurately the free surface and the static-flowing transition region, in both the dynamical and final resting phases. More precisely, they behave very similarly and both underestimate the flowing threshold, leading to an increased flowability and broader deposits.

\begin{figure}
  \includegraphics[width=\textwidth]{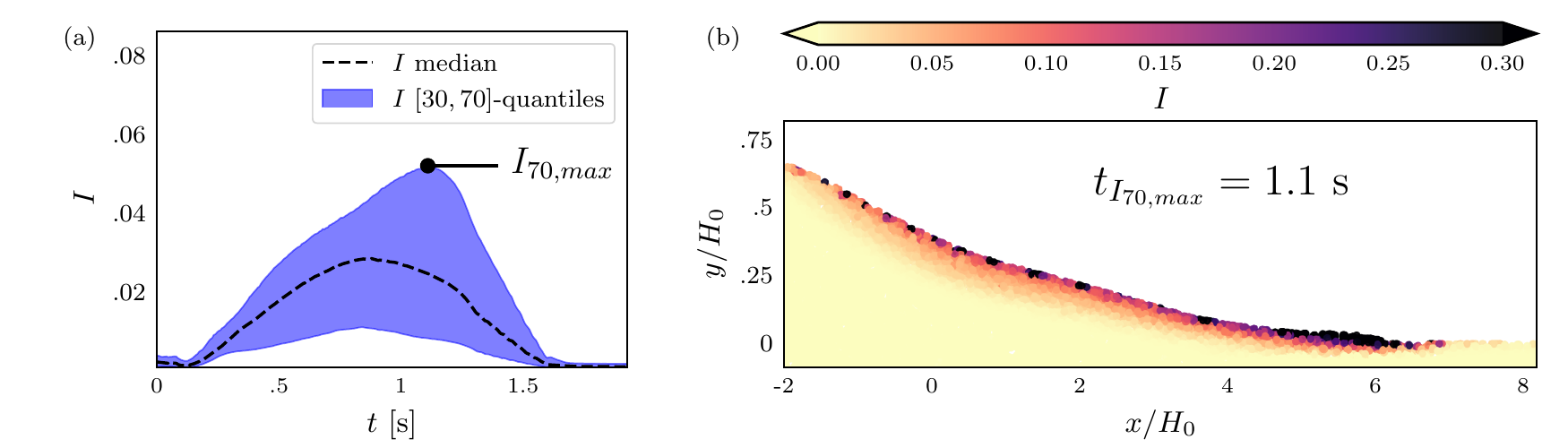}
  \caption{Inertial number $I$ during the simulated $\SI{15}{\degree}$ bead collapse (B15) with $\mu=\mu_a=0.44$: (a) Evolution of $I$ as a function of time. Median, $30$- and $70$-quantiles are estimated in the regions where $I \geq 5 \times 10^{-4}$ thereby filtering out static parts of the collapse; (b) Instantaneous inertial number heatmap at $t_{I_{70,max}} = \SI{1.1}{s}$, i.e. the time for which the $70$-quantile of $I$ is maximal. $I$ values remain significantly lower than $I_0 \sim 0.3$ during our collapses explaining why the increase of the friction coefficient from $\mu_{stop}$ to $\mu_2$ in the $\mu(I)$ rheology \cref{eq:muIrh} has a negligible effect. (see movie~5 in the online supplementary material for an animated evolution of the inertial number heatmap)}
  \label{fig:muIEvolution}
\end{figure}

The similarity observed between the $\mu(I)$ and constant $\mu_{stop}$ rheologies is coherent with the systematic low measured inertial numbers $I$, as shown in \cref{fig:muIEvolution}: except for marginal surface points, the inertial numbers are in the order of at most $I \sim 10^{-2}$ during all the collapse, which is much lower than the characteristic value $I_0 = 0.279$ of the $\mu(I)$ rheology. 
The additional viscous dissipation introduced by the $\mu(I)$ law compared to a constant $\mu_{stop}$ is therefore negligible for such collapses, as was also observed by \citet{valette2019sensitivity} for horizontal column collapses with a regularised $\mu(I)$ rheology.

The necessary use of a friction coefficient larger than the ``stop'' friction coefficient $\mu_{stop}$ measured from steady inclined plane experiments to quantitatively match experiments was also discussed by \cite{lagree2011granular}, or \citet{ionescu2015viscoplastic} and \citet{martin2017continuum}. The last two explain their increased friction coefficient by the role of the lateral walls, an effect that we shall however discard below (c.f. \cref{sec:walls}). 

\subsection{Experimental validation of wall friction effect}
\label{subsec:wallsExpValidation}
The validation of the numerical method is ended by evaluating the simulation of the lateral walls as such boundary conditions are crucial when modelling 3-D configurations \citep{jop2005crucial,lajeunesse2005granular,ionescu2015viscoplastic,martin2017continuum}.
For that purpose, we perform dedicated collapse experiments with two different channel widths, and compare the corresponding height profiles with the simulated ones. For simplicity, the granular collapses are performed in a rough horizontal -- \SI{0}{\degree} -- channel, with the \SI{0.5}{mm} glass beads described above. We consider two different channel widths: $W = \SI{1}{cm}$ and $W = \SI{4}{cm}$, which thus correspond to $W/d = 20$ and $80$ respectively. Experimental conditions are similar to the aforementioned \SI{0}{\degree} collapse, but with initial column dimensions of $H_0 = L_0 = \SI{12}{cm}$, giving an aspect ratio $a=1$, which allows us to make the effects of the walls more visible on the final upslope height (c.f. \citet{balmforth2005granular} or \citet{zhang2021three}), thereby providing a simple measure to compare the effect of lateral friction, which we shall use to determine the corresponding two-dimensional effective friction in \cref{subsec:twoDimensionalEffectiveFrictionCoefficient}.
The corresponding numerical collapses are carried out consistently, using the constant bulk friction coefficient $\mu = 0.44$ and a friction coefficient between the material and the walls $\mu_w = 0.23$.

\begin{figure}
	\centering
	\includegraphics[scale=1.3]{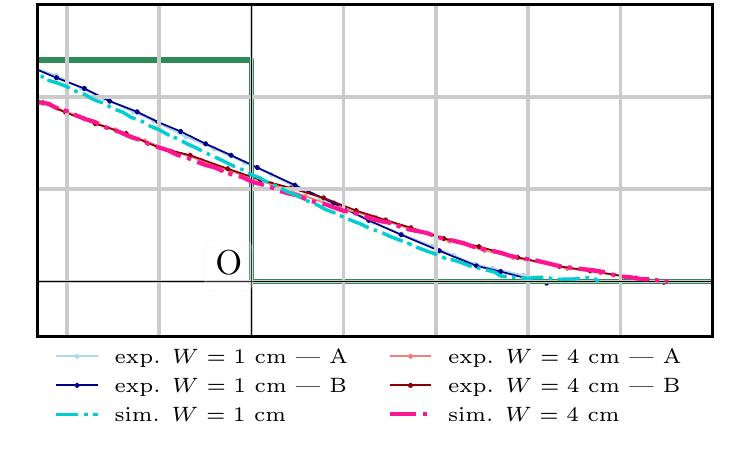}
	\caption{Influence of the sidewalls: rest-state height profiles of simulated and experimental collapses for $a=1$ and two different widths ($W=\SI{1}{cm}$ and $W=\SI{4}{cm}$) at \SI{0}{\degree} slope. Two experimental replicates (A and B) are compared for each width run. The green solid line delineates the initial column with size $0.12 \times 0.12$ $\si{cm}^2$. Numerical collapses are performed with a constant internal friction coefficient $\mu=0.44$ and a friction coefficient between the glass walls and the material $\mu_w=0.23$.}
	\label{fig:expWallsB00}
\end{figure}

\Cref{fig:expWallsB00} shows the rest-state height profiles for the simulations and experiments, with two independent experimental runs for each width in order to ensure reliability of the results. We observe that the simulated profiles compare noticeably well with the experimental ones, which underlines the validity of our numerical method, and indicates that a plain Drucker--Prager rheology, with -- independently measured -- friction coefficients $\mu$ and $\mu_w$ can accurately account for confinement effects in such transient flows. Note that the good agreement between experiments and simulation observed in the narrow case $W/d = 20$ also suggests that continuum modelling remains valid in this range, at least for such collapse flows.

\section{Revisiting features of granular collapses with 3-D simulations in light of a constant friction coefficient}
\label{sec:walls}

As demonstrated in \cref{sec:validation}, the plain Drucker--Prager rheology, with a constant friction coefficient set from the avalanche angle of the material, can accurately predict experimental column collapses on inclined beds, for a large range of inclinations. 
Our validated three-dimensional non-smooth numerical model therefore enables to explore and revisit phenomenological features of transient collapses, which have been previously interpreted in the framework of a steady viscoplastic rheology. 
While viscous rheological contributions -- as introduced by the $\mu(I)$ rheology -- are undeniably crucial to represent inertial granular flows, notably to explain the existence of steady inclined flows, the range of applicability of the fully plastic Drucker--Prager rheology has, to the best of our knowledge, never been thoroughly framed, in particular in the context of transient flows at low strain-rates. It thus appears essential to elucidate the different ingredients responsible for the various features of granular collapses, in order to disentangle the respective roles of frictional or viscous effects, and clarify the links between steady and transient geometries in three-dimensional configurations.
In the following, we first investigate this question by referring to established scaling laws relating the final pile run-out and height to the initial column aspect ratio \citep{lajeunesse2005granular,lube2005,balmforth2005granular, lagree2011granular,dunatunga2015continuum}.
We then propose a comprehensive analysis of the impact of sidewalls in granular collapses, in light of the steady law proposed by \citet{savage1979gravity,taberlet2003superstable,jop2005crucial} and transposed to the transient case by \citet{ionescu2015viscoplastic}.

Our results notably highlight the weak role played by viscosity in transient granular collapses by demonstrating that established scaling laws are well accounted for by the plain Drucker--Prager rheology, and show that the effect of sidewall friction, while possessing a similar scaling behaviour than in the steady case, has much less impact on  transient collapses than previously assumed in~\cite{ionescu2015viscoplastic,martin2017continuum}.

\subsection{Aspect ratio scaling laws in wide and narrow configurations}

\begin{figure}
  \centering
  \includegraphics[width=\textwidth]{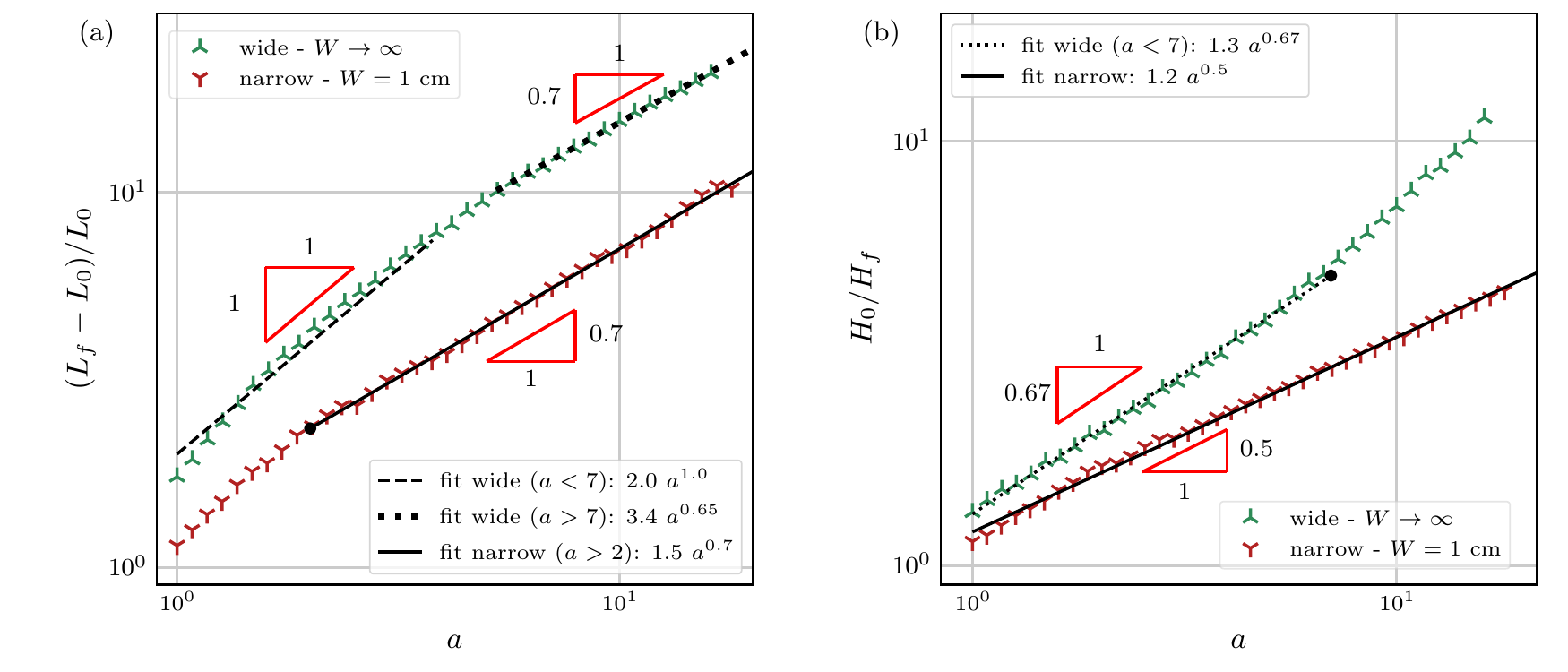}
  \caption{(a) Inverse normalised final upslope height ($H_0/H_{f}$) and (b) normalised run-out ($L_{f}-L_0)/L_0$  as a function of the aspect ratio of the initial column $a$. $L_{f}$ is obtained using the position the MPM particle that goes the furthest at the final state.}

  \label{fig:scalingsB00}
\end{figure}

We perform simulations in wide and narrow channels using different aspect ratios for the initial granular column, and compare the final run-out and upslope height with previously proposed scaling laws. 
The normalised rest state run-out $L_{f}$ and upslope height $H_{f}$ are shown in \cref{fig:scalingsB00} for $40$ simulated collapses in both wide and narrow configurations with aspect ratios $a = H_0 / L_0$ varying logarithmically from $1$ to $20$. The wide configuration is obtained from 3-D collapses by setting the wall friction to $0$ (which is equivalent to the 2 dimensional collapse situation). The narrow situation corresponds to a $\SI{1}{cm}$-wide channel with a wall friction coefficient set to $\mu_w=0.23$. The granular material considered is again the glass beads, with a bulk friction coefficient of $\mu=0.44$.

The power-law fits obtained for our data (summarised in \cref{tab:aspectRatioScalings}) exhibit scaling behaviours similar to the previous experimental and numerical studies from \citet{lajeunesse2004spreading,balmforth2005granular,lagree2011granular,dunatunga2015continuum}, and in particular feature the change of regime between the short ($a \lesssim 7$), and high ($a > 7$) columns in the \textit{wide} configuration. Note that we also observe the theoretical scaling exponents from the shallow 2-D model of \citet{kerswell2005dam} and \citet{balmforth2005granular}, which predicts $H_{0}/H_{f} \sim a^{0.69}$ in our aspect ratio range for the \textit{wide} configuration and $H_{0}/H_{f} \sim a^{0.5}$ for the narrow one. 
In comparison with the work of \cite{lagree2011granular,dunatunga2015continuum}, our results thus not only confirm the robustness of the power-law scalings observed by \citet{lajeunesse2005granular,balmforth2005granular,lube2005}, but also highlight that this scaling behaviour of collapse flows is well described by the plain Drucker--Prager rheology with a constant friction coefficient, for a large range of aspect ratios -- up to $a\sim 10$.

\begin{table*}
\begin{center}
\begin{tabular}{c@{\hskip 0.3in}|c@{\hskip 0.3in}c@{\hskip 0.3in}}
  & $\displaystyle \frac{L_{f}-L_0}{L_0}$ & $\displaystyle \frac{H_{0}}{H_{f}}$ 
  \\
  \hline
  \textit{wide} channel 
  &
    $\displaystyle \left\{\begin{aligned}
      &\,2.0\,a &\: \textrm{for } a \lesssim 7 \\
      &\,3.4\,a^{0.65} &\: \textrm{for } a \gtrsim 7
    \end{aligned}\right.$
  &
    $1.3\,a^{0.67} \quad \textrm{for } a \lesssim 7$
  \\[0.3in]
  \textit{narrow} channel 
  &
    $1.4\,a^{0.7}$
  &
    $1.2\,a^{0.5}$
\end{tabular}
\end{center}
\caption{Power-law scaling laws for the run-out and upslope height of the final state of the collapse depending on the initial aspect ratio $a$.}
\label{tab:aspectRatioScalings}
\end{table*}

\subsection{Role of sidewall friction}
\label{subsec:twoDimensionalEffectiveFrictionCoefficient}

A linear relation between the flow thickness and the channel width has been reported by \citet{savage1979gravity,taberlet2003superstable,jop2005crucial} in the context of steady inclined granular flows, stating that the presence of friction with the lateral walls can be accounted for by rescaling the friction coefficient as
\begin{equation}
  \mu_{eq} = \mu + \mu_w \frac{h}{W}
  \label{eq:taberletScaling}
\end{equation} 
where $W$ is the flow width, $h$ the steady flow thickness and $\mu_w$ the friction coefficient with the walls. This empirical law, which can be interpreted from force balance principles in the steady flow configuration, has also been used in the context of transient granular collapses by \citet{ionescu2015viscoplastic} to support an increase of the effective 2-D friction coefficient, albeit without proper definition of a flow thickness in such a framework. In the following, we use our validated numerical simulator to explore the relevance of such linear scaling in the context of unsteady flows.

\subsubsection{Impact of the channel width in transient collapses}
\label{subsubsec:widthImpactInTransientCollapses}

\begin{figure}
  \centering
  \includegraphics[width=\textwidth]{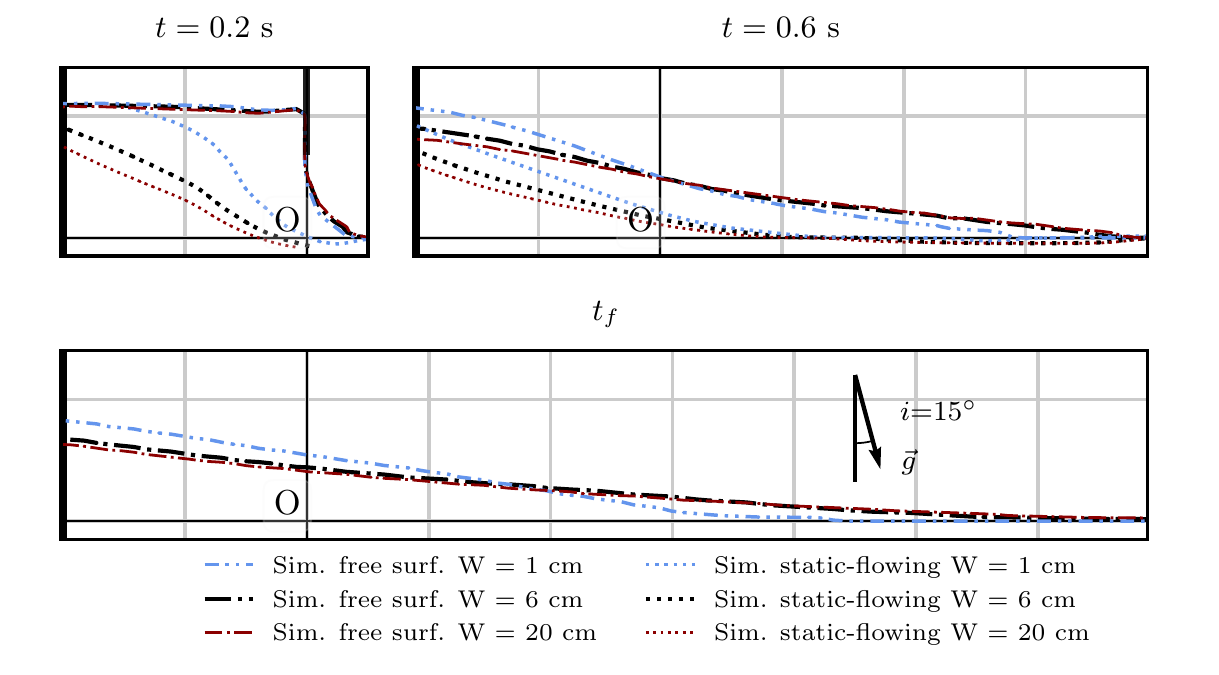}
  \caption{Influence of the sidewalls: comparison of 3-D simulations of the $\SI{15}{\degree}$ bead collapse with several channel widths, and a wall friction coefficient  $\mu_w = 0.23$.}
  \label{fig:comparisonWallsB15}
\end{figure}

We investigate the role of sidewalls friction on collapse dynamics by considering the \SI{15}{\degree}- inclined granular collapse case introduced in section~\ref{subsec:collapseSetup}, and perform several 3-D simulations with various channel widths ranging from $W=\SI{1}{cm}$ to $W=\SI{20}{cm}$.
The results for transient and final states are shown in \cref{fig:comparisonWallsB15}.
While the early collapse dynamics, outlined by the static-flowing transition contours, seems impacted for narrow channels, the influence of the width on the free surface height appears negligible for $W \geq \SI{6}{cm}$, and wall friction only starts to play a significant role on the final deposit for very narrow channels $W \lesssim \SI{1}{cm}$.

Note that this weak impact of the channel width on the collapse rest state was also observed in the recent Discrete Element Method simulations of \citet{zhang2021three}, which report negligible impact of the width for $W \gtrsim 20d = \SI{5}{cm}$ for the collapse of granular columns similar to ours.

\subsubsection{Equivalent 2-D friction coefficient}

To further investigate the linear relation~(\ref{eq:taberletScaling}) in the context of unsteady flows, we perform multiple simulations of 3-D collapses with different channel widths, and determine for each 3-D simulation (with a given width $W$) the best equivalent friction coefficient able to reproduce the corresponding 3-D collapse with a 2-D -- unconfined -- simulation. Comparison to determine this 2-D equivalent coefficient $\mu_{2D,eq}$ is performed on the rest state upslope height $H_{f}$, and as such does admittedly not fully allow to replace the 3-D simulation by a 2-D one, but still provides insight into the role of lateral confinement and friction.

\begin{figure}
  \centering
  \includegraphics[width=\textwidth]{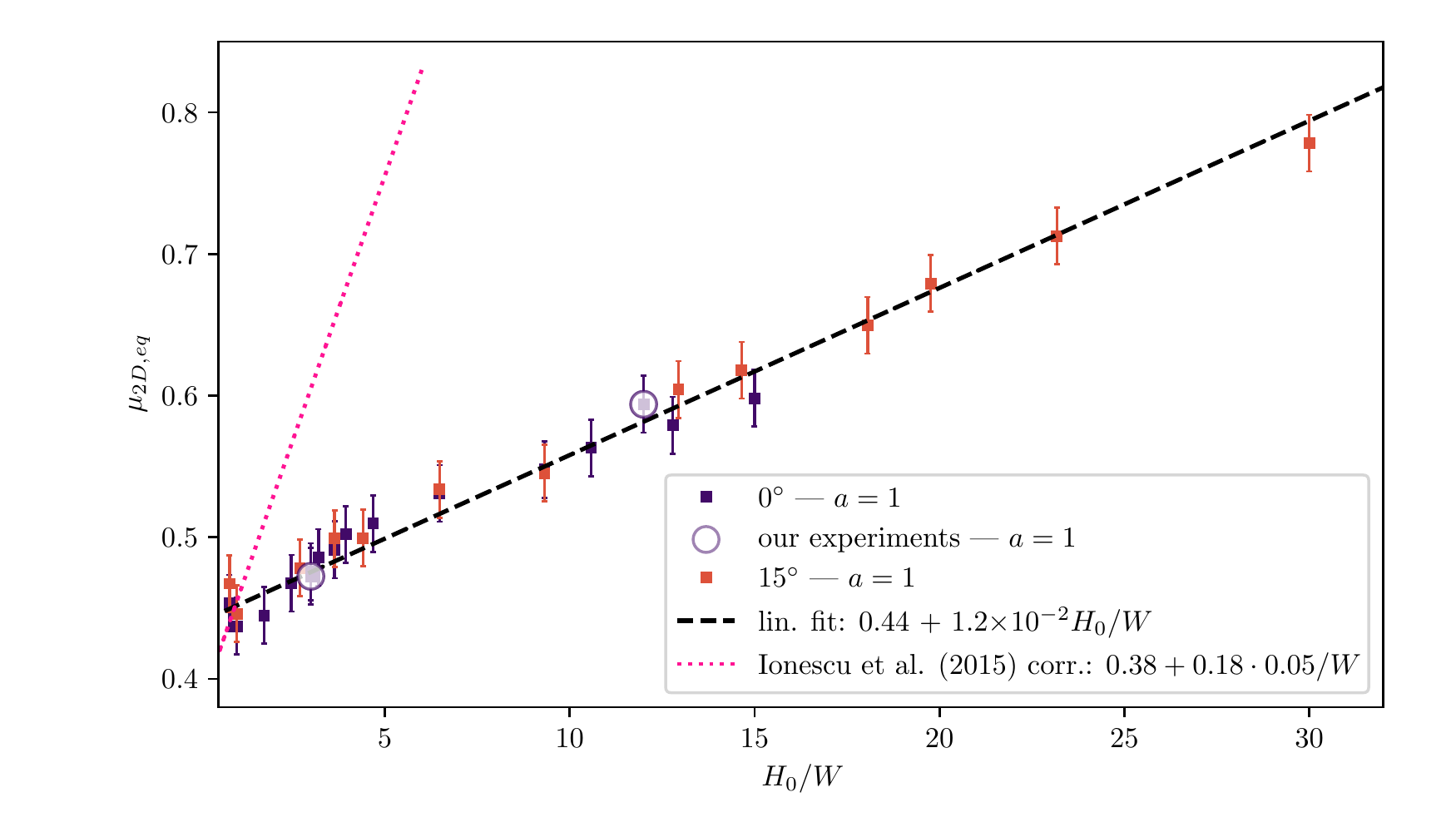}
  \caption{2-D equivalent friction coefficient determined to best match the corresponding 3-D final free-surface, as a function of the channel width, for the $\SI{0}{\degree}$ and $\SI{15}{\degree}$ beads collapses. Our results seem compatible with a linear dependency, which we illustrate by the dashed fit. We also report the linear scaling (pink dotted line) used by \citet{ionescu2015viscoplastic}.}
\label{fig:mu2DEquivalent}
\end{figure}

We consider horizontal and $\SI{15}{\degree}$-inclined collapses in order to test the robustness of the scaling. As mentioned above, we use an initial column aspect ratio of $a=1$ to increase the effect of wall friction on $H_{f}$. In fact, as already observed by for example \citet{balmforth2005granular} or \citet{zhang2021three}, the upslope part of the column remains static for aspect ratios $a$ below $0.5$ in the horizontal case, so that the final upslope height cannot help assess the role of lateral confinement for such stocky collapses.

\Cref{fig:mu2DEquivalent} gives the results for channel widths ranging from $W=\SI{20}{cm}$ down to $W=\SI{0.4}{cm}$. Note that we use the initial column height $H_0$ as a length scale, and consider the non-dimensional width $W/H_0$, since the flow thickness $h$ is not relevant in our transient case. To compare our data with the linear relation \cref{eq:taberletScaling}, we plot $\mu_{2D,eq}$ as a function of the inverse non-dimensional width $H_0/W$.

For the sake of comparison, we also show the linear scaling hypothesis used by \cite{ionescu2015viscoplastic} to obtain the 2-D friction coefficient corresponding to the 3-D $\mu(I)$ law with an effective flowing thickness  at $h \sim \SI{0.05}{m}$ estimated using the maximum flowing thickness for collapses with $H_0 \sim \SI{0.1}{m}$ (similar to our collapses). This hypothesis was used to account for the confinement effects in collapses within 2-D simulations, and typically led to a rescaling of $\mu_1$ from $0.38$ to $0.38 + \mu_w 0.05/W \simeq 0.48$ for $\SI{10}{cm}$-wide collapses.

We can however observe that our 3-D simulations do not support this hypothesis, and exhibit a significantly weaker impact of the lateral walls, as was also suggested in \cref{subsubsec:widthImpactInTransientCollapses}. Our equivalent 2-D friction coefficient nevertheless still appears linearly dependent on $H_0/W$, with fitted coefficients corresponding to
\begin{equation*}
  \mu_{2D,eq} = 0.44 + 1.2 \times 10^{-2} \,\frac{H_0}{W}
\end{equation*}

Interpreted from \cref{eq:taberletScaling}, it would correspond to an effective flow height $h \simeq 1.2 \times 10^{-2} H_0 / \mu_w \simeq \SI{6.3}{mm}$. This value, which is about one order of magnitude lower than the maximum flowing thickness used by \citet{ionescu2015viscoplastic} to estimate an effective flow thickness, illustrates that special care is required to estimate an effective flow thickness in order to deduce the 2-D equivalent friction coefficient.

\section{May an hysteresis phenomenology explain the collapse onset behaviour?}
\label{sec:hysteresis}

\begin{figure}
  \begin{center}
    \includegraphics[scale=1]{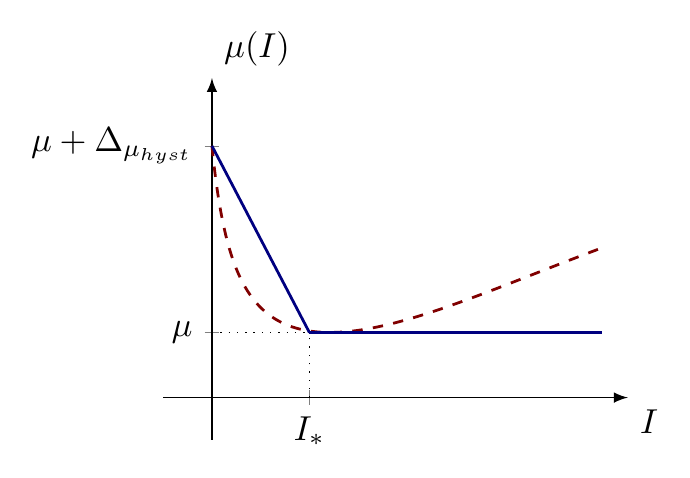}
    \caption{Friction law as a function of the inertial number $I$. The dash-dotted red curve represents a non-monotonic friction law as expected for granular material \citep{degiuli2016phase}. The blue solid line is the simplified linear law implemented with a hysteresis gap $\Delta\mu_{hyst}$.}
    \label{fig:muHyst}
  \end{center}
\end{figure}

Our non-smooth granular model can accurately predict the material profiles of granular collapses using a single friction coefficient inferred from measured avalanche angles. 
While generally consistent with experimental measures, the simulated flow dynamics however exhibits significant differences in the early stages of the collapse, as illustrated by the static-flowing transition contours in \cref{fig:validationInclinationsBeads}.
This early deviation, also observed using the $\mu(I)$ rheology by \citet{martin2017continuum}, suggests that more complex rheological effects are at play during the onset of the flow, which naturally involves low velocity and hysteresis phenomenology \citep{pouliquen2002friction}.
As a first step to highlight the role of a static friction coefficient and a potential hysteresis between the solid and flowing phases, we implement a non-constant friction coefficient law, somehow reminiscent of the well-known static-dynamic transition in solid friction:

\begin{equation}
  \mu_{hyst} (I) = \left\{
  \begin{aligned}
    &\mu + \Delta\mu_{hyst} \left( 1 - \frac{I}{I_*} \right) && \textrm{if } I < I_{*}
    \\
    &\mu && \textrm{if } I \geq I_{*},
  \end{aligned}
  \right.
  \label{eq:muHyst}
\end{equation}

and illustrated in \cref{fig:muHyst}.

The friction coefficient $\mu(I=0)$, which describes the static-flowing transition, is then $\mu + \Delta\mu_{hyst}$, and $I_{*}$ characterises the transition from ``static'' to ``dynamic'' friction. While simpler than the non-monotonic law discussed by \citet{degiuli2017friction}, our two-valued law still allows for hysteretic instability due to the decrease of the friction coefficient between $I=0$ and $I_*$, and is chosen in this context to illustrate the potential effects of such phenomenology on transient collapses, paving the way to more elaborate analyses.
Note that the increase of the effective friction coefficient at larger $I$ due to collisional dissipation, as predicted by the $\mu(I)$ rheology is also not accounted for in \cref{eq:muHyst} as we consider only low inertial number flows.

\begin{figure}
  \begin{center}
    \includegraphics[width=\textwidth]{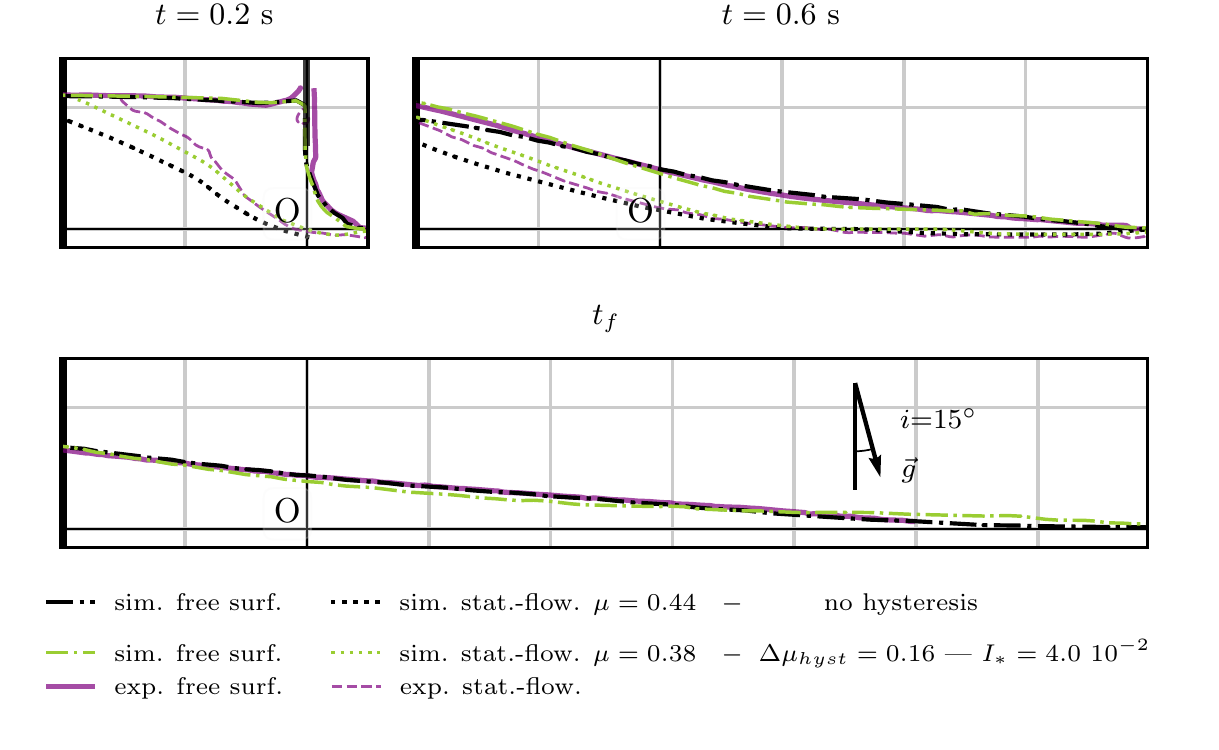}
    \caption{Hysteresis effect on the \SI{15}{\degree}-collapse (B15): simulation results obtained with the hysteretic law \cref{eq:muHyst} against experimental profiles and static-flowing transition contours. The non-hysteretic simulation results from \cref{fig:validationInclinationsBeads} are recalled for reference, and the hysteretic simulation corresponds to $\mu = \mu_{stop} = 0.38$, $\Delta\mu_{hyst} = 0.16$ and $I_*= 4 \times 10^{-2}$. See movie~6 in the online supplementary material for a visualisation of the collapse in time.}
    \label{fig:B15hyst}
  \end{center}
\end{figure}

\Cref{fig:B15hyst} shows the results for the $\SI{15}{\degree}$-inclined collapse, using the hysteretic law \labelcref{eq:muHyst} with $\mu=\mu_{stop}=0.38$, $\Delta\mu_{hyst}=0.16$ and a transition inertial number $I_{*} = 4 \times 10^{-2}$, consistent with the study of \citet{degiuli2017friction}. 
Despite the simplicity of the law, we observe that the increase of the friction coefficient for very low inertial numbers, characterised by the static friction coefficient $\mu + \Delta\mu_{hyst} = 0.54$, definitely improves the prediction of the flow dynamics at early stages, while only weakly affecting the final rest state.

We should stress that the corresponding increase of the ``static'' friction coefficient (up to $\mu + \Delta\mu_{hyst} = 0.38 + 0.16 = 0.54$) provided here is only phenomenological, and does not correspond to an independent experimental measurement. However, it suggests that the actual rest-to-flow yield transition, characterised by $\mu(I = 0)$, does not directly correspond to the experimental avalanche friction coefficient $\mu_a$, which measurement is highly sensitive to the mechanical noise and could thus incorporate nucleation effects \citep{degiuli2017friction, de2022scaling} close to the free-surface, where the pressure conditions can be impacted by preparation effects and small irregularities.
Within our approach, the measured $\mu_a$ would thus correspond to some small but non-zero $I$, somewhere between $0$ and $I_*$, more representative of the characteristic collapse inertial numbers than the smallest ones typically used to extract $\mu_{stop}$ from steady inclined flows, which are of order $I \lesssim I_0 = 0.279$~\citep{pouliquen1999scaling,jop2005crucial}.
Note that setting a higher $\mu(I=0)$ than $\mu_a$ also appears consistent with the higher friction coefficient values deduced from uni-axial or triaxial compression tests on similar micro-metric glass bead materials~\citep{ancey2001dry,adjemian2004experimental,cui2017stick} where friction coefficient values range from $\tan \SI{26.5}{\degree} \sim 0.5$ to $\tan \SI{30}{\degree} \sim 0.58$, much larger than $\mu_a=0.44$, illustrating the sensitivity of avalanche onsets to pressure conditions. 
Despite its simplicity, this ad-hoc hysteretic model thus supports the prominence of low inertial number effects in granular collapses, which appear mostly driven by solid-liquid transitions in the transient case, where the slope is not sufficient to sustain steady flow.

\section{Conclusion}
\label{sec:conclusion}
%

Our two- and three-dimensional non-smooth numerical model can faithfully simulate granular collapses, and quantitatively predict the final deposits for a wide range of bed inclinations, channel widths and column aspect ratios using a fully plastic model with a plain Drucker--Prager rheology.
In contrast with previous numerical investigations advocating more complex rheologies \citep{lagree2011granular,mast2015simulating,dunatunga2015continuum}, this suggests that transient granular flows are mostly driven by transitions from rest to flow, and that the final stable states can accurately be described by a unique constant bulk friction coefficient.

Comparisons with experimental collapses furthermore support the use of a friction coefficient corresponding to the avalanche angle, as opposed to the stop angle measured from steady experiments. This observation is not restricted to our numerical study, and was for example also noted by \cite{ionescu2015viscoplastic}, albeit not interpreted in this way: while the correction initially attributed to the role of the lateral walls cannot hold in the light of the study provided in \cref{sec:walls}, the effective friction coefficient they use turns out to coincide with the avalanche angle measured on their experiment.

Owing to the low inertial numbers involved in the friction-dominated collapse flows, the $\mu(I)$ rheology, which accounts for additional viscous collisional dissipation, is not able to quantitatively improve flow predictions. 
Furthermore, adjustment of the parameters of the $\mu(I)$ law using steady inclined flow experiments, in particular the \textit{stop} friction coefficient $\mu_{stop}$, appears unable to accurately predict experimental collapse profiles, and systematically underestimates frictional dissipation within the material.

Studying the effect of friction on the lateral walls on three-dimensional simulated collapses in narrow flumes with various widths, we have shown that a linear rescaling of the bulk friction coefficient for predicting two-dimensional collapses is still valid, but with an effective flow thickness much smaller than the maximum flow thickness used in \citet{ionescu2015viscoplastic}.

As a first step to explore low-velocity extensions of the rheology, we have implemented a simple hysteretic law inspired by the work of \citet{pouliquen2002friction} and \citet{degiuli2017friction}. This two-valued rheology, reminiscent of the static and dynamic regimes of Coulomb friction, significantly improves the prediction of the collapse dynamics, especially at the onset of the flow, while only weakly affecting the final state and run-out.
Again, this points out the crucial role played by static-flowing transitions in such transient configurations, which cannot be reproduced quantitatively by parameters measured in steady flow experiments. The extraction of solid-liquid friction coefficients is however very sensitive to mechanical noise and preparation effects, as the decreasing dependence of friction at low inertial numbers is a large source of instability, which can prevent access to the true $I=0$ limit in free-surface geometries.
We should also mention that the solid-liquid transition is strongly affected by non-local effects, as mechanical interactions at the grains scale are precisely responsible for nucleation or strengthening effects \citep{perrin2021nonlocal,kamrin2019non}, and improving our non-smooth numerical model to account for a non-local rheology would definitely provide additional insights into the role of spatial inhomogeneities in transient flows~\citep{mowlavi2021interplay}.

Our non-smooth solver provide good predictions of complex 3-D granular flows involving high granular deformations and interactions with frictional surfaces. However, the full potential of {\sc Sand6} supporting cohesion and dynamic interaction with complex objects remains to be fully exploited. A next step would be to simulate full 3-D steady flows with well controlled frictional boundary conditions and compare them to experiments in order to accurately bridge the gap between the constant friction rheology and the $\mu(I)$ rheology.

\backsection[Supplementary data]{\label{SupMat}In this preprint version movies are available at \\ \href{https://drive.switch.ch/index.php/s/xnxlGVw40Eh0KUt}{https://drive.switch.ch/index.php/s/xnxlGVw40Eh0KUt}}

\backsection[Acknowledgements]{We would like to thank Olivier Pouliquen, Pierre-Yves Lagrée, Ioan Ionescu, Hugo Perrin, Tom de Geus and Matthieu Wyart for their insights during this research study. We are grateful to Christophe Ancey, who graciously let us make experiments in the Environmental Hydraulic Laboratory at EPFL (Switzerland). We extend our appreciation to the anonymous reviewers for their constructive feedback, which has notably enhanced the quality of this paper.}

\backsection[Funding]{ This research was supported by EPFL, Inria, the ERC grant GEM (StG-2014-639139), and TU Wien.}

\backsection[Declaration of interests]{The authors report no conflict of interest.}

\backsection[Data availability statement]{Data and Python scripts used for plotting figures are archived at Zenodo \href{https://doi.org/10.5281/zenodo.7288829}{https://doi.org/10.5281/zenodo.7288829}}

\backsection[Author contributions]{G. R. and H. R. conducted the granular experiments. G. D., G. R. and T. M. adapted the Sand6 software to perform numerical collapses. G. R. and T. M. conducted the numerical experiments. G. R., T. M. and F. B.-D. analysed the comparisons results and performed the main scientific investigations. G. R., T. M. and F. B.-D. wrote the paper. All authors proofread the paper.}

\appendix

\section{Numerical method}
\label{appendix:numericalMethod}

\subsection{Modified Fischer-Burmeister function for the Drucker--Prager rheology}
\label{appendix:fischerBurmeisterFunction}

In this appendix, we give some details regarding the modified second-order cone Fischer-Burmeister complementarity function $f_\textrm{MFB}$ mentioned in \cref{subsec:numericalMethod} to impose the non-smooth Drucker--Prager rheology \labelcref{eq:fullDruckerPragerRheology}.
Note that, while we use the \textsc{Sand6} implementation from \citep{daviet2016semi} for our simulations, the presentation given here differs from the original one (\citep{daviet2016semi,daviet2016nonsmooth,daviet2011hybrid}) and avoids the need to introduce parallels between Coulomb friction and the Drucker--Prager rheology.

The first step to reformulate the Drucker--Prager rheology \labelcref{eq:fullDruckerPragerRheology} as a root-finding problem is to recast it as second-order cone complementarity problem (SOCCP).
We introduce the second-order cone
\begin{equation}
  \mathcal{K}_{\mu} = 
  \left\{ \tensor{\tau} \in S(\Dim); \: \lVert\Dev{\tensor{\tau}}\rVert \leq \mu \frac{\Trace{\tensor{\tau}}}{\sqrt{2\Dim}} \right\}
\end{equation}
where $S(\Dim)$ denotes the space of $\Dim \times \Dim$ symmetric rank-$2$ tensors, with dimension $s_\Dim = \frac{\Dim(\Dim+1)}{2}$, which can be decomposed as an orthogonal sum between the space generated by the unit basis vector $\tensor{\iota}_\Dim \equiv \sqrt{\frac{2}{\Dim}} \ID$ and the space of traceless symmetric tensors, namely $S(\Dim) = \mathrm{Span}\left\{\tensor{\iota}_\Dim\right\} \oplus T(\Dim)$. 

As shown in \citep{daviet2016nonsmooth}, the Drucker--Prager rheology \labelcref{eq:fullDruckerPragerRheology} is equivalent to the SOCCP
\begin{equation}
  \left( \tensor{\gamma}, \tensor{\lambda} \right) \in \mathcal{DP}(\mu) \iff
  \mathcal{K}_{\tilde{\mu}} \ni \tensor{\lambda} \perp \tilde{\tensor{\gamma}} \in \mathcal{K}_{\frac{1}{\tilde{\mu}}}
  \label{eq:SOCCP}
\end{equation}
with 
\begin{equation}
  \tilde{\mu} = \sqrt{\frac{2}{\Dim}} \, \mu
\end{equation}
and
\begin{equation}
  \begin{aligned}
    \tilde{\tensor{\gamma}} 
    &= \tensor{\gamma} + \tilde{\mu} \, \lVert\Dev{\tensor{\gamma}}\rVert \, \tensor{\iota}_\Dim
    \\
    &= \tensor{\gamma} + \tilde{\mu} \sqrt{\frac{2}{\Dim}} \, \lVert\Dev{\tensor{\gamma}}\rVert \, \ID.
  \end{aligned}
\end{equation}
Note that the SOCCP is expressed in terms of $\tilde{\mu}$ instead of $\mu$ to exhibit symmetry in the SOCCP inclusions in $\mathcal{K}_{\tilde{\mu}}$ and its dual cone $\mathcal{K}_{\frac{1}{\tilde{\mu}}}$.

\Cref{eq:SOCCP} can be further symmetrised and put in a canonical self-dual form with the additional change of variable 
\begin{equation}
  \begin{aligned}
    \hat{\tensor{\lambda}} 
    &= \tilde{\mu} \, \frac{\Trace{\tensor{\lambda}}}{\sqrt{2\Dim}} \,\tensor{\iota}_\Dim + \Dev{\tensor{\lambda}} \\
    &= \tilde{\mu} \, \frac{\Trace{\tensor{\lambda}}}{\Dim} \,\ID + \Dev{\tensor{\lambda}}
    \\
    \hat{\tilde{\tensor{\gamma}}} 
    &= \frac{\Trace{\tilde{\tensor{\gamma}}}}{\sqrt{2\Dim}} \,\tensor{\iota}_\Dim + \tilde{\mu} \, \Dev{\tilde{\tensor{\gamma}}}\\
    &= \frac{\Trace{\tilde{\tensor{\gamma}}}}{\Dim} \,\ID + \tilde{\mu} \, \Dev{\tilde{\tensor{\gamma}}}
  \end{aligned}
\end{equation}
so that
\begin{equation}
  \left( \tensor{\gamma}, \tensor{\lambda} \right) \in \mathcal{DP}(\mu) \iff
  \mathcal{K}_{1} \ni \hat{\tensor{\lambda}} \perp \hat{\tilde{\tensor{\gamma}}} \in \mathcal{K}_{1}.
  \label{eq:symmetrisedSOCCP}
\end{equation}
where $\mathcal{K}_{1}=\left\{ \tensor{\tau} \in S(\Dim); \: \lVert\Dev{\tensor{\tau}}\rVert \leq \frac{\Trace{\tensor{\tau}}}{\sqrt{2\Dim}} \right\}$ can also be identified as the self-dual Lorentz cone in $\mathbb{R}^{s_\Dim - 1}$ using the natural orthonormal isomorphism between $\left( S(\Dim) = \mathrm{Span}\left\{\tensor{\iota}_\Dim\right\} \oplus T(\Dim), \langle \bullet, \bullet \rangle \right)$ and $\left( \mathbb{R} \oplus \mathbb{R}^{s_\Dim - 1}, \bullet \cdot \bullet \right)$. 

The new symmetric SOCCP \labelcref{eq:symmetrisedSOCCP} now falls directly in the framework of \citet{fukushima2002smoothing}, which allows to rewrite it as a root-finding problem on a \emph{modified} Fischer-Burmeister function
\begin{equation}
  \left( \tensor{\gamma}, \tensor{\lambda} \right) \in \mathcal{DP}(\mu) \iff
  f_{\mathrm{MFB}}(\tensor{\gamma},\tensor{\lambda}) = 0.
\end{equation}
with
\begin{equation}
  \begin{aligned}
    f_{\mathrm{MFB}}: \;
    &S(\Dim) \times S(\Dim) \to S(\Dim)
    \\
    &(\tensor{\gamma},\tensor{\lambda}) \mapsto 
    f_{\mathrm{FB}}(\hat{\tilde{\tensor{\gamma}}}, \hat{\tensor{\lambda}})
  \end{aligned}
\end{equation}
where
\begin{equation}
  \begin{aligned} 
    f_{\mathrm{FB}}(\hat{\tilde{\tensor{\gamma}}}, \hat{\tensor{\lambda}}) &=
    \left( \Trace{\hat{\tilde{\tensor{\gamma}}}} + \Trace{\hat{\tensor{\lambda}}} - s \right) \, \frac{\ID}{\Dim} 
    + \left( 1 - \frac{\Trace{\hat{\tilde{\tensor{\gamma}}}}}{s} \right) \,\Dev{\hat{\tilde{\tensor{\gamma}}}}
    + \left( 1 - \frac{\Trace{\hat{\tensor{\lambda}}}}{s} \right) \,\Dev{\hat{\tensor{\lambda}}}
    \\
    s &= \sqrt{ \Dim \left( 
      \lVert \hat{\tilde{\tensor{\gamma}}} \rVert^2 + \lVert \hat{\tensor{\lambda}} \rVert^2 + \sqrt{ \left( \lVert \hat{\tilde{\tensor{\gamma}}} \rVert^2 + \lVert \hat{\tensor{\lambda}} \rVert^2 \right)^2 - \frac{2}{\Dim} \left\lVert \Trace{\hat{\tilde{\tensor{\gamma}}}} \Dev{\hat{\tilde{\tensor{\gamma}}}} + \Trace{\hat{\tensor{\lambda}}} \Dev{\hat{\tensor{\lambda}}} \right\rVert^2 }
      \right) }.
  \end{aligned}
\end{equation}

\subsection{Spatial discretization}
\label{appendix:spatialDiscretization}

We discretise the conservations equations~(\ref{eq:massConservation} -- \ref{eq:momentumConservation}) using the material point method (MPM)~\citep{sulsky1995application, bardenhagen2000material},
which leverages both an Eulerian grid to enforce momentum conservation and a Lagrangian particle representation to resolve transport terms.

The volume fraction field $\phi$ is thus approximated as a set of $N$ material points $(\vec{x}_p)_{1 \leq p \leq N}$ with finite material volume $V_p$ and velocity $\vec{v}_p$, following the mathematical distribution 
\begin{equation}
  \phi(\vec{x}, t) = \sum_p{V_p \,\delta\left(\vec{x} - \vec{x}_p(t)\right)}.
  \label{eq:phiMPMDiscretisation}
\end{equation}
Discrete-time mass conservation~\labelcref{eq:discreteMassConservation} is then achieved by advecting the particles over each time-step 
in a semi-implicit way as $\vec{x}_p^{(n+1)} = \vec{x}_p^{(n)} + \delta t \,\vec{v}_p^{(n+1)}$, with $\vec{v}_p^{(n+1)}$ 
sampled from the continuous velocity field $\vec{u}$ as 
$\vec{v}_p^{(n+1)} = \vec{u}^{(n+1)}\left(\vec{x}_p^{(n)}\right)$. 

The material derivative $\frac{D\vec{u}}{D t}$ in the momentum conservation equation~(\ref{eq:massConservation}) 
is first discretised in time as $\frac{\vec{u}^{(n+1)} - \vec{u}^{(*)} }{\delta t}$, where $\vec{u}^{(*)}$ is the velocity field 
recovered by transferring back the particle velocity from the previous time-step to the grid. In practice, we use the 
APIC velocity transfer scheme from \citet{jiang2015affine}.

In order to discretise the momentum conservation equation in space, 
 we first rewrite the Cauchy stress without loss of generality as $\tensor{\sigma} = \phi \hat{\tensor{\sigma}}$, and similarly, 
$\tensor{\lambda} = \phi \hat{\tensor{\lambda}}$ and $\vec{f} = \phi \hat{\vec{f}}$. We recall that the Drucker--Prager rheology is 
invariant with respect to a positive scaling factor on the stress~\citep{daviet2016semi}, so that 
$(\tensor{\gamma},\tensor{\lambda}) \in \mathcal{DP}(\mu) \iff (\tensor{\gamma},\hat{\tensor{\lambda}}) \in \mathcal{DP}(\mu)$.

Now, let $V \subset H^1_0(\Omega)$ be a discrete space of square-integrable velocity fields with square-integrable gradients over $\Omega$, and 
$T \subset L^2(\Omega)$ a space of square-integrable symmetric tensor fields. Note that in practice, we use the space of trilinear shape functions over a regular Cartesian grid 
for both $U$ and $T$.
\Cref{eq:momentumConservation} corresponds to the variational formulation
\begin{equation*}
  \rho_g \int_{\Omega}  \phi \vec{w}^T \frac{\vec{u}^{(n+1)} - \vec{u}^{*} }{\delta t}  - \int_{\Omega}{\vec{w}^T \nabla \cdot \left[ \phi \hat{\tensor{\lambda}}\right]  }  = 
  \int_{\Omega} \phi \vec{w}^T \hat{\vec{f}},  \quad \forall \vec{w} \in V,
\end{equation*}
or after integration by parts,
\begin{equation}
  \label{eq:weakMassConservation}
  \rho_g \int_{\Omega}  \phi \vec{w}^T \frac{\vec{u}^{(n+1)} - \vec{u}^{*} }{\delta t}  + \int_{\Omega}{ \nabla \vec{w} : \left[ \phi \hat{\tensor{\lambda}} \right]  }  = 
  \int_{\Omega} \phi \vec{w}^T \hat{\vec{f}} -  \int_{\partial \Omega}{\phi \vec{w}^T \hat{\tensor{\lambda}} \vec{n}  }, \quad \forall \vec{w} \in V.
\end{equation}
Note that the boundary term vanishes if either Dirichlet boundary conditions are used or the domain extends sufficiently
far away from the material such that $\phi_{\vert \partial \Omega} = 0$.

Using the discrete expression for $\phi$~\labelcref{eq:phiMPMDiscretisation}, we can rewrite the variational mass conservation~\labelcref{eq:weakMassConservation} as 
\begin{equation}
  a(\vec{u}, \vec{w}) = b(\vec{u}, \vec{w}) + l(\vec{w})  \quad \forall \vec{w} \in V
\end{equation}
with 
\begin{align*}
a(\vec{u}, \vec{w}) &:= \frac {\rho_g}{\delta t} \sum_p V_p \vec{w}(\vec{x}_p)^T \vec{u}(\vec{x}_p) \\
b(\hat{\tensor{\lambda}}, \vec{w}) &:= \sum_p V_p (\nabla {w})(\vec{x}_p) : \hat{\tensor{\lambda}}(\vec{x}_p) \\
l(\vec{w}) &:=  \sum_p V_p \left[ \hat{f} + \vec{w}(\vec{x}_p)^T \vec{u}^{(*)}(\vec{x}_p) \right].\\
\end{align*}

Similarly, we write the definition of the auxiliary strain rate tensor $\tensor{\gamma}$ in a variational form as 
\begin{equation*}
\int_{\Omega} \tensor{\gamma} : \tensor{\tau} = \int_{\Omega}{\phi \nabla \vec{u} : \tensor{\tau}} + \frac 1 {\delta t}\int_{\Omega}{ \frac{\left( \phi - \phi^c \right)}{\Dim}\, \ID : \tensor{\tau}} \quad \forall \tensor{\tau} \in T,
\end{equation*}
or equivalently
\begin{equation}
  s(\tensor{\gamma}, \tensor{\tau}) = b(\tensor{\tau}, \vec{w}) + k(\tensor{\tau})  \quad \forall \tensor{\tau} \in T
\end{equation}
with 
\begin{align*}
s(\tensor{\gamma}, \tensor{\tau}) &:= \int_{\Omega} {\tensor{\gamma}} : \tensor{\tau} \\
k(\tensor{\tau}) &:=  \frac 1 {\Dim \delta t} \sum_p V_p \mathbb{I} : \tensor{\tau} - \frac 1 {\Dim \delta t} \int_{\Omega} \phi^c \mathbb{I} : \tensor{\tau}
\end{align*}

The discrete system is then obtained by assembling the matrices $M, B, S$ and vectors $\vec{l}, \vec{k}$ corresponding to the bilinear forms $m, b ,s$  and linear forms $a, b, s$ and linear forms $l, k$, respectively.
Note however that we follow the suggestions from~\citep{daviet2016semi, daviet2016nonsmooth} to replace the 
matrix $M$ with its lumped diagonal version $\hat{M}$ defined as that $\hat{M}_{i,j} = \delta_i^j \sum_k M_{i,k}$ 
(which is consistent with our use of the APIC particle--grid transfer scheme)
and to replace $S$ with the identity matrix (which amounts to perform numerical integration of $s$ using the trapezoidal rule). 

This finally leads to the algebraic problem
\begin{equation}
  \begin{gathered}
  \textrm{Find } \vec{u}, \tensor{\gamma}, \hat{\tensor{\lambda}} \in \mathbb{R}^{\Dim} \times \mathbb{R}^{s_\Dim} \times \mathbb{R}^{s_\Dim} \textrm{ s.t.}
  \\
  \left\lbrace\begin{aligned}
    &\hat{M} \vec{u} = B^T \hat{\tensor{\lambda}} + \vec{l}\\
    &\tensor{\gamma} = B \vec{u} + \vec{k}
    \\
    &(\tensor{\gamma},\tensor{\lambda}) \in \mathcal{DP}(\mu) \textrm{,}
  \end{aligned}\right.
  \end{gathered}
  \label{eq:weakDDPP}
\end{equation}
from which we can eliminate the velocity variable $\vec{u}$ by introducting the 
Schur complement $W := B \hat{M}^{-1} B^T$, yielding problem~(\ref{eq:dual}).

Note that the constrained algebraic problem \cref{eq:weakDDPP} can naturally be extended to incorporate rigid body dynamics, with two-way frictional boundary interaction with the granular material obeying a Coulomb-like condition. Details regarding the coupling with rigid bodies are provided in \cite{daviet2016semi}.

\subsection{Gauss-Seidel algorithm}
\label{appendix:gaussSeidelAlgorithm}

Problem~(\ref{eq:dual}) could be solved with any technique able to address 
discrete Coulomb friction problems; here we follow the method of \citet{daviet2011hybrid},
which is itself a variant of the Non-Smooth Contact Dynamics~\citep{jean1999nonsmooth} algorithm.

In this approach, the contacts (here, the instances of the Drucker--Prager condition~\ref{eq:fullDruckerPragerRheology}) are 
repeatedly solved one by one in a Gauss-Seidel approach:
at the $k^{\text{th}}$ iteration of the algorithm, and for each discrete degree of freedom $i$ of our 
tensor fields, we solve for the local stress $\tensor{\lambda}^{[k+1]}_i$ assuming that all other stress degrees of freedom are frozen, i.e.

\begin{equation}
  \begin{gathered}
  \textrm{Find } \tensor{\gamma}_i^{[k+1]}, \tensor{\lambda}_i^{[k+1]} \in \mathbb{R}^{s_\Dim} \times \mathbb{R}^{s_\Dim} \textrm{ s.t.}
  \\
  \left\lbrace\begin{aligned}
    &\tensor{\gamma}_i^{[k+1]} = \tensor{W_{ii}} \tensor{\lambda_i}^{[k+1]} + \underbrace{\tensor{b}_i + \sum_{j < i}\tensor{W_{ij}} \tensor{\lambda_j}^{[k+1]}  + \sum_{i > j}\tensor{W_{ij}} \tensor{\lambda_j}^{[k]}}_{\tensor{b}_i^{[k+1]}}
    \\
    &(\tensor{\gamma}_i^{[k+1]},\tensor{\lambda}_i^{[k+1]}) \in \mathcal{DP}(\mu). 
  \end{aligned}\right.
  \end{gathered}
  \label{eq:local}
\end{equation}
iterating (on $k$) until convergence.

The local problem~\ref{eq:local} is equivalent to solving
$$
  f_{\mathrm{MFB}}(\tensor{W_{ii}} \tensor{\lambda_i}^{[k+1]} + \tensor{b}_i^{[k+1]},\tensor{\lambda_i}^{[k+1]}) = 0,
$$
which we do using a generalised (non-smooth) Newton algorithm.

Note that for $\Dim=2$, the dimension of $S(\Dim)$ is $3$, which means the problem has 
a structure similar to that of discrete Coulomb friction. In this case, we 
directly reuse the solver from~\citep{daviet2011hybrid}, which combines 
the Newton-based optimization problem with an analytical solver based on finding the roots of a degree-$4$ polynomial.
For $\Dim=3$, the dimension of $S(\Dim)$ is $6$, and to the best of our knowledge, no analytical solver is available.
We thus use the Newton-based solver only.

\section{Additional results}
\label{appendix:additionalResults}

\subsection{Comparison of the different rheologies}
\label{appendix:frictionCoefficientComparison}

\begin{figure}
  \centering
  \begin{subfigure}[b]{\textwidth}
    \includegraphics[width=\textwidth]{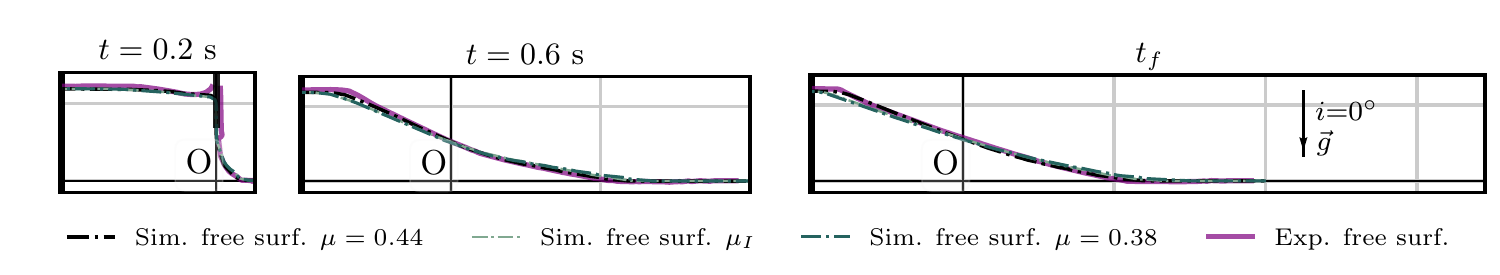}
  \end{subfigure}

  \begin{subfigure}[b]{\textwidth}
    \includegraphics[width=\textwidth]{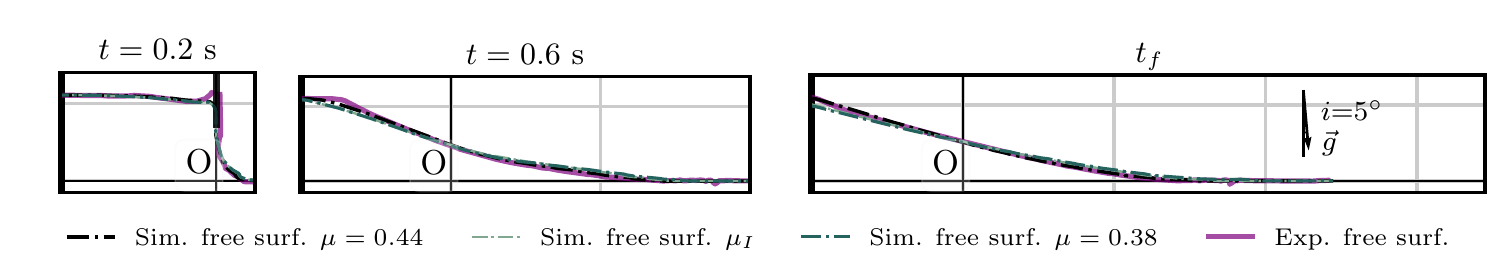}
  \end{subfigure}

  \begin{subfigure}[b]{\textwidth}
    \includegraphics[width=\textwidth]{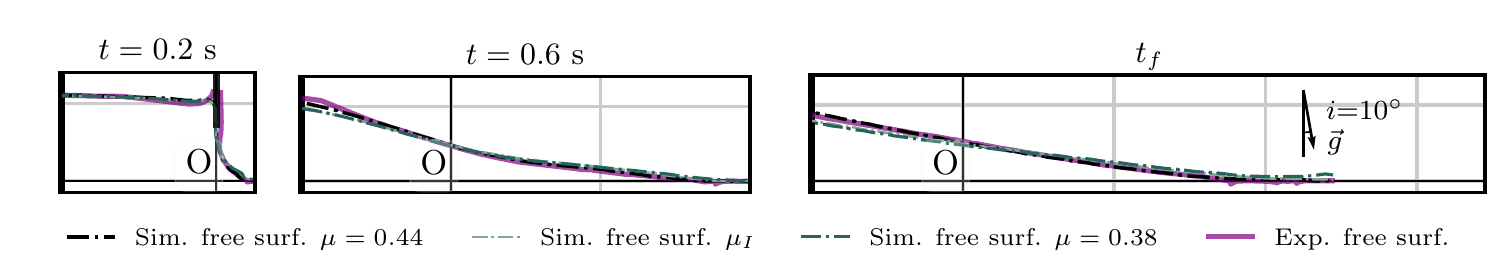}
  \end{subfigure}

  \begin{subfigure}[b]{\textwidth}
    \includegraphics[width=\textwidth]{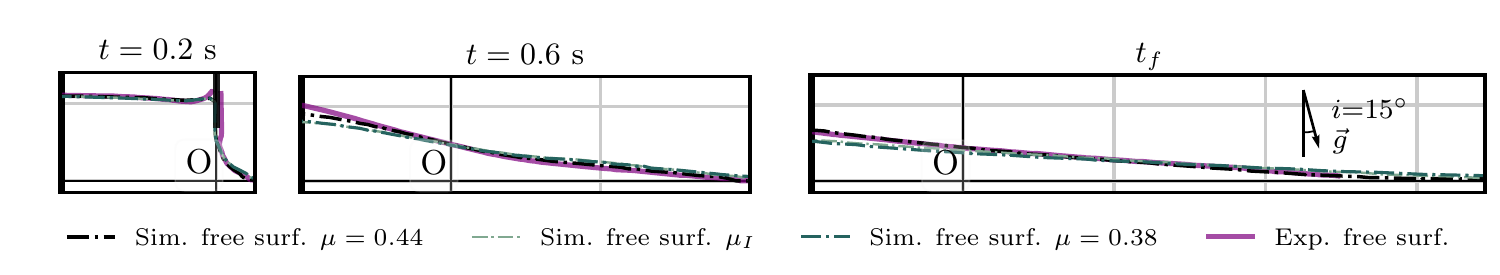}
  \end{subfigure}

  \begin{subfigure}[b]{\textwidth}
    \includegraphics[width=\textwidth]{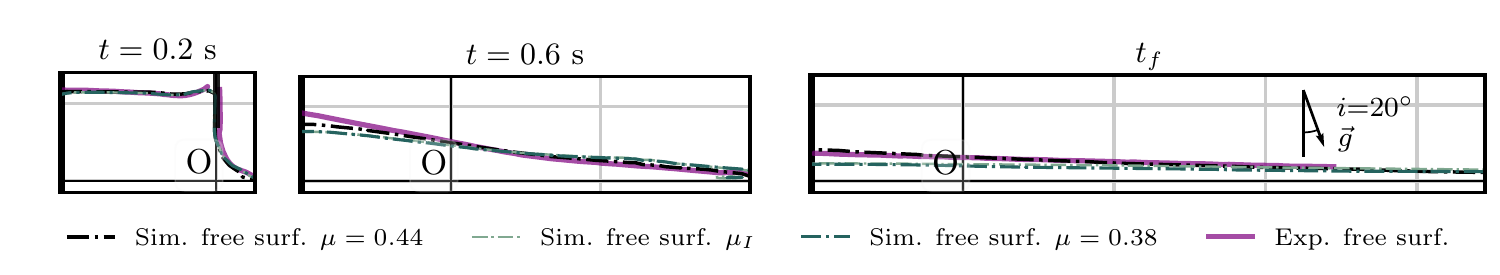}
  \end{subfigure}

  \caption{Comparison of 3-D simulations performed with $\mu = \mu_a = 0.44$ (black lines), $\mu=\mu(I)$ and $\mu = \mu_{stop} = 0.38$ (blue lines) for the all the bead collapses at $t=\SI{0.2}{s}$, $\SI{0.6}{s}$ and $t_{f}$. Experimental curves (solid pink line) are provided for reference, and we show both the free-surface lines (resp. plain and dash-dotted) and the static-flowing transition contours (resp. dashed and dotted).}  
  \label{fig:comparisonMuAMuIMuSAllInclinations}
\end{figure}

\Cref{fig:comparisonMuAMuIMuSAllInclinations} collects the exhaustive comparisons between the $\mu = \mu_a$, the $\mu = \mu_{stop}$ and $\mu = \mu(I)$ rheologies on bead collapses for all inclinations.

It supports the observations of \cref{subsec:comparisonMuIRheology}, highlighting the systematic underestimation of the internal friction by the $\mu_{stop}$ and $\mu(I)$ rheologies, which give very similar results.

\subsection{Impact of the lifting gate}
\label{subsec:impactLiftingGate}

\begin{figure}
  \centering
  \includegraphics[width=\textwidth]{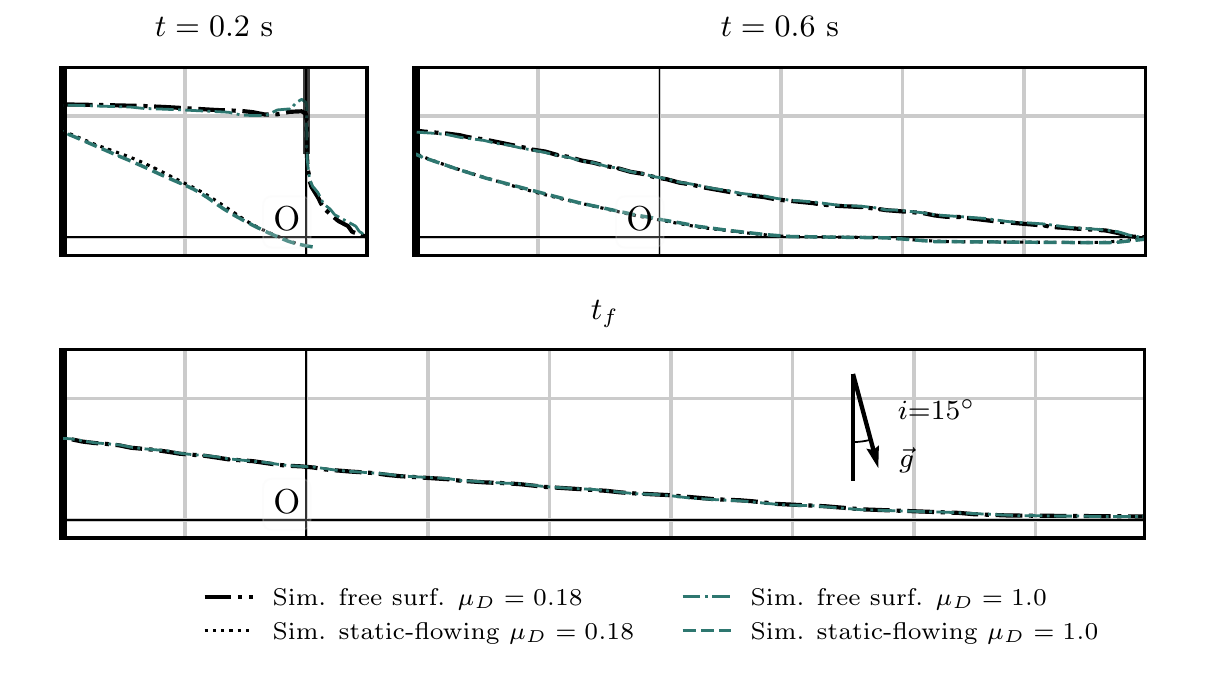}
  \caption{Impact of the lifting gate friction: comparison of simulations of the $15^{\circ}$ bead collapse with the standard friction coefficient between the material and the door $\mu_D = 0.18$ (black lines) and a non-realistic high coefficient $\mu_D = 1$ (green lines).}
  \label{fig:doorFrictionComparison}
\end{figure}

In order to check the weak impact of friction between the granular material and the lifting gate, we have run a simulation with a non-realistically high granular-gate friction coefficient $\mu_D = 1$, and compared with the value used in the paper ($\mu_D = 0.18$).
Both simulations use the same bulk coefficient corresponding to the bead material ($\mu = 0.44$) and the same numerical parameters. The resulting height profiles and static-flowing transition contours in \cref{fig:doorFrictionComparison} show that while friction with the gate can indeed affect very locally the profile close to the door for early times, it does not impact the flow once the gate is fully lifted, and gives the same collapse free-surface and overall dynamics.

\bibliographystyle{jfm}

\clearpage

\begin{thebibliography}{69}
  \expandafter\ifx\csname natexlab\endcsname\relax\def\natexlab#1{#1}\fi
  \def\au#1{#1} \def\ed#1{#1} \def\yr#1{#1}\def\at#1{#1}\def\jt#1{\textit{#1}}
    \def\bt#1{#1}\def\bvol#1{\textbf{#1}} \def\vol#1{#1} \def\pg#1{#1}
    \def\publ#1{#1}\def\arxiv#1{#1}\def\org#1{#1}\def\st#1{\textit{#1}}
  
  \bibitem[Adjemian \& Evesque(2004)]{adjemian2004experimental}
  {\sc \au{Adjemian, F.} \& \au{Evesque, P.}} \yr{2004}  \at{Experimental study
    of stick-slip behaviour}.  \jt{Int. J. Numer. Analyt. Meth. Geomech.}
    \bvol{28}~(6),  \pg{501--530}.
  
  \bibitem[Ancey(2001)]{ancey2001dry}
  {\sc \au{Ancey, C.}} \yr{2001}  \at{Dry granular flows down an inclined
    channel: Experimental investigations on the frictional-collisional regime}.
    \jt{Phys. Rev. E}  \bvol{65}~(1),  \pg{011304}.
  
  \bibitem[Andreotti {\em et~al.\/}(2013)Andreotti, Forterre \&
    Pouliquen]{andreotti2013granular}
  {\sc \au{Andreotti, B.}, \au{Forterre, Y.} \& \au{Pouliquen, O.}} \yr{2013}
    {\em Granular media: between fluid and solid\/}.  \publ{Cambridge University
    Press}.
  
  \bibitem[Artoni {\em et~al.\/}(2011)Artoni, Santomaso \&
    Canu]{artoni2011hysteresis}
  {\sc \au{Artoni, R.}, \au{Santomaso, A.} \& \au{Canu, P.}} \yr{2011}
    \at{Hysteresis in a hydrodynamic model of dense granular flows}.  \jt{Phys.
    Rev. E}  \bvol{83}~(5),  \pg{051304}.
  
  \bibitem[Az{\'e}ma \& Radjai(2014)]{azema2014internal}
  {\sc \au{Az{\'e}ma, E.} \& \au{Radjai, F.}} \yr{2014}  \at{Internal structure
    of inertial granular flows}.  \jt{Phys. Rev. Lett.}  \bvol{112}~(7),
    \pg{078001}.
  
  \bibitem[Bagnold(1954)]{bagnold1954experiments}
  {\sc \au{Bagnold, R.A.}} \yr{1954}  \at{Experiments on a gravity-free
    dispersion of large solid spheres in a newtonian fluid under shear}.
    \jt{Proc. R. Soc. London ser. A}  \bvol{225}~(1160),  \pg{49--63}.
  
  \bibitem[Balmforth \& Kerswell(2005)]{balmforth2005granular}
  {\sc \au{Balmforth, N.J.} \& \au{Kerswell, R.R.}} \yr{2005}  \at{Granular
    collapse in two dimensions}.  \jt{J. Fluid Mech.}  \bvol{538},
    \pg{399--428}.
  
  \bibitem[Bardenhagen {\em et~al.\/}(2000)Bardenhagen, Brackbill \&
    Sulsky]{bardenhagen2000material}
  {\sc \au{Bardenhagen, S.G.}, \au{Brackbill, J.U.} \& \au{Sulsky, D.}} \yr{2000}
     \at{The material-point method for granular materials}.  \jt{Comput. Methods
    Appl. Mech. Eng.}  \bvol{187}~(3-4),  \pg{529--541}.
  
  \bibitem[Boutreux \& de~Gennes(1997)]{boutreux1997evolution}
  {\sc \au{Boutreux, Thomas} \& \au{de~Gennes, Pierre-Gilles}} \yr{1997}
    \at{Evolution of a step in a granular material: the sinai problem}.
    \jt{Comptes Rendus de l'Academie des Sciences Series IIB Mechanics Physics
    Chemistry Astronomy}  \bvol{2}~(325),  \pg{85--89}.
  
  \bibitem[Chauchat \& M{\'e}dale(2014)]{chauchat2014three}
  {\sc \au{Chauchat, Julien} \& \au{M{\'e}dale, Marc}} \yr{2014}  \at{A
    three-dimensional numerical model for dense granular flows based on the $\mu$
    (i) rheology}.  \jt{J. Comput. Phys.}  \bvol{256},  \pg{696--712}.
  
  \bibitem[Chupin {\em et~al.\/}(2021)Chupin, Dubois, Phan \&
    Roche]{chupin2021pressure}
  {\sc \au{Chupin, L.}, \au{Dubois, T.}, \au{Phan, M.} \& \au{Roche, O.}}
    \yr{2021}  \at{Pressure-dependent threshold in a granular flow: Numerical
    modeling and experimental validation}.  \jt{J. Non-Newtonian Fluid Mech.}
    \bvol{291},  \pg{104529}.
  
  \bibitem[Cui {\em et~al.\/}(2017)Cui, Wu, Xiang, Doanh, Chen, Wang, Liu \&
    Wang]{cui2017stick}
  {\sc \au{Cui, D.}, \au{Wu, W.}, \au{Xiang, W.}, \au{Doanh, T.}, \au{Chen, Q.},
    \au{Wang, S.}, \au{Liu, Q.} \& \au{Wang, J.}} \yr{2017}  \at{Stick-slip
    behaviours of dry glass beads in triaxial compression}.  \jt{Granul. Matter}
    \bvol{19}~(1),  \pg{1--18}.
  
  \bibitem[Da~Cruz {\em et~al.\/}(2002)Da~Cruz, Chevoir, Bonn \&
    Coussot]{da2002viscosity}
  {\sc \au{Da~Cruz, F}, \au{Chevoir, F}, \au{Bonn, Daniel} \& \au{Coussot, Ph}}
    \yr{2002}  \at{Viscosity bifurcation in granular materials, foams, and
    emulsions}.  \jt{Phys. Rev. E}  \bvol{66}~(5),  \pg{051305}.
  
  \bibitem[Da~Cruz {\em et~al.\/}(2005)Da~Cruz, Emam, Prochnow, Roux \&
    Chevoir]{da2005rheophysics}
  {\sc \au{Da~Cruz, F.}, \au{Emam, S.}, \au{Prochnow, M.}, \au{Roux, J.-N.} \&
    \au{Chevoir, F.}} \yr{2005}  \at{Rheophysics of dense granular materials:
    Discrete simulation of plane shear flows}.  \jt{Phys. Rev. E}  \bvol{72}~(2),
     \pg{021309}.
  
  \bibitem[Daerr \& Douady(1999{\natexlab{{\em a\/}}})]{daerr1999sensitivity}
  {\sc \au{Daerr, A} \& \au{Douady, St{\'e}phane}} \yr{1999{\natexlab{{\em
    a\/}}}}  \at{Sensitivity of granular surface flows to preparation}.
    \jt{Europhys. Lett.}  \bvol{47}~(3),  \pg{324}.
  
  \bibitem[Daerr \& Douady(1999{\natexlab{{\em b\/}}})]{daerr1999two}
  {\sc \au{Daerr, A.} \& \au{Douady, S.}} \yr{1999{\natexlab{{\em b\/}}}}
    \at{Two types of avalanche behaviour in granular media}.  \jt{Nature}
    \bvol{399}~(6733),  \pg{241--243}.
  
  \bibitem[Daviet \& Bertails-Descoubes(2016{\natexlab{{\em
    a\/}}})]{daviet2016nonsmooth}
  {\sc \au{Daviet, G.} \& \au{Bertails-Descoubes, F.}} \yr{2016{\natexlab{{\em
    a\/}}}}  \at{Nonsmooth simulation of dense granular flows with
    pressure-dependent yield stress}.  \jt{J. Non-Newtonian Fluid Mech.}
    \bvol{234},  \pg{15--35}.
  
  \bibitem[Daviet \& Bertails-Descoubes(2016{\natexlab{{\em
    b\/}}})]{daviet2016semi}
  {\sc \au{Daviet, G.} \& \au{Bertails-Descoubes, F.}} \yr{2016{\natexlab{{\em
    b\/}}}}  \at{A semi-implicit material point method for the continuum
    simulation of granular materials}.  \jt{ACM Trans. Graph.}  \bvol{35}~(4),
    \pg{102}.
  
  \bibitem[Daviet {\em et~al.\/}(2011)Daviet, Bertails-Descoubes \&
    Boissieux]{daviet2011hybrid}
  {\sc \au{Daviet, G.}, \au{Bertails-Descoubes, F.} \& \au{Boissieux, L.}}
    \yr{2011} A hybrid iterative solver for robustly capturing coulomb friction
    in hair dynamics.  \bt{In {\em Proceedings of the 2011 SIGGRAPH Asia
    Conference\/}},  \pg{pp. 1--12}.
  
  \bibitem[DeGiuli {\em et~al.\/}(2016)DeGiuli, McElwaine \&
    Wyart]{degiuli2016phase}
  {\sc \au{DeGiuli, E.}, \au{McElwaine, J.~N.} \& \au{Wyart, M.}} \yr{2016}
    \at{Phase diagram for inertial granular flows}.  \jt{Phys. Rev. E}
    \bvol{94}~(1),  \pg{012904}.
  
  \bibitem[DeGiuli \& Wyart(2017)]{degiuli2017friction}
  {\sc \au{DeGiuli, E.} \& \au{Wyart, M.}} \yr{2017}  \at{Friction law and
    hysteresis in granular materials}.  \jt{Proc. Natl. Acad. Sci. U.S.A.}
    \bvol{114}~(35),  \pg{9284--9289}.
  
  \bibitem[Drucker \& Prager(1952)]{drucker1952soil}
  {\sc \au{Drucker, D.~C.} \& \au{Prager, W.}} \yr{1952}  \at{Soil mechanics and
    plastic analysis or limit design}.  \jt{Quarterly of applied mathematics}
    \bvol{10}~(2),  \pg{157--165}.
  
  \bibitem[Dunatunga \& Kamrin(2015)]{dunatunga2015continuum}
  {\sc \au{Dunatunga, S.} \& \au{Kamrin, K.}} \yr{2015}  \at{Continuum modelling
    and simulation of granular flows through their many phases}.  \jt{J. Fluid
    Mech.}  \bvol{779},  \pg{483--513}.
  
  \bibitem[Farin {\em et~al.\/}(2014)Farin, Mangeney \&
    Roche]{farin2014fundamental}
  {\sc \au{Farin, M.}, \au{Mangeney, A.} \& \au{Roche, O.}} \yr{2014}
    \at{Fundamental changes of granular flow dynamics, deposition, and erosion
    processes at high slope angles: insights from laboratory experiments}.
    \jt{J. Geophys. Res Earth Surf.}  \bvol{119}~(3),  \pg{504--532}.
  
  \bibitem[Forterre \& Pouliquen(2003)]{forterre2003long}
  {\sc \au{Forterre, Y.} \& \au{Pouliquen, O.}} \yr{2003}  \at{Long-surface-wave
    instability in dense granular flows}.  \jt{J. Fluid Mech.}  \bvol{486},
    \pg{21--50}.
  
  \bibitem[Franci \& Cremonesi(2019)]{franci20193d}
  {\sc \au{Franci, A.} \& \au{Cremonesi, M.}} \yr{2019}  \at{3d regularized $\mu$
    ({I})-rheology for granular flows simulation}.  \jt{J. Comput. Phys.}
    \bvol{378},  \pg{257--277}.
  
  \bibitem[Fukushima {\em et~al.\/}(2002)Fukushima, Luo \&
    Tseng]{fukushima2002smoothing}
  {\sc \au{Fukushima, M.}, \au{Luo, Z.-Q.} \& \au{Tseng, P.}} \yr{2002}
    \at{Smoothing functions for second-order-cone complementarity problems}.
    \jt{SIAM Journal on optimization}  \bvol{12}~(2),  \pg{436--460}.
  
  \bibitem[Gaume {\em et~al.\/}(2018)Gaume, Gast, Teran, van Herwijnen \&
    Jiang]{gaume2018dynamic}
  {\sc \au{Gaume, J.}, \au{Gast, T.}, \au{Teran, J.}, \au{van Herwijnen, A.} \&
    \au{Jiang, C.}} \yr{2018}  \at{Dynamic anticrack propagation in snow}.
    \jt{Nature communications}  \bvol{9}~(1),  \pg{3047}.
  
  \bibitem[de~Geus \& Wyart(2022)]{de2022scaling}
  {\sc \au{de~Geus, Tom~WJ} \& \au{Wyart, Matthieu}} \yr{2022}  \at{Scaling
    theory for the statistics of slip at frictional interfaces}.  \jt{Phys. Rev.
    E}  \bvol{106}~(6),  \pg{065001}.
  
  \bibitem[Hutter \& Koch(1991)]{hutter1991motion}
  {\sc \au{Hutter, K.} \& \au{Koch, T.}} \yr{1991}  \at{Motion of a granular
    avalanche in an exponentially curved chute: experiments and theoretical
    predictions}.  \jt{Phil. Trans. Roy. Soc. London A}  \bvol{334}~(1633),
    \pg{93--138}.
  
  \bibitem[Ionescu {\em et~al.\/}(2015)Ionescu, Mangeney, Bouchut \&
    Roche]{ionescu2015viscoplastic}
  {\sc \au{Ionescu, I.R.}, \au{Mangeney, A.}, \au{Bouchut, F.} \& \au{Roche, O.}}
    \yr{2015}  \at{Viscoplastic modeling of granular column collapse with
    pressure-dependent rheology}.  \jt{J. Non-Newtonian Fluid Mech.}  \bvol{219},
     \pg{1--18}.
  
  \bibitem[Jean(1999)]{jean1999nonsmooth}
  {\sc \au{Jean, M.}} \yr{1999}  \at{The non-smooth contact dynamics method}.
    \jt{Computer Methods in Applied Mech, and Engineering}  \bvol{177}~(3),
    \pg{235--257}.
  
  \bibitem[Jiang {\em et~al.\/}(2015)Jiang, Schroeder, Selle, Teran \&
    Stomakhin]{jiang2015affine}
  {\sc \au{Jiang, C.}, \au{Schroeder, C.}, \au{Selle, A.}, \au{Teran, J.} \&
    \au{Stomakhin, A.}} \yr{2015}  \at{The affine particle-in-cell method}.
    \jt{ACM Trans. Graph.}  \bvol{34}~(4),  \pg{51}.
  
  \bibitem[Jop {\em et~al.\/}(2005)Jop, Forterre \& Pouliquen]{jop2005crucial}
  {\sc \au{Jop, P.}, \au{Forterre, Y.} \& \au{Pouliquen, O.}} \yr{2005}
    \at{Crucial role of sidewalls in granular surface flows: consequences for the
    rheology}.  \jt{J. Fluid Mech.}  \bvol{541},  \pg{167--192}.
  
  \bibitem[Jop {\em et~al.\/}(2006)Jop, Forterre \&
    Pouliquen]{jop2006constitutive}
  {\sc \au{Jop, P.}, \au{Forterre, Y.} \& \au{Pouliquen, O.}} \yr{2006}  \at{A
    constitutive law for dense granular flows}.  \jt{Nature}  \bvol{441}~(7094),
    \pg{727--30}.
  
  \bibitem[Kamrin(2019)]{kamrin2019non}
  {\sc \au{Kamrin, Ken}} \yr{2019}  \at{Non-locality in granular flow:
    Phenomenology and modeling approaches}.  \jt{Frontiers in Phys.}  \bvol{7},
    \pg{116}.
  
  \bibitem[Kerswell(2005)]{kerswell2005dam}
  {\sc \au{Kerswell, R.R.}} \yr{2005}  \at{Dam break with coulomb friction: A
    model for granular slumping?}  \jt{Phys. Fluids}  \bvol{17}~(5),
    \pg{057101}.
  
  \bibitem[Kl{\'a}r {\em et~al.\/}(2016)Kl{\'a}r, Gast, Pradhana, Fu, Jiang \&
    Teran]{klar2016drucker}
  {\sc \au{Kl{\'a}r, G.}, \au{Gast, Th.}, \au{Pradhana, A.}, \au{Fu,
    C.and~Schroeder, C.}, \au{Jiang, C.} \& \au{Teran, J.}} \yr{2016}
    \at{Drucker--Prager elastoplasticity for sand animation}.  \jt{ACM Trans.
    Graph.}  \bvol{35}~(4),  \pg{1--12}.
  
  \bibitem[Lacaze \& Kerswell(2009)]{lacaze2009axisymmetric}
  {\sc \au{Lacaze, L.} \& \au{Kerswell, R.R.}} \yr{2009}  \at{Axisymmetric
    granular collapse: a transient 3d flow test of viscoplasticity}.  \jt{Phys.
    Rev. Lett.}  \bvol{102}~(10),  \pg{108305}.
  
  \bibitem[Lacaze {\em et~al.\/}(2008)Lacaze, Phillips \&
    Kerswell]{lacaze2008planar}
  {\sc \au{Lacaze, L.}, \au{Phillips, J.C.} \& \au{Kerswell, R.R.}} \yr{2008}
    \at{Planar collapse of a granular column: Experiments and discrete element
    simulations}.  \jt{Phys. Fluids}  \bvol{20}~(6),  \pg{063302}.
  
  \bibitem[Lagr{\'e}e {\em et~al.\/}(2011)Lagr{\'e}e, Staron \&
    Popinet]{lagree2011granular}
  {\sc \au{Lagr{\'e}e, P.-Y.}, \au{Staron, L.} \& \au{Popinet, S.}} \yr{2011}
    \at{The granular column collapse as a continuum: validity of a
    two-dimensional navier--stokes model with a $\mu$ ({I})-rheology}.  \jt{J.
    Fluid Mech.}  \bvol{686},  \pg{378--408}.
  
  \bibitem[Lajeunesse {\em et~al.\/}(2004)Lajeunesse, Mangeney-Castelnau \&
    Vilotte]{lajeunesse2004spreading}
  {\sc \au{Lajeunesse, E.}, \au{Mangeney-Castelnau, A.} \& \au{Vilotte, J.P.}}
    \yr{2004}  \at{Spreading of a granular mass on a horizontal plane}.
    \jt{Phys. Fluids}  \bvol{16}~(7),  \pg{2371--2381}.
  
  \bibitem[Lajeunesse {\em et~al.\/}(2005)Lajeunesse, Monnier \&
    Homsy]{lajeunesse2005granular}
  {\sc \au{Lajeunesse, E.}, \au{Monnier, J.B.} \& \au{Homsy, G.M.}} \yr{2005}
    \at{Granular slumping on a horizontal surface}.  \jt{Phys. Fluids}
    \bvol{17}~(10),  \pg{103302}.
  
  \bibitem[Lube {\em et~al.\/}(2005)Lube, Huppert, Sparks \& Freundt]{lube2005}
  {\sc \au{Lube, Gert}, \au{Huppert, Herbert~E.}, \au{Sparks, R. Stephen~J.} \&
    \au{Freundt, Armin}} \yr{2005}  \at{Collapses of two-dimensional granular
    columns}.  \jt{Phys. Rev. E}  \bvol{72},  \pg{041301}.
  
  \bibitem[Mangeney-Castelnau {\em et~al.\/}(2005)Mangeney-Castelnau, Bouchut,
    Vilotte, Lajeunesse, Aubertin \& Pirulli]{mangeney2005use}
  {\sc \au{Mangeney-Castelnau, A.}, \au{Bouchut, F.}, \au{Vilotte, J.P.},
    \au{Lajeunesse, E.}, \au{Aubertin, A.} \& \au{Pirulli, M.}} \yr{2005}  \at{On
    the use of saint venant equations to simulate the spreading of a granular
    mass}.  \jt{Journal of Geophysical Research: Solid Earth}  \bvol{110}~(B9).
  
  \bibitem[Martin {\em et~al.\/}(2017)Martin, Ionescu, Mangeney, Bouchut \&
    Farin]{martin2017continuum}
  {\sc \au{Martin, N.}, \au{Ionescu, I.R.}, \au{Mangeney, A.}, \au{Bouchut, F.}
    \& \au{Farin, M.}} \yr{2017}  \at{Continuum viscoplastic simulation of a
    granular column collapse on large slopes: $\mu$ ({I}) rheology and lateral
    wall effects}.  \jt{Phys. Fluids}  \bvol{29}~(1),  \pg{013301}.
  
  \bibitem[Mast {\em et~al.\/}(2015)Mast, Arduino, Mackenzie-Helnwein \&
    Miller]{mast2015simulating}
  {\sc \au{Mast, C.M.}, \au{Arduino, P.}, \au{Mackenzie-Helnwein, P.} \&
    \au{Miller, G.R.}} \yr{2015}  \at{Simulating granular column collapse using
    the material point method}.  \jt{Acta Geotechnica}  \bvol{10}~(1),
    \pg{101--116}.
  
  \bibitem[MiDi(2004)]{midi2004dense}
  {\sc \au{MiDi, GDR}} \yr{2004}  \at{On dense granular flows}.  \jt{Eur. Phys.
    J. E}  \bvol{14}~(4).
  
  \bibitem[Miozzi {\em et~al.\/}(2008)Miozzi, Jacob \&
    Olivieri]{miozzi2008performances}
  {\sc \au{Miozzi, M.}, \au{Jacob, B.} \& \au{Olivieri, A.}} \yr{2008}
    \at{Performances of feature tracking in turbulent boundary layer
    investigation}.  \jt{Exper. Fluids}  \bvol{45}~(4),  \pg{765--780}.
  
  \bibitem[Moreau(1994)]{moreau1994some}
  {\sc \au{Moreau, J.J.}} \yr{1994}  \at{Some numerical methods in multibody
    dynamics: application to granular materials}.  \jt{European J. of Mech.
    A/Solids}  \bvol{13}~(4-suppl),  \pg{93--114}.
  
  \bibitem[Moretti {\em et~al.\/}(2012)Moretti, Mangeney, Capdeville, Stutzmann,
    Huggel, Schneider \& Bouchut]{moretti2012numerical}
  {\sc \au{Moretti, L.}, \au{Mangeney, A.}, \au{Capdeville, Y.}, \au{Stutzmann,
    E.}, \au{Huggel, C.}, \au{Schneider, D.} \& \au{Bouchut, F.}} \yr{2012}
    \at{Numerical modeling of the mount steller landslide flow history and of the
    generated long period seismic waves}.  \jt{Geophys. Res. Lett.}
    \bvol{39}~(16).
  
  \bibitem[Mowlavi \& Kamrin(2021)]{mowlavi2021interplay}
  {\sc \au{Mowlavi, Saviz} \& \au{Kamrin, Ken}} \yr{2021}  \at{Interplay between
    hysteresis and nonlocality during onset and arrest of flow in granular
    materials}.  \jt{Soft Matter}  \bvol{17}~(31),  \pg{7359--7375}.
  
  \bibitem[Naaim {\em et~al.\/}(2003)Naaim, Faug \& Naaim-Bouvet]{naaim2003dry}
  {\sc \au{Naaim, M.}, \au{Faug, T.} \& \au{Naaim-Bouvet, F.}} \yr{2003}  \at{Dry
    granular flow modelling including erosion and deposition}.  \jt{Surv.
    Geophys.}  \bvol{24}~(5),  \pg{569--585}.
  
  \bibitem[Narain {\em et~al.\/}(2010)Narain, Golas \& Lin]{narain2010free}
  {\sc \au{Narain, R.}, \au{Golas, A.} \& \au{Lin, M.~C.}} \yr{2010} Free-flowing
    granular materials with two-way solid coupling.  \bt{In {\em ACM Trans.
    Graph.\/}}, ,  \vol{vol.~29},  \pg{p. 173}. ACM.
  
  \bibitem[Perrin {\em et~al.\/}(2019)Perrin, Clavaud, Wyart, Metzger \&
    Forterre]{perrin2019interparticle}
  {\sc \au{Perrin, Hugo}, \au{Clavaud, C{\'e}cile}, \au{Wyart, Matthieu},
    \au{Metzger, Bloen} \& \au{Forterre, Yo{\"e}l}} \yr{2019}  \at{Interparticle
    friction leads to nonmonotonic flow curves and hysteresis in viscous
    suspensions}.  \jt{Phys. Rev. X}  \bvol{9}~(3),  \pg{031027}.
  
  \bibitem[Perrin {\em et~al.\/}(2021)Perrin, Wyart, Metzger \&
    Forterre]{perrin2021nonlocal}
  {\sc \au{Perrin, Hugo}, \au{Wyart, Matthieu}, \au{Metzger, Bloen} \&
    \au{Forterre, Yoel}} \yr{2021}  \at{Nonlocal effects reflect the jamming
    criticality in frictionless granular flows down inclines}.  \jt{Phys. Rev.
    Lett.}  \bvol{126}~(22),  \pg{228002}.
  
  \bibitem[Pouliquen(1999)]{pouliquen1999scaling}
  {\sc \au{Pouliquen, O.}} \yr{1999}  \at{Scaling laws in granular flows down
    rough inclined planes}.  \jt{Phys. Fluids}  \bvol{11}~(3),  \pg{542--548}.
  
  \bibitem[Pouliquen \& Forterre(2002)]{pouliquen2002friction}
  {\sc \au{Pouliquen, O.} \& \au{Forterre, Y.}} \yr{2002}  \at{Friction law for
    dense granular flows: application to the motion of a mass down a rough
    inclined plane}.  \jt{J. Fluid Mech.}  \bvol{453},  \pg{133--151}.
  
  \bibitem[Rauter {\em et~al.\/}(2020)Rauter, Barker \&
    Fellin]{rauter2020granular}
  {\sc \au{Rauter, Matthias}, \au{Barker, Thomas} \& \au{Fellin, Wolfgang}}
    \yr{2020}  \at{Granular viscosity from plastic yield surfaces: the role of
    the deformation type in granular flows}.  \jt{Computers and Geotechnics}
    \bvol{122},  \pg{103492}.
  
  \bibitem[Rousseau \& Ancey(2020)]{rousseau2020scanning}
  {\sc \au{Rousseau, G.} \& \au{Ancey, C.}} \yr{2020}  \at{Scanning {PIV} of
    turbulent flows over and through rough porous beds using refractive index
    matching}.  \jt{Exper. Fluids}  \bvol{61}~(8),  \pg{1--24}.
  
  \bibitem[Russell {\em et~al.\/}(2019)Russell, Johnson, Viroulet, Rocha \&
    Gray]{russell2019retrogressive}
  {\sc \au{Russell, A.S.}, \au{Johnson, C.G. .and~Edwards, A.N.}, \au{Viroulet,
    S.}, \au{Rocha, F.M.} \& \au{Gray, J.M.N.T.}} \yr{2019}  \at{Retrogressive
    failure of a static granular layer on an inclined plane}.  \jt{J. Fluid
    Mech.}  \bvol{869},  \pg{313--340}.
  
  \bibitem[Savage(1979)]{savage1979gravity}
  {\sc \au{Savage, S.~B.}} \yr{1979}  \at{Gravity flow of cohesionless granular
    materials in chutes and channels}.  \jt{J. Fluid Mech.}  \bvol{92}~(1),
    \pg{53--96}.
  
  \bibitem[Shi(1994)]{shi1994good}
  {\sc \au{Shi, Jianbo}} \yr{1994} Good features to track.  \bt{In {\em 1994
    Proceedings of IEEE conference on computer vision and pattern
    recognition\/}},  \pg{pp. 593--600}. IEEE.
  
  \bibitem[Silbert {\em et~al.\/}(2001)Silbert, Erta{\c{s}}, Grest, Halsey,
    Levine \& Plimpton]{silbert2001granular}
  {\sc \au{Silbert, L.E.}, \au{Erta{\c{s}}, D.}, \au{Grest, G.S.}, \au{Halsey,
    T.C.}, \au{Levine, D.} \& \au{Plimpton, S.J.}} \yr{2001}  \at{Granular flow
    down an inclined plane: Bagnold scaling and rheology}.  \jt{Phys. Rev. E}
    \bvol{64}~(5),  \pg{051302}.
  
  \bibitem[Staron \& Hinch(2005)]{staron2005study}
  {\sc \au{Staron, L.} \& \au{Hinch, E.J.}} \yr{2005}  \at{Study of the collapse
    of granular columns using two-dimensional discrete-grain simulation}.  \jt{J.
    Fluid Mech.}  \bvol{545},  \pg{1--27}.
  
  \bibitem[Sulsky {\em et~al.\/}(1995)Sulsky, Zhou \&
    Schreyer]{sulsky1995application}
  {\sc \au{Sulsky, D.}, \au{Zhou, S.-J.} \& \au{Schreyer, H.~L.}} \yr{1995}
    \at{Application of a particle-in-cell method to solid mechanics}.
    \jt{Comput. Phys. communications}  \bvol{87}~(1-2),  \pg{236--252}.
  
  \bibitem[Taberlet {\em et~al.\/}(2003)Taberlet, Richard, Valance, Losert,
    Pasini, Jenkins \& Delannay]{taberlet2003superstable}
  {\sc \au{Taberlet, N.}, \au{Richard, P.}, \au{Valance, A.}, \au{Losert, W.},
    \au{Pasini, J.~M.}, \au{Jenkins, J.~T.} \& \au{Delannay, R.}} \yr{2003}
    \at{Superstable granular heap in a thin channel}.  \jt{Phys. Rev. Lett.}
    \bvol{91}~(26),  \pg{264301}.
  
  \bibitem[Valette {\em et~al.\/}(2019)Valette, Riber, Sardo, Castellani, Costes,
    Vriend \& Hachem]{valette2019sensitivity}
  {\sc \au{Valette, R.}, \au{Riber, S.}, \au{Sardo, L.}, \au{Castellani, R.},
    \au{Costes, F.}, \au{Vriend, N.} \& \au{Hachem, E.}} \yr{2019}
    \at{Sensitivity to the rheology and geometry of granular collapses by using
    the $\mu$ ({I}) rheology}.  \jt{Computers \& Fluids}  \bvol{191},
    \pg{104260}.
  
  \bibitem[Zhang {\em et~al.\/}(2021)Zhang, Su, Lei \& Chen]{zhang2021three}
  {\sc \au{Zhang, R.}, \au{Su, D.}, \au{Lei, G.} \& \au{Chen, X.}} \yr{2021}
    \at{Three-dimensional granular column collapse: Impact of column thickness}.
    \jt{Powder Technol.}  \bvol{389},  \pg{328--338}.
  
  \end{thebibliography}

\end{document}